\documentclass[11pt,letterpaper]{article}
\usepackage[T1]{fontenc}
\usepackage[utf8]{inputenc}
\usepackage{lmodern}
\usepackage[margin=1in]{geometry}
\usepackage{microtype}
\usepackage{amsmath,amssymb,amsfonts,amsthm,bm,mathtools}
\usepackage{graphicx,booktabs,array}
\usepackage[authoryear,round]{natbib}
\usepackage{tikz}
\usetikzlibrary{positioning}
\usepackage{xcolor,enumitem,float,xurl}
\usepackage[font=small,labelfont=bf,justification=raggedright]{caption}
\usepackage{etoc}
\usepackage[hidelinks,unicode]{hyperref}

\hypersetup{
  pdftitle={Bayesian Latent Functional Mediation Analysis with Application to Population-Scale Single-Cell Data},
  pdfauthor={Zehua Li, Yang Ni, Weixuan Zhu},
  pdfsubject={Bayesian latent functional mediation analysis},
  pdfkeywords={Bayesian hierarchical model, Functional mediation, Latent cellular state, Single-cell genomics}
}

\theoremstyle{plain}
\newtheorem{proposition}{Proposition}

\newcommand{\R}{\mathbb{R}}
\newcommand{\E}{\operatorname{E}}
\newcommand{\Var}{\operatorname{Var}}

\newcommand{\diag}{\operatorname{diag}}
\newcommand{\TN}{\operatorname{TN}}
\newcommand{\IG}{\operatorname{IG}}
\newcommand{\GIG}{\operatorname{GIG}}
\newcommand{\ME}{\operatorname{ME}}
\newcommand{\DE}{\operatorname{DE}}
\newcommand{\TE}{\operatorname{TE}}
\newcommand{\IME}{\operatorname{IME}}
\newcommand{\IDE}{\operatorname{IDE}}
\newcommand{\ITE}{\operatorname{ITE}}

\newcommand{\suppfig}[2][]{\includegraphics[#1]{#2}}

\title{Bayesian Latent Functional Mediation Analysis with Application to Population-Scale Single-Cell Data}
\author{%
  Zehua Li$^{1}$, Yang Ni$^{3,*}$, and Weixuan Zhu$^{1,2,*}$\\[0.7em]
  \small\textit{$^{1}$Department of Statistics and Data Science, School of Economics,}\\
  \small\textit{Xiamen University, Xiamen, China}\\[0.3em]
  \small\textit{$^{2}$Wang Yanan Institute for Studies in Economics,}\\
  \small\textit{Xiamen University, Xiamen, China}\\[0.3em]
  \small\textit{$^{3}$Department of Statistics and Data Sciences,}\\
  \small\textit{The University of Texas at Austin, Austin, TX, USA}\\[0.3em]
  \small $^{*}$Corresponding authors
}
\date{}

\begin{document}
\maketitle
\etocdepthtag.toc{main}

\begin{abstract}
Population-scale single-cell RNA sequencing studies pair cell-level gene expression with donor-level genotypes, enabling investigation of genetic regulation across cellular states. Mediation analysis in this setting must account for cells nested within donors and for unobserved cellular states. We propose a Bayesian latent functional mediation model that decomposes the association between genotype and distal gene expression into a component mediated through a candidate cis gene and a remaining direct component. The model represents expression as donor-specific trajectories over a shared latent state coordinate constrained by coarse cell annotations. Joint inference propagates uncertainty in latent states, trajectories, and mediation parameters into state-specific effect functions and their integrated summaries. Simulations show that effect recovery improves with the number of donors and that averaging expression within donors can change the mediated and direct components even when the total association is preserved. In 403,087 CD4+ T cells from 961 OneK1K donors, higher rs1131017 G-allele dosage was associated with higher RPS26 expression across the fitted state domain. The estimated RPS26-mediated component of the association with BAX expression was negative over most of the domain and was partly offset by a positive direct component, yielding a smaller negative integrated total association. In sensitivity analyses conditional on estimated donor trajectories, all three integrated effects retained their signs, with 95\% credible intervals excluding zero, at the five examined coefficient-level residual correlations with $|\rho_s|\leq 0.3$. The mediated and direct effects reversed signs at $\rho_s=-0.5$, while the integrated total association remained negative across all seven tested settings.
\end{abstract}


\section{Introduction}
\label{sec:introduction}

Expression quantitative trait locus (eQTL) studies link genetic
variants to the expression of nearby (cis) and distal (trans) genes.
A possible mechanism for a trans association is a regulatory cascade
in which a variant alters the expression of a cis-gene, which in turn
affects a distal gene. The variant may also influence the distal gene
through pathways that do not involve that cis-gene. Mediation analysis
provides a framework for separating the component transmitted through
a candidate mediator from the remaining direct component
\citep{RobinsGreenland1992,Pearl2001,ImaiKeeleYamamoto2010}.
Genomic mediation studies have used this framework to investigate
cis-gene mediation of trans-eQTL associations and identify candidate
regulatory intermediaries across human tissues
\citep{Pierce2014,Yang2017}. Such analyses help characterize how
genetic variation affects downstream expression programs and
prioritize genes and pathways for experimental investigation.
Mediated and direct components can coexist and have opposite signs,
and hence a small total association may conceal larger contributions from
the individual pathways.

Population-scale single-cell RNA sequencing (scRNA-seq) studies
provide an opportunity to examine these regulatory relationships
across cellular states by pairing cell-level expression measurements
with donor-level genotypes \citep{Kang2018,Yazar2022}.
Genetic associations can vary along cellular states
\citep{Cuomo2020}, suggesting that the contributions of mediated and
direct pathways may also depend on cell state. This setting poses
two related challenges for mediation analysis: cells are nested
within donors and share the same genetic exposure, while their
expression measurements reflect heterogeneous cellular states.
Treating cells as independent observations ignores the donor-level
sampling structure, whereas averaging expression within each donor (i.e., pseudo-bulking) discards
state variation and can change the mediated--direct decomposition,
even when the total association is preserved. Multilevel mediation
methods accommodate hierarchical observations \citep{Preacher2010},
but studying regulation across cellular states additionally requires
a representation of how expression and mediation relationships vary
along a common, unobserved state coordinate.

A natural way to retain this variation is to represent each donor's
mediator and outcome expression as functions of cellular state.
Functional mediation methods have been developed for mediators and
outcomes indexed by an observed functional domain, such as measurement
time \citep{Lindquist2012,Zhao2018}. In single-cell studies, however,
the measurements are cross-sectional snapshots, and each cell's
position along a continuous state coordinate is unknown. Related
single-cell methods infer cell positions from molecular measurements
and biological information
\citep{Trapnell2014,Street2018,Saelens2019}.
Estimating this coordinate first and then treating it as known in a
functional mediation analysis fails to propagate uncertainty in the
cells' latent states; errors in those states can also affect the
estimated trajectories and mediation effects. The consequences of
ignoring uncertainty in inferred cell ordering have been demonstrated
for downstream differential expression analysis \citep{CampbellYau2016}.
These considerations motivate a model that estimates the latent
coordinate together with the expression trajectories and their
mediation relationships.

We propose a Bayesian latent functional mediation (BLaFM) model that
integrates these components in a joint hierarchical fashion.
Building on functional regression \citep{RamsaySilverman2005,Morris2015}
and latent trajectory embedding \citep{Zhou2023}, we model cell-level
mediator and outcome measurements as noisy observations of
donor-specific trajectories over a shared latent state coordinate.
A donor-level hierarchy links genotype to the mediator trajectory and links both
genotype and the mediator trajectory to the outcome trajectory.
This construction retains donors as the independent sampling units
and accommodates unequal cell counts across donors. Coarse cell-state annotations
constrain the latent states and anchor the orientation of the
trajectory. Posterior inference jointly estimates cellular states,
donor-specific trajectories, and population mediation parameters,
propagating their uncertainty into mediated, direct, and total effect
functions and their integrated summaries. These functions describe
how the components of the genotype--outcome relationship vary across
the same standardized cellular-state coordinate. %

Through simulations, we evaluate finite-sample recovery,
compare joint and two-stage estimation, examine sensitivity
to functional basis complexity, and study the consequences
of donor averaging. Our application examines the rs1131017--RPS26--BAX
relationship in 403,087 CD4$^+$ T cells from 961 OneK1K donors
\citep{Yazar2022}. Cross-cohort eQTL evidence motivates the
rs1131017--RPS26 pair, and independent perturbation evidence linking
RPS26 to pro-apoptotic gene expression motivates BAX as the outcome
\citep{Coulter2025distQTL,Chen2021RPS26}. The analysis reports
opposing mediated and direct components whose relative strengths
vary across cellular states and whose partial cancellation yields a
smaller total association. Conditional on estimated donor trajectories,
a residual-dependence analysis examines the sensitivity of this
decomposition to correlation between the mediator and outcome
coefficient residuals.
Together, these analyses illustrate the additional information about
genotype-associated expression patterns provided by resolving
mediation across cellular states.

\section{Model and effect decomposition}
\label{sec:model}
\label{sec:problem_setup}

The BLaFM model connects the cell-level measurements to the donor-level
exposure through latent expression trajectories. Cells from the same
donor inform that donor's mediator and outcome trajectories, while
variation across donors informs the relationships among exposure and
trajectories. We first describe the observation model and the shared
cellular-state coordinate, and then specify the donor-level mediation
hierarchy and derive the resulting effect functions.

\subsection{Cell-level observations and donor-specific trajectories}
\label{subsec:sampling_model}

Let $i=1,\ldots,n$ index donors and $c=1,\ldots,C_i$ index cells
within donor $i$. The observed data are
\[
\mathcal D_i=
\left(Z_i,\{X_{ic},Y_{ic},S_{ic}:c=1,\ldots,C_i\}\right),
\]
where $Z_i$ is the donor-level \emph{exposure}, $X_{ic}$ and $Y_{ic}$
are \emph{mediator} and \emph{outcome} measurements from the same cell, and
$S_{ic}$ is an observed coarse cell-state annotation. In the
application, $Z_i$ is additive genotype dosage, $X_{ic}$ and
$Y_{ic}$ are normalized expression measurements on the SCTransform
Pearson-residual scale \citep{HafemeisterSatija2019}, and $S_{ic}$ is
binary, with $S_{ic}=0$ for naive CD4$^+$ T cells and $S_{ic}=1$ for
central-memory or effector-memory CD4$^+$ T cells. The genetic
exposure is shared by all cells within a donor, and donors are the
independent sampling units.

Let $t_{ic}\in[0,1]$ denote the latent state of cell $c$ from donor
$i$, expressed on a standardized coordinate specific to the analyzed
mediator--outcome gene pair. This coordinate indexes variation in
the paired expression measurements, with its support constrained by
coarse cell-state annotations. For donor $i$, let $g_i(t)$
and $h_i(t)$ describe expected mediator and outcome expression at
state $t$. These trajectories represent expression variation across
cells at different states. We specify
\begin{align}
X_{ic}&=g_i(t_{ic})+\epsilon_{ic},
&\epsilon_{ic}&\sim N(0,\sigma_i^2),\label{eq:x_obs}\\
Y_{ic}&=h_i(t_{ic})+\delta_{ic},
&\delta_{ic}&\sim N(0,\tau_i^2).\label{eq:y_obs}
\end{align}
Both measurements from a cell are evaluated at the same latent
state. The observation errors are mutually independent
conditional on the trajectories, latent states, and donor-specific
variances. The variances $\sigma_i^2$ and $\tau_i^2$ allow residual
variation to differ across donors and between the two genes, while
remaining constant across states within a donor.

Using the common cubic B-spline basis
$\phi(t)=(\phi_1(t),\ldots,\phi_K(t))^\top$, we write
\begin{align*}
g_i(t)&=\phi(t)^\top\boldsymbol\alpha_i,\\
h_i(t)&=\phi(t)^\top\boldsymbol\beta_i.
\end{align*}
The coefficient vectors $\boldsymbol\alpha_i,\boldsymbol\beta_i
\in\mathbb R^K$ provide finite-dimensional representations of the
entire donor-specific curves. Using the same basis for all donors
and both genes allows their trajectory features to be related
through a common regression model. Cubic B-splines provide smooth
curves through overlapping basis functions with local support
\citep{deBoor1978,EilersMarx1996}. The dimension $K$ controls the
flexibility of the representation; its selection is described in
Section~\ref{subsec:basis_selection}.

\subsection{Latent cellular states and annotation constraints}
\label{subsec:latent_state}

The latent cell states $t_{ic}$ are estimated together with the
expression trajectories. We assume that a given value of
$t$ represents a comparable standardized cellular state across
donors, and hence genotype groups can be compared along the same coordinate.
Coarse annotations restrict cells to group-specific regions of the
coordinate without specifying their exact locations within those regions:
\begin{equation}
\mathcal I_{ic}=
\begin{cases}
[0,0.5],&S_{ic}=0,\\
[0.5,1],&S_{ic}=1,
\end{cases}
\qquad
t_{ic}\mid S_{ic}\sim\operatorname{Unif}(\mathcal I_{ic}).
\label{eq:t_prior}
\end{equation}
Conditional on the annotations, the latent cell states are independent
a priori. The uniform distribution assigns equal prior density to
all states within each allowed interval; both mediator and outcome
expression inform the posterior distribution of each cell's latent
state through
Equations~\eqref{eq:x_obs}--\eqref{eq:y_obs}.

These restrictions also orient the shared coordinate. With the
symmetric B-spline basis used here, reflecting all latent states
from $t$ to $1-t$ and reversing the corresponding basis coefficients
leaves the fitted expression values unchanged. Assigning the two
annotation groups to the lower and upper intervals sets an
orientation convention that anchors this global reflection ambiguity.
Further details are
provided in Supplementary Material, Section A.

\subsection{Donor-level mediation hierarchy}
\label{subsec:donor_model}

Mediation is modeled through the donor-specific coefficient vectors.
The mediator trajectory depends on exposure, and the outcome
trajectory depends on both exposure and the mediator trajectory.
We specify
\begin{align}
\boldsymbol\alpha_i
&=\boldsymbol\alpha_0+AZ_i+\boldsymbol\eta_i,
&\boldsymbol\eta_i&\sim N(\boldsymbol0,\Sigma),
\label{eq:alpha_model}\\
\boldsymbol\beta_i
&=\boldsymbol\beta_0+CZ_i+B\boldsymbol\alpha_i+\boldsymbol\xi_i,
&\boldsymbol\xi_i&\sim N(\boldsymbol0,\Lambda),
\label{eq:beta_model}
\end{align}
with $\boldsymbol\alpha_0,\boldsymbol\beta_0,A,C\in\mathbb R^K$
and $B\in\mathbb R^{K\times K}$. The vector
$\boldsymbol\alpha_0$ gives the mean mediator coefficients at
$Z_i=0$, and $A=(a_1,\ldots,a_K)^\top$ describes their change per
unit increase in exposure. The matrix $B$ maps mediator coefficients
to outcome coefficients. The vector $C=(c_1,\ldots,c_K)^\top$
describes the remaining exposure association with the outcome
coefficients when the mediator trajectory is held fixed.
The mean outcome coefficients at $Z_i=0$ are
$\boldsymbol\beta_0+B\boldsymbol\alpha_0$.

The donor deviations $\boldsymbol\eta_i$ and $\boldsymbol\xi_i$
represent between-donor trajectory variation beyond these regression
relationships. The baseline model assumes that they are independent
of each other, independent across donors, and independent of $Z_i$.
Their covariance matrices $\Sigma$ and $\Lambda$ are diagonal, and
we restrict $B=\operatorname{diag}(b_1,\ldots,b_K)$. Thus, each
outcome coefficient depends on the corresponding mediator coefficient,
reducing the number of mediator--outcome regression parameters from
$K^2$ to $K$.

This restriction still allows the relationship to vary
across basis components. Since basis functions overlap, $b_k$ is
an association between trajectory coefficients, not a
mediator--outcome slope at a single state. Likewise, diagonal
coefficient covariance matrices still induce dependence between
expression values at different states through the shared
coefficients. These diagonal restrictions are specific to the chosen
basis: changing $K$ changes both the function space and the
coefficient-level model.

The coefficient map also has a function-on-function representation.
Let
\[
G_\phi=\int_0^1\phi(s)\phi(s)^\top\,ds
\]
be the Gram matrix of the linearly independent basis functions.
This matrix is positive definite, and hence we can define
\begin{equation}
D=G_\phi^{-1}B^\top,
\qquad
\zeta(s,t)=\phi(s)^\top D\phi(t).
\label{eq:kernel_main}
\end{equation}
Then Equation~\eqref{eq:beta_model} is equivalent to
\begin{equation*}
h_i(t)=\phi(t)^\top(\boldsymbol\beta_0+CZ_i)
+\int_0^1\zeta(s,t)g_i(s)\,ds
+\phi(t)^\top\boldsymbol\xi_i.
\end{equation*}
Indeed, the integral equals
$\boldsymbol\alpha_i^\top G_\phi D\phi(t)
=\phi(t)^\top B\boldsymbol\alpha_i$.
Thus the outcome at state $t$ can depend on the mediator trajectory
across states. Because B-spline basis functions
overlap, $G_\phi$ is generally not diagonal, and even a diagonal $B$
can yield $\zeta(s,t)\ne0$ for $s\ne t$. In general, matching basis coefficients
is distinct from specifying a pointwise mediator--outcome slope.
The kernel represents the same coefficient map within the chosen
spline space and introduces no additional parameters.

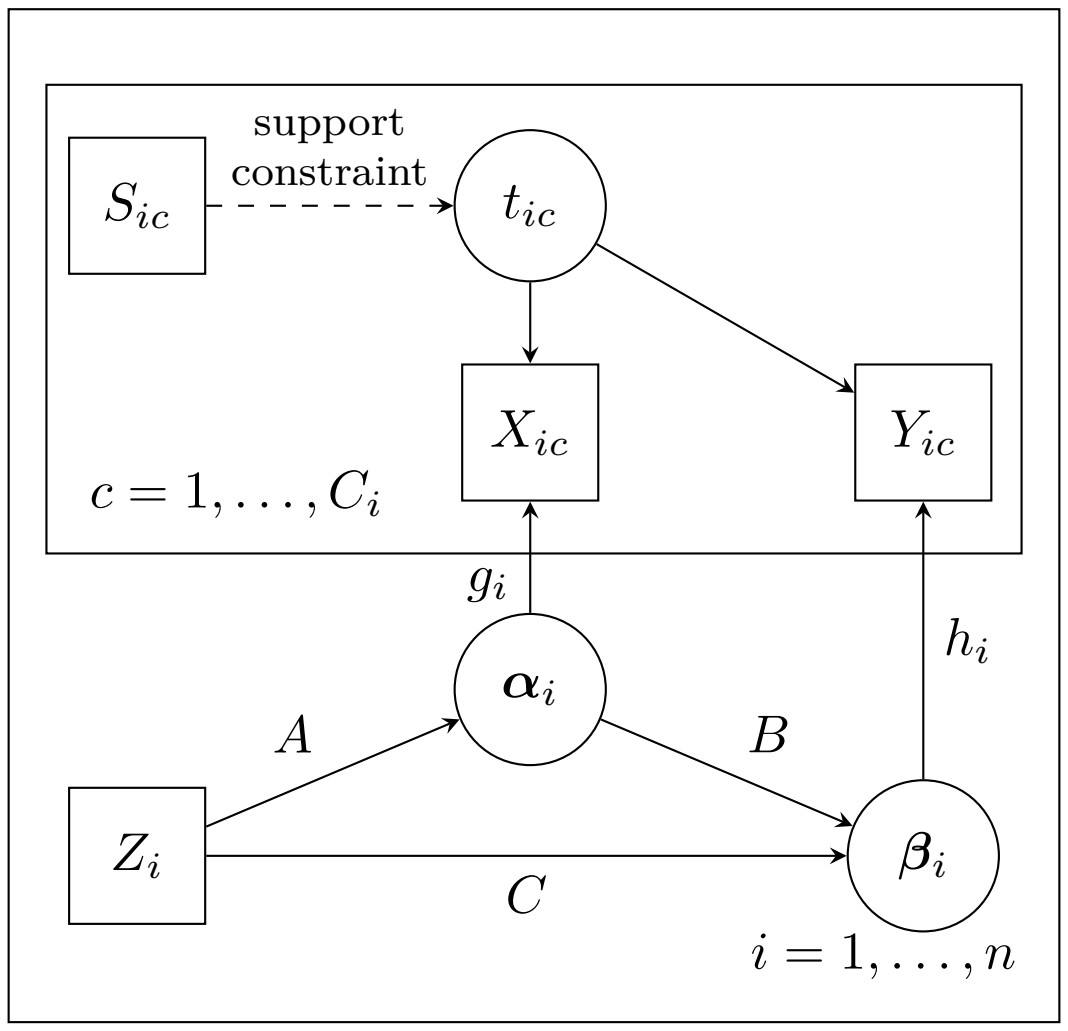
\begin{figure}[htbp]
\centering
\begin{tikzpicture}[>=stealth,
  every node/.style={font=\small},
  obs/.style={rectangle,draw,minimum width=9mm,minimum height=9mm,inner sep=1pt},
  latent/.style={circle,draw,minimum size=10mm}]
\node[obs] (Z) at (0,0) {$Z_i$};
\node[latent] (A) at (2.6,1.1) {$\boldsymbol\alpha_i$};
\node[latent] (B) at (5.2,0) {$\boldsymbol\beta_i$};
\node[obs] (X) at (2.6,2.8) {$X_{ic}$};
\node[obs] (Y) at (5.2,2.8) {$Y_{ic}$};
\node[latent] (T) at (2.6,4.3) {$t_{ic}$};
\node[obs] (S) at (0,4.3) {$S_{ic}$};
\draw[->] (Z)--node[above left] {$A$}(A);
\draw[->] (Z)--node[below] {$C$}(B);
\draw[->] (A)--node[above right] {$B$}(B);
\draw[->] (A)--node[left,pos=0.25] {$g_i$}(X);
\draw[->] (B)--node[right] {$h_i$}(Y);
\draw[->] (T)--(X);
\draw[->] (T)--(Y);
\draw[->,dashed] (S)--node[above,align=center,font=\scriptsize] {support\\constraint}(T);
\draw (-0.6,2.0) rectangle (5.85,5.1);
\node[anchor=south west] at (-0.45,2.1) {$c=1,\ldots,C_i$};
\draw (-0.85,-1.1) rectangle (6.1,5.6);
\node[anchor=south east] at (5.95,-0.95) {$i=1,\ldots,n$};
\end{tikzpicture}
\caption{Hierarchical structure of the BLaFM model.
Squares denote observed variables and circles denote latent variables.
The outer plate indexes donors ($i=1,\ldots,n$), and the inner plate
indexes cells within donor $i$ ($c=1,\ldots,C_i$).
Genotype and trajectory coefficients are shared by cells from the
same donor. Each cell's mediator and outcome measurements are
evaluated at its latent state. The dashed arrow represents the
annotation-informed restriction on the latent-state support.}
\label{fig:causal_diagram}
\end{figure}

Within donor $i$, the vector $\boldsymbol\alpha_i$ determines the
mediator trajectory $g_i$, and $\boldsymbol\beta_i$ determines the
outcome trajectory $h_i$. Every cell from that donor is an observation
of these two trajectories at its own latent state $t_{ic}$, with
cell-specific observation errors. All of the donor's cells therefore
jointly inform $\boldsymbol\alpha_i$ and $\boldsymbol\beta_i$,
while the population hierarchy shares information across donors.
Figure~\ref{fig:causal_diagram}
summarizes how the cell-level observations, latent states, and
donor-level regression paths are connected.

\subsection{State-specific effects and integrated summaries}
\label{subsec:tv_effects}

The hierarchy yields a decomposition of the exposure association
with the outcome trajectory. Substituting
Equation~\eqref{eq:alpha_model} into Equation~\eqref{eq:beta_model}
gives
\[
\boldsymbol\beta_i
=\boldsymbol\beta_0+B\boldsymbol\alpha_0
+(BA+C)Z_i+B\boldsymbol\eta_i+\boldsymbol\xi_i.
\]
The coefficient $BA+C$ therefore describes the total change in
mean outcome coefficients per unit increase in exposure. Taking
expectations over the donor deviations gives the population
conditional mean trajectories
\begin{align*}
\mu_X(t\mid z)
&=\phi(t)^\top(\boldsymbol\alpha_0+Az),
\\
\mu_Y(t\mid z)
&=\phi(t)^\top\{\boldsymbol\beta_0+Cz+
B(\boldsymbol\alpha_0+Az)\}.
\end{align*}

For an exposure contrast $\Delta z=z_1-z_0$, the mean mediator
coefficient vector changes by $A\Delta z$. Passing this change
through the mediator--outcome map $B$ gives a change of
$BA\Delta z$ in outcome coefficients. The separate exposure path
contributes $C\Delta z$. Evaluating these coefficient changes at
state $t$ yields the mediated, direct, and total effect functions:
\begin{align}
\mathrm{ME}(t;z_1,z_0)&=\phi(t)^\top BA\,\Delta z,
\label{eq:me_curve}\\
\mathrm{DE}(t;z_1,z_0)&=\phi(t)^\top C\,\Delta z,
\label{eq:de_curve}\\
\mathrm{TE}(t;z_1,z_0)&=\phi(t)^\top(BA+C)\,\Delta z.
\label{eq:te_curve}
\end{align}
Thus $\mathrm{TE}(t;z_1,z_0)=\mu_Y(t\mid z_1)-\mu_Y(t\mid z_0)$
and $\mathrm{TE}=\mathrm{ME}+\mathrm{DE}$ at each state.
The mediated component operates through a change in the entire
mediator trajectory represented by its coefficients; the direct
component collects the exposure association not transmitted through
that trajectory. Under diagonal $B$, the mediated effect is
$\sum_{k=1}^K\phi_k(t)b_ka_k\Delta z$. Here $a_k$ describes the
exposure association with mediator coefficient $\alpha_{ik}$, and
$b_k$ describes the association of $\alpha_{ik}$ with outcome
coefficient $\beta_{ik}$ conditional on exposure. Their product
$b_ka_k\Delta z$ is the mediated contribution through basis
component $k$; weighting these contributions by $\phi_k(t)$ and
summing over $k$ gives the mediated effect at state $t$.

Unless indicated otherwise, we report a unit exposure contrast and
write $\mathrm{ME}(t)$, $\mathrm{DE}(t)$, and $\mathrm{TE}(t)$.
In the application, this contrast corresponds to one additional
G allele at rs1131017.
Mediated and direct
effects may reinforce or oppose one another, and their relative
strengths may vary with $t$. Examining the component curves can
therefore reveal substantial associations that cancel in the total
curve.

For an overall summary on the common state scale, we integrate
uniformly over the state coordinate:
\begin{equation}
\mathrm{IME}=\int_0^1\mathrm{ME}(t)\,dt,\quad
\mathrm{IDE}=\int_0^1\mathrm{DE}(t)\,dt,\quad
\mathrm{ITE}=\mathrm{IME}+\mathrm{IDE}.
\label{eq:integrated_effects_main}
\end{equation}
These scalar summaries are convenient but do not replace
the effect curves: positive and negative portions of an individual
curve can cancel upon integration, in addition to cancellation
between mediated and direct components at the same state.

We use Rubin's potential-outcomes framework to define natural direct
and indirect effects for expression trajectories
\citep{RobinsGreenland1992,Pearl2001}. Let $\boldsymbol\alpha_i(z)$
denote the mediator coefficients donor $i$ would have under exposure
$z$. Let $H_i^{z,\boldsymbol a}(t)$ denote that donor's outcome
expression trajectory evaluated at state $t$ if exposure were set to
$z$ and the entire mediator trajectory were set to
$\phi(\cdot)^\top\boldsymbol a$. The structural equations for these
two quantities are
\begin{align}
\boldsymbol\alpha_i(z)
&=\boldsymbol\alpha_0+Az+\boldsymbol\eta_i,
\label{eq:potential_alpha_main}\\
H_i^{z,\boldsymbol a}(t)
&=\phi(t)^\top
  (\boldsymbol\beta_0+Cz+B\boldsymbol a+\boldsymbol\xi_i).
\label{eq:potential_outcome_main}
\end{align}
Substituting $\boldsymbol\alpha_i(z')$ for $\boldsymbol a$ gives
the \emph{nested potential outcome} $H_i^{z,\boldsymbol\alpha_i(z')}(t)$.
Here $z$ sets the exposure in the outcome equation, while $z'$
determines which of the donor's potential mediator trajectories
enters that equation. In particular, $H_i^{z_1,\boldsymbol\alpha_i(z_0)}(t)$
provides the intermediate comparison needed to decompose the
total effect. Changing the exposure from $z_0$ to $z_1$ while
holding the mediator at its $z_0$ trajectory gives the direct
contrast. Then changing the mediator to its $z_1$ trajectory
while holding exposure at $z_1$ gives the indirect contrast.
Averaging these contrasts over donors at a fixed state $t$ gives
the natural indirect, direct, and total effects
\begin{align*}
\operatorname{NIE}(t;z_1,z_0)
&=\mathbb E\!\left[
H_i^{z_1,\boldsymbol\alpha_i(z_1)}(t)
-H_i^{z_1,\boldsymbol\alpha_i(z_0)}(t)\right],
\\
\operatorname{NDE}(t;z_1,z_0)
&=\mathbb E\!\left[
H_i^{z_1,\boldsymbol\alpha_i(z_0)}(t)
-H_i^{z_0,\boldsymbol\alpha_i(z_0)}(t)\right],
\\
\operatorname{NTE}(t;z_1,z_0)
&=\mathbb E\!\left[
H_i^{z_1,\boldsymbol\alpha_i(z_1)}(t)
-H_i^{z_0,\boldsymbol\alpha_i(z_0)}(t)\right].
\end{align*}

\begin{proposition}
\label{prop:functional_natural_effects_main}
Suppose the potential trajectories satisfy the structural equations
\eqref{eq:potential_alpha_main}--\eqref{eq:potential_outcome_main} on a common cellular-state coordinate. Under assumptions
(A1)--(A8) in Supplementary Material, Section~A, the model-based
mediated, direct, and total effects coincide with the corresponding
functional natural effects for the exposure contrast $(z_1,z_0)$,
\begin{align*}
\operatorname{NIE}(t;z_1,z_0)&=\mathrm{ME}(t;z_1,z_0),\\
\operatorname{NDE}(t;z_1,z_0)&=\mathrm{ME}(t;z_1,z_0),\\
\operatorname{NTE}(t;z_1,z_0)&=\mathrm{TE}(t;z_1,z_0).\\
\end{align*}
\end{proposition}

This result follows by substitution into the structural equations.
Interpreting the fitted decomposition causally requires consistency,
positivity, no interference across donors, exchangeability for the
exposure and mediator--outcome relations, and absence of exposure-induced
mediator--outcome confounding
\citep{RobinsGreenland1992,Pearl2001,ImaiKeeleYamamoto2010}.
In this setting we also assume that the functional model adequately
describes the expression trajectories and their relationships, and
that a given $t$ represents a comparable cellular state across
donors and exposure levels.
Supplementary Material,
Section A gives the full assumptions and proof.

\subsection{Prior specification}
\label{subsec:prior_model}

The remaining priors describe variation at the cell and donor levels
and regularize the population regression coefficients. We assign
independent inverse-gamma priors to the observation variances
$\sigma_i^2,\tau_i^2$ and to the diagonal entries
$\Sigma_{kk},\Lambda_{kk}$ of the donor-level covariance matrices.
For all four variance families, we use
$\operatorname{IG}(2,0.1)$ with the shape--scale parameterization.
The observation variances govern scatter around each donor's
trajectories, whereas the coefficient variances govern departures
from the population regression relationships.

Within basis component $k$, collect the mediator and outcome
regression coefficients into
\[
\boldsymbol d_k^{(\alpha)}=(\alpha_{0k},a_k)^\top,\qquad
\boldsymbol d_k^{(\beta)}=(\beta_{0k},c_k,b_k)^\top.
\]
We assign the conditional Gaussian priors
\begin{align*}
\boldsymbol d_k^{(\alpha)}\mid\Sigma_{kk},\tau_a
&\sim N(\boldsymbol0,\tau_a\Sigma_{kk}I_2),
\\
\boldsymbol d_k^{(\beta)}\mid\Lambda_{kk},\tau_d
&\sim N(\boldsymbol0,\tau_d\Lambda_{kk}I_3),
\end{align*}
independently across blocks conditional on their variances and shared
scales. The variance multipliers $\tau_a$ and $\tau_d$ control prior
dispersion across the mediator and outcome regressions, respectively.
Smaller values concentrate the coefficients more strongly around
zero, relative to the corresponding donor-level residual variance.
We use independent $\operatorname{Gamma}(10,0.1)$ priors for these
multipliers, with the shape--rate parameterization. The same prior
specification is used in the simulation and application analyses,
including the conditional residual-correlation sensitivity analysis.
Supplementary Material, Section B, gives the distributional
parameterizations and posterior computation details.

\section{Posterior inference}
\label{sec:posterior}

For a fixed basis dimension $K$, we sample jointly from the
posterior of the latent cellular states, donor-specific trajectory
coefficients, and population-level parameters. We first describe
the sampler and the resulting effect summaries, and then give the
criterion used to select $K$. Throughout, let
$\mathcal D=\{\mathcal D_i\}_{i=1}^n$ denote the observed data.

\subsection{Posterior sampling}
\label{subsec:posterior_sampling}

We use a Metropolis--within--Gibbs sampler for the model in
Section~\ref{sec:model}. Conditional on the current latent states,
the basis evaluations $\phi(t_{ic})$ are known, and the observation
and donor-level models form a Gaussian regression hierarchy.
Let $\boldsymbol x_i=(X_{i1},\ldots,X_{iC_i})^\top$ and
$\boldsymbol y_i=(Y_{i1},\ldots,Y_{iC_i})^\top$, and let
$\Phi_i$ be the $C_i\times K$ matrix with $c$th row
$\phi(t_{ic})^\top$. Write $F_i=\Phi_i^\top\Phi_i$ and define
\[
\boldsymbol m_{\alpha i}=\boldsymbol\alpha_0+AZ_i,
\qquad
\boldsymbol m_{\beta i}=\boldsymbol\beta_0+CZ_i.
\]
The full conditional distribution of the two trajectory coefficient
vectors is
\begin{equation*}
\begin{pmatrix}\boldsymbol\alpha_i\\\boldsymbol\beta_i\end{pmatrix}
\,\bigg|\,-\ \sim
N\!\left(Q_i^{-1}\boldsymbol q_i,Q_i^{-1}\right),
\end{equation*}
where $-$ denotes the observed data and all remaining unknown
quantities, and
\begin{align*}
Q_i&=
\begin{pmatrix}
\Sigma^{-1}+\sigma_i^{-2}F_i+B^\top\Lambda^{-1}B
&-B^\top\Lambda^{-1}\\
-\Lambda^{-1}B&\Lambda^{-1}+\tau_i^{-2}F_i
\end{pmatrix},
\\
\boldsymbol q_i&=
\begin{pmatrix}
\sigma_i^{-2}\Phi_i^\top\boldsymbol x_i
+\Sigma^{-1}\boldsymbol m_{\alpha i}
-B^\top\Lambda^{-1}\boldsymbol m_{\beta i}\\
\tau_i^{-2}\Phi_i^\top\boldsymbol y_i
+\Lambda^{-1}\boldsymbol m_{\beta i}
\end{pmatrix}.
\end{align*}
The off-diagonal blocks of $Q_i$ encode the dependence between
mediator and outcome coefficients induced by $B$. Consequently,
both expression measurements and the population mediation hierarchy
inform the donor's trajectories in a single update. The terms
$\sigma_i^{-2}F_i$ and $\tau_i^{-2}F_i$ summarize the information
from that donor's cells, while the remaining precision terms
come from the donor-level hierarchy. We use a Cholesky factorization
to sample this $2K$-dimensional Gaussian block without explicitly
forming $Q_i^{-1}$.

The remaining regression and variance parameters also permit Gibbs
updates. The population coefficient blocks
$\boldsymbol d_k^{(\alpha)}$ and $\boldsymbol d_k^{(\beta)}$
have multivariate normal full conditionals. The observation
variances $\sigma_i^2,\tau_i^2$ and coefficient variances
$\Sigma_{kk},\Lambda_{kk}$ have inverse-gamma full conditionals,
while the shared prior variance multipliers $\tau_a,\tau_d$
have generalized inverse Gaussian full conditionals.

We update each latent state using a Metropolis--Hastings step.
Its full conditional density is
\begin{equation*}
p(t_{ic}\mid-)
\propto \mathbf{1}\{t_{ic}\in\mathcal I_{ic}\}
\exp\left\{
-\frac{[X_{ic}-g_i(t_{ic})]^2}{2\sigma_i^2}
-\frac{[Y_{ic}-h_i(t_{ic})]^2}{2\tau_i^2}
\right\}.
\end{equation*}
Both genes inform the latent state, subject to the support
constraint supplied by the cell annotation. We propose a new state
from a normal distribution centered at the current value and
truncated to $\mathcal I_{ic}$. The acceptance probability includes
the proposal-density ratio, since truncation makes the random-walk
proposal asymmetric near the interval boundaries. Alternating
these state updates with the coefficient and variance updates
propagates uncertainty in the state coordinate to the trajectories
and mediation parameters. Full conditional distributions and
implementation settings are provided in Supplementary Material,
Section B.

\subsection{Posterior effect summaries}
\label{subsec:posterior_effects}

For each retained posterior draw $m$, we evaluate
Equations~\eqref{eq:me_curve}--\eqref{eq:te_curve} on a common grid
over $[0,1]$. For a unit exposure contrast, the mediated effect is
$\mathrm{ME}^{(m)}(t)=\phi(t)^\top B^{(m)}A^{(m)}$;
the direct and total effects are computed from the same draw.
Forming products and sums before summarization preserves the
posterior dependence among $A$, $B$, and $C$. Although the effect
formulas involve only these population parameters, their posterior
distribution incorporates uncertainty in the latent states and
donor-specific trajectories through the joint model.

At each grid point, we report the posterior mean and the
2.5th and 97.5th percentiles of the sampled effect values.
These give pointwise 95\% credible intervals for the population
effect functions. We also integrate each sampled effect curve as in
Equation~\eqref{eq:integrated_effects_main}, obtaining posterior
draws of $\mathrm{IME}$, $\mathrm{IDE}$, and $\mathrm{ITE}$.
Their posterior means and credible intervals summarize the
overall effects and their uncertainty on the common state scale.

Under the identification assumptions in
Section~\ref{subsec:tv_effects}, the same posterior draws provide
model-based estimates of the natural indirect, direct, and total
effect functions for a unit exposure contrast.
Their uniformly integrated summaries are therefore given by
$\mathrm{IME}$, $\mathrm{IDE}$, and $\mathrm{ITE}$, respectively.

\subsection{Selection of the basis dimension}
\label{subsec:basis_selection}

The basis dimension controls trajectory flexibility. Because the
effects compare trajectories at a common state, we select $K$
using a conditional plug-in criterion that evaluates expression
prediction at estimated states. Integrating a held-out cell's state
over its annotation interval would instead assess a mixture of
expression distributions, which can fit well without preserving
the correspondence between expression and state.

A simple example illustrates the distinction. Suppose
$T\sim\operatorname{Unif}[u,v]$, where $\ell=v-u>0$ is the
interval length, and $X=T+\epsilon$, with
$\epsilon\sim N(0,\sigma^2)$ independent of $T$.
The true trajectory is therefore $g_0(t)=t$. Compare a shifted trajectory
$g_P(t)=t+d$, with $d>0$, and a reflected trajectory
$g_Q(t)=u+v-t$. Evaluating both candidates at the same realized
state $T$ gives the expected conditional log-score difference
\begin{equation*}
\mathbb E_0\!\left[
\log\frac{f_N\{X;g_P(T),\sigma^2\}}
{f_N\{X;g_Q(T),\sigma^2\}}\right]
=\frac{\ell^2/3-d^2}{2\sigma^2}.
\end{equation*}
Here $f_N(x;\mu,s^2)$ is the normal density with mean $\mu$ and
variance $s^2$, and $\mathbb E_0$ averages under the true joint
distribution. The shifted and reflected trajectories have mean
squared deviations $d^2$ and $\ell^2/3$ from $g_0(T)$,
respectively. Consequently, the conditional score favors $P$
when $d<\ell/\sqrt3$.
After integrating over
$T$, however, reflection preserves the uniform distribution within
the interval: the marginal density under $Q$ equals the true density
$p_0(x)$, whereas that under $P$ is $p_0(x-d)$.
Marginal log scoring therefore favors $Q$, even though it reverses
the expression--state relationship within the annotation interval.
Conditional scoring assesses the expression--state relationship
by evaluating both candidates at the same state for each cell, whereas marginal scoring assesses the expression distribution after
averaging over states. Supplementary Material, Section B provides the
marginal-score calculation and a graphical illustration.

Let
$\boldsymbol t=(t_{ic})_{i,c}$ be the collection of latent states
and collect all remaining unknown quantities into
\[
\begin{aligned}
\theta=\bigl(&\{\boldsymbol\alpha_i,\boldsymbol\beta_i,
\sigma_i^2,\tau_i^2\}_{i=1}^n,
\boldsymbol\alpha_0,\boldsymbol\beta_0,A,B,C,
\Sigma,\Lambda,\tau_a,\tau_d\bigr).
\end{aligned}
\]
For each candidate $K\in\mathcal K$, we first fit the joint model
$p(\theta,\boldsymbol t\mid\mathcal D,K)$ to the full data and
obtain
\begin{equation*}
\widehat{\boldsymbol t}_K=(\widehat t_{ic,K})_{i,c}
\text{~~with~~}
\widehat t_{ic,K}=\operatorname{E}(t_{ic}\mid\mathcal D,K).
\end{equation*}
Each candidate has its own estimated state vector. This vector
is held fixed throughout the subsequent refits and scoring for
that candidate.

We partition cells into five folds, stratified within donor and
coarse cell-state group, and use the same fold assignments for
all candidates. Let $\mathcal F_f$ contain the cell indices in
fold $f$, and let $\mathcal D_{-f}$ denote the data with the paired
expression measurements $\{(X_{ic},Y_{ic}):(i,c)\in\mathcal F_f\}$
removed; donor exposures and cell annotations remain observed.
The refit excluding fold $f$ targets the conditional posterior
\begin{equation*}
\begin{aligned}
\pi_{K,-f}(\theta)
&=p\!\left(\theta\mid\mathcal D_{-f},
\boldsymbol t=\widehat{\boldsymbol t}_K,K\right)\\
&\propto p(\theta\mid\boldsymbol Z,K)
\prod_{(i,c)\notin\mathcal F_f}
p\!\left(X_{ic},Y_{ic}\mid\widehat t_{ic,K},\theta,K\right),
\end{aligned}
\end{equation*}
where $\boldsymbol Z=(Z_1,\ldots,Z_n)^\top$ and
$p(\theta\mid\boldsymbol Z,K)$ is the hierarchical prior,
including the donor coefficient models conditional on exposure.
Thus the refit samples every component of $\theta$ using only the
training expression likelihood, while fixing
$t_{ic}=\widehat t_{ic,K}$ for every cell. In particular, all
basis evaluations $\phi(\widehat t_{ic,K})$ are fixed, but the
donor-specific trajectory coefficients and variances are updated.

Let $\theta_{K,-f}^{(m)}$, $m=1,\ldots,M_f$, denote the retained
MCMC draws targeting $\pi_{K,-f}$. For a held-out cell $(i,c)$
in fold $f=f(ic)$, we evaluate its sampling density at the same
fixed state $\widehat t_{ic,K}$ and calculate the log predictive density,
\begin{equation*}
\widehat{\operatorname{lpd}}^{XY}_{ic,K}
=\log\left\{\frac1{M_f}\sum_{m=1}^{M_f}
p\!\left(X_{ic},Y_{ic}\mid
\widehat t_{ic,K},\theta_{K,-f}^{(m)},K\right)\right\}.
\end{equation*}
Here $p$ is the joint sampling density from
Equations~\eqref{eq:x_obs}--\eqref{eq:y_obs}.
The state $\widehat t_{ic,K}$ does not vary across draws, whereas
the coefficients and variances entering the density come from
$\theta_{K,-f}^{(m)}$. The density therefore retains observation
noise and averages over uncertainty in $\theta$ conditional on
the fixed states.
Each cell is scored once in its held-out fold. We select
\begin{equation*}
\widehat K=\min\operatorname*{arg\,max}_{K\in\mathcal K}
\left\{\frac1n\sum_{i=1}^n\frac1{C_i}
\sum_{c=1}^{C_i}\widehat{\operatorname{lpd}}^{XY}_{ic,K}\right\},
\end{equation*}
which averages cell scores within each donor and weights donors
equally.%

The use of full-data reference states addresses the fact that exact
cellular states are not independently observed. For a held-out cell
$(i,c)\in\mathcal F_f$, both expression measurements are absent
from $\mathcal D_{-f}$. If this state is left unknown, the
independent prior in Equation~\eqref{eq:t_prior} implies
\[
t_{ic}\mid\mathcal D_{-f},K
\sim\operatorname{Unif}(\mathcal I_{ic}).
\]
A fully held-out predictive density consequently averages over
this interval and evaluates the expression mixture described
above. In contrast, using $\widehat{\boldsymbol t}_K$ provides
one fitted assignment of cells to states for all folds of a given
$K$. The conditional refits assess the trajectory representation
on this fixed assignment, omitting each scored cell's expression
likelihood from the corresponding refit. This targets the
expression--state relationship that underlies the effect curves.

We use this procedure as a practical tuning criterion for
state-specific analysis within the observed donors. Because the
reference states use the full data, held-out expression still
contributes indirectly through $\widehat{\boldsymbol t}_K$.
The scores therefore do not estimate independent predictive
performance of the entire procedure. The example above motivates
preserving the expression--state relationship during scoring, but does not establish
optimality of this criterion when states are estimated separately
for each candidate dimension.

After selecting
$\widehat K$, we obtain effect summaries from
$p(\theta,\boldsymbol t\mid\mathcal D,\widehat K)$, with all
expression measurements included and the states again treated as
unknown. Thus fixing the states is only a device for basis
selection. The resulting credible intervals condition on
$\widehat K$ and do not include uncertainty in basis selection.

\section{Simulation studies}
\label{sec:simulation}

Three experiments assess recovery with increasing donor sample
size, sensitivity to the working representation, and changes
induced by donor averaging. An additional comparison of joint and two-stage estimation
reuses the finite-sample datasets at $n=1000$.
Recovery and predictive criteria
are summarized over $R=10$ independent datasets per experiment.
Complete data-generating mechanisms, evaluation definitions,
cross-validation construction, and additional results are provided
in Supplementary Material, Section C.

\subsection{Finite-sample recovery}
\label{subsec:finite_sample}

For each of $R=10$ independent master replicates, we generated
1000 donors. Exposures followed $Z_i\sim N(0,1)$, independently of
cell counts $C_i\sim\operatorname{DiscreteUniform}\{10,\ldots,200\}$.
Randomly permuting the donor indices produced nested samples
of size $n=200,500,1000$, retaining all cells from each donor.
Both data generation and fitting used $K=5$. True cell
positions formed equally spaced grids on $[0,1]$; fitting
treated them as latent and used only their lower--upper interval
restrictions.

Coefficient recovery was summarized by block RMSE for
$A$, $\operatorname{diag}(B)$, and $C$, while recovery of
the mediated, direct, and total effect functions was evaluated
by root integrated squared error (RISE). Results were summarized
over the $R=10$ independent master replicates, with Monte Carlo
standard errors reported in parentheses.

\begin{table}[htbp]
\centering
\scriptsize
\renewcommand{\arraystretch}{1.12}
\setlength{\tabcolsep}{4.5pt}
\caption{Finite-sample recovery under the fixed working basis
dimension $K=5$. Coefficient RMSE pools squared errors across the
10 simulation replicates and the $K=5$ components within each
parameter block; its Monte Carlo standard error is obtained using a
delete-one-replicate jackknife. Effect-function RISE values are means
across replicates, with Monte Carlo standard errors in parentheses.
All RMSE and RISE values are multiplied by $10^2$.}
\label{tab:finite_sample_main}
\begin{tabular}{ccccccc}
\toprule
&
\multicolumn{3}{c}{Coefficient RMSE}
&
\multicolumn{3}{c}{Effect-function RISE}
\\
\cmidrule(lr){2-4}
\cmidrule(lr){5-7}
$n$
& $A$
& $\operatorname{diag}(B)$
& $C$
& $\mathrm{ME}$
& $\mathrm{DE}$
& $\mathrm{TE}$
\\
\midrule

$200$
& $1.063\;(0.079)$
& $6.819\;(0.515)$
& $5.843\;(0.350)$
& $4.130\;(0.440)$
& $4.093\;(0.403)$
& $0.628\;(0.031)$
\\

$500$
& $0.826\;(0.085)$
& $4.014\;(0.284)$
& $3.754\;(0.216)$
& $2.389\;(0.260)$
& $2.420\;(0.234)$
& $0.507\;(0.032)$
\\

$1000$
& $0.676\;(0.037)$
& $3.384\;(0.258)$
& $3.157\;(0.200)$
& $1.863\;(0.227)$
& $1.908\;(0.213)$
& $0.470\;(0.014)$
\\

\bottomrule
\end{tabular}
\end{table}

Table~\ref{tab:finite_sample_main} shows improved block-level
recovery with increasing donor sample size. From $n=200$ to
$n=1000$, RMSE decreased from 0.0106 to 0.0068 for $A$, from
0.0682 to 0.0338 for $\operatorname{diag}(B)$, and from 0.0584
to 0.0316 for $C$. The mean RISE approximately halved for the
mediated and direct effects, while the total-effect function
was more accurately recovered at every sample size.

\subsection{Sensitivity to functional basis complexity}
\label{subsec:basis_robustness}

We generated $R=10$ independent datasets with $n=500$ donors.
The functional truths were defined analytically with smooth
donor-specific Fourier deviations, independently of the B-spline
representation used for fitting. Thus none of the candidate
dimensions represented a data-generating value of $K$. Each
dataset was fitted with $K=4,\ldots,9$, using the same dataset
across dimensions within each replicate.

For each candidate dimension, we evaluated full-model recovery of
the mediated, direct, and total effect functions using root
integrated squared error (RISE). The effect estimates were obtained
from the full-data latent-state model, in which cell states
and the remaining parameters were jointly estimated. These
results assess how the working basis affects recovery of the
effect functions that are the primary targets of the analysis.

Basis selection used the conditional plug-in criterion in
Section~\ref{subsec:basis_selection}. Within each replicate,
all candidates used common five-fold assignments, their own
full-data state estimates, and donor-balanced joint scores.
The maximizing dimension $\widehat K_r$ was selected separately
in each replicate. The reported effect-function errors use the
corresponding full latent-state fits.

\begin{table}[htbp]
\centering
\small
\renewcommand{\arraystretch}{1.12}
\setlength{\tabcolsep}{5pt}
\caption{Full-model effect recovery and conditional plug-in basis
selection over 10 paired simulation replicates. Entries are means
with simulation Monte Carlo standard errors in parentheses.
All RISE values are calculated from full-data fits with latent states
and multiplied by $10^2$; lower is better.
Conditional CVLPD is the donor-balanced joint score evaluated
at fixed, candidate-specific state estimates; higher is
better.
Boldface marks the best mean among the fixed-$K$ analyses.}
\label{tab:basis_robustness_main}
\begin{tabular}{ccccc}
\toprule
&\multicolumn{3}{c}{Full-model effect-function RISE}
&Conditional\\
\cmidrule(lr){2-4}
$K$&$\mathrm{ME}$&$\mathrm{DE}$&$\mathrm{TE}$&CVLPD\\
\midrule
4&$10.71\;(0.90)$&$7.97\;(0.86)$&$5.17\;(0.26)$
 &$1.5843\;(0.0045)$\\
5&$\mathbf{5.81\;(0.65)}$&$\mathbf{5.62\;(0.46)}$
 &$\mathbf{3.04\;(0.26)}$&$\mathbf{1.6247\;(0.0060)}$\\
6&$11.60\;(1.02)$&$9.78\;(0.94)$&$3.91\;(0.27)$
 &$1.6187\;(0.0077)$\\
7&$12.42\;(0.55)$&$12.00\;(0.57)$&$3.98\;(0.35)$
 &$1.4134\;(0.0116)$\\
8&$13.26\;(1.68)$&$12.47\;(1.41)$&$3.94\;(0.24)$
 &$1.5611\;(0.0088)$\\
9&$17.53\;(0.88)$&$17.02\;(1.09)$&$5.03\;(0.31)$
 &$1.3824\;(0.0179)$\\
\bottomrule
\end{tabular}
\end{table}

Among the examined representations, $K=5$ gave both the smallest
mean full-model RISE for all three effect functions and the
largest mean conditional CVLPD
(Table~\ref{tab:basis_robustness_main}). Increasing the basis
dimension beyond $K=5$ did not improve mean effect recovery
in this experiment, with larger changes in ME and DE errors
than in TE errors. The conditional scores at $K=5$ and $K=6$
were close: their paired mean difference, $K=5$ minus $K=6$,
was 0.0059, with simulation MCSE 0.0089.

\subsection{Consequences of donor-level aggregation}
\label{subsec:aggregation_stress}

We generated 10 independent datasets with $n=1000$ donors,
standard-normal exposure, and the same cell-count distribution.
A symmetric $K=4$ data-generating construction produced a
mediated effect that was positive at smaller $t$ and negative at larger $t$,
with a positive direct effect. Its uniform integrated targets
were $(\mathrm{IME}^0,\mathrm{IDE}^0,\mathrm{ITE}^0)
=(0.15,0.20,0.35)$.

We compared BLaFM, an oracle analysis using
the true latent states, and product-of-coefficients mediation
applied to donor-average expression. Both functional fits
used $K=4$ and the prior hyperparameters specified above.
The symmetric design
makes the scalar exposure--mediator slope zero, while
preserving an exposure--outcome slope of 0.35.

\begin{table}[htbp]
\centering
\small
\renewcommand{\arraystretch}{1.12}
\setlength{\tabcolsep}{7pt}
\caption{Integrated effect summaries in the aggregation-stress
experiment. Entries for fitted procedures are mean estimates across
10 simulation replicates, with Monte Carlo standard errors in
parentheses.}
\label{tab:aggregation_main}
\begin{tabular}{lccc}
\toprule
& Mediated
& Direct
& Total \\
\midrule

Functional target
& $0.150$
& $0.200$
& $0.350$ \\

BLaFM
& $0.122\;(0.0014)$
& $0.229\;(0.0015)$
& $0.351\;(0.0006)$ \\

Oracle with known states
& $0.130\;(0.0014)$
& $0.221\;(0.0013)$
& $0.350\;(0.0006)$ \\

Donor-average scalar mediation
& $0.0004\;(0.0004)$
& $0.350\;(0.0004)$
& $0.350\;(0.0006)$ \\

\bottomrule
\end{tabular}
\end{table}

\begin{figure}[htbp]
\centering
\includegraphics[width=\linewidth]
{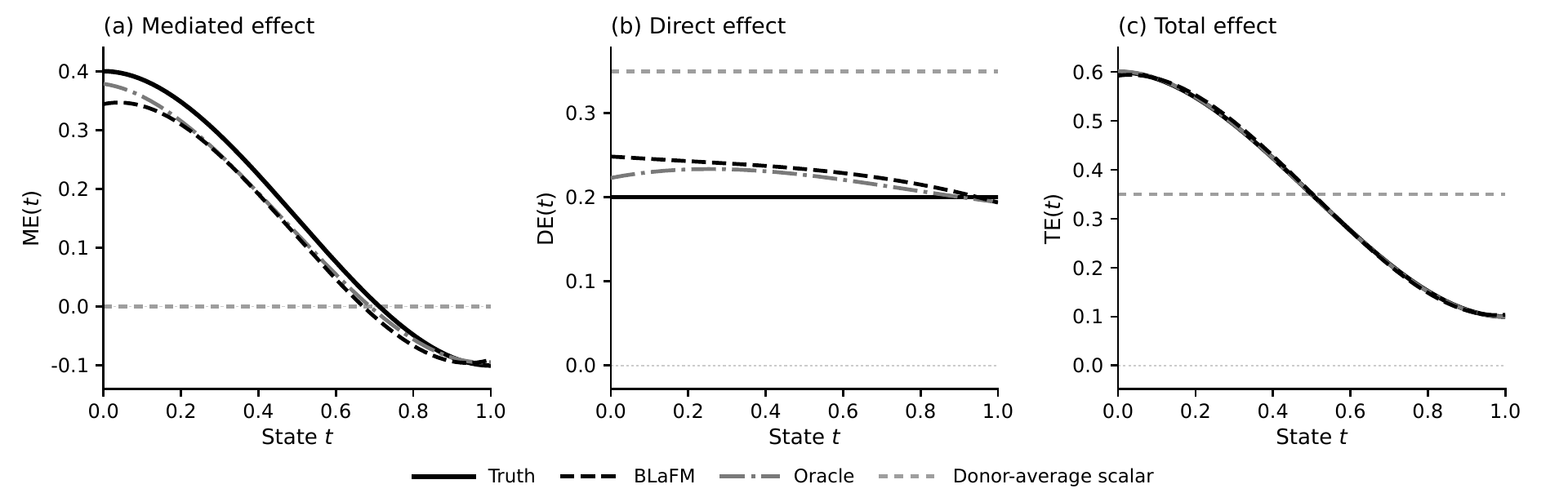}
\caption{Comparison of mediated, direct, and total effects in the
aggregation-stress experiment. Solid dark curves denote the true
state-specific effects. Dashed black and dash-dotted gray curves
denote posterior mean effect functions averaged across the 10
simulation replicates for BLaFM and the oracle analysis,
respectively. Dashed light-gray horizontal lines denote donor-average
scalar summaries and are included only as reference levels.}
\label{fig:aggregation_comparison_effects}
\end{figure}

Both functional analyses closely recovered the integrated total
effect. BLaFM and the oracle analysis underestimated IME by
approximately 0.028 and 0.020, respectively, with compensating
direct-effect errors. Residual component error therefore
remained even when the true latent states were supplied. The
donor-average procedure yielded a near-zero indirect component,
but this scalar summary did not capture the positive-to-negative
pattern across the state coordinate
(Figure~\ref{fig:aggregation_comparison_effects}).
This experiment illustrates a change in the target
decomposition after averaging.

\subsection{Comparison with two-stage estimation}
\label{subsec:two_stage_comparison}

We used the same $R=10$ master replicates at $n=1000$ from the
finite-sample experiment to compare BLaFM,
two two-stage procedures, and an oracle analysis. All four analyses
used $K=5$. The two-stage procedures shared a first-stage fit to the
paired expression measurements and annotation constraints, with
separate hierarchical normal priors for the mediator and outcome
trajectory coefficients. Their population means and diagonal
variances were estimated without using exposure or imposing the
donor-level mediation regressions.

The two-stage procedure with estimated states fixed the states at
their first-stage posterior means and re-estimated all remaining parameters of the BLaFM model, including both sets of donor-specific trajectory coefficients
and observation variances. The two-stage procedure with estimated
coefficients instead fixed both coefficient vectors at their
first-stage posterior means and fitted only the donor-level
mediation regressions. It did not reuse the cell-level likelihood
in the second stage. The oracle analysis fixed the true states and
re-estimated all remaining parameters of the BLaFM model. Effect-function recovery was
evaluated by RISE within each dataset and summarized across the
10 paired replicates.

\begin{table}[htbp]
\centering
\small
\renewcommand{\arraystretch}{1.12}
\setlength{\tabcolsep}{4pt}
\caption{Effect-function recovery under joint and two-stage
estimation at $K=5$ and $n=1000$. Entries are mean RISE across
$R=10$ paired master replicates, with simulation MCSEs in parentheses.
Both quantities are multiplied by $10^2$; lower is better.
The two-stage procedures fix either first-stage posterior mean
states or first-stage posterior mean trajectory coefficients.
The oracle analysis fixes the true states.}
\label{tab:two_stage_comparison}
\begin{tabular}{lccc}
\toprule
&\multicolumn{3}{c}{Effect-function RISE}\\
\cmidrule(lr){2-4}
Method & $\mathrm{ME}$ & $\mathrm{DE}$ & $\mathrm{TE}$ \\
\midrule
BLaFM & $1.863\;(0.227)$ & $1.908\;(0.213)$ & $0.470\;(0.014)$ \\
Two-stage with estimated states & $11.002\;(1.179)$ & $15.516\;(0.440)$ & $18.209\;(0.585)$ \\
Two-stage with estimated coefficients & $17.900\;(0.845)$ & $20.280\;(0.900)$ & $4.727\;(0.201)$ \\
Oracle with known states & $1.644\;(0.159)$ & $1.609\;(0.161)$ & $0.132\;(0.022)$ \\
\bottomrule
\end{tabular}
\end{table}

Table~\ref{tab:two_stage_comparison} shows smaller mean RISE for
BLaFM than for either two-stage procedure
for all three effect functions. The procedure with estimated
coefficients had larger mediated- and direct-effect errors than
the procedure with estimated states, but a smaller total-effect
error. Thus, the ranking of the two-stage procedures depended on
the target, and more accurate recovery of the total effect did
not imply more accurate recovery of its components. BLaFM had mediated- and direct-effect errors close to those
of the oracle, while the oracle had the smallest mean RISE for
all three functions.

\begin{figure}[htbp]
\centering
\includegraphics[width=\linewidth]
{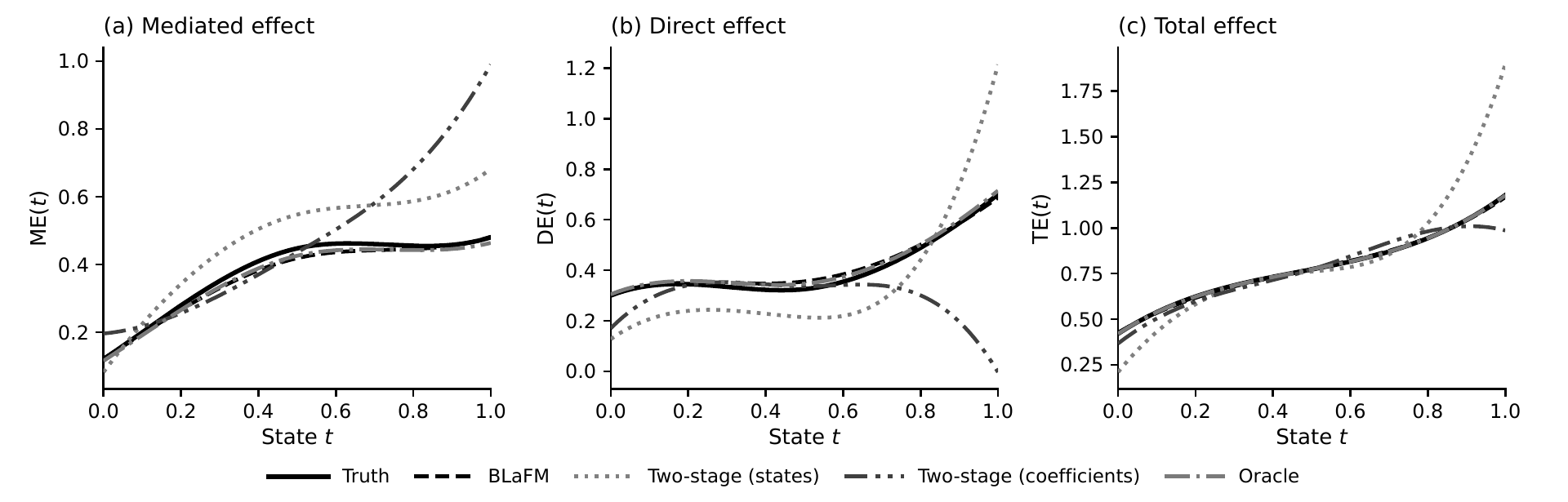}
\caption{Comparison of mediated, direct, and total effects at
$K=5$ and $n=1000$. Solid black curves denote the true
state-specific effects. The remaining curves are posterior mean
effect functions averaged across $R=10$ master replicates for
BLaFM, the two-stage procedure with estimated states, the two-stage
procedure with estimated coefficients, and the oracle analysis, as identified in
the legend. The two-stage procedures use the same first-stage fit.}
\label{fig:two_stage_effects}
\end{figure}

Figure~\ref{fig:two_stage_effects} compares the mean estimated
effect functions. In this setting, joint estimation improved
effect recovery relative to both procedures that conditioned on
first-stage estimates. These comparisons reflect the first-stage
model and the quantities held fixed in the second stage; they do
not isolate the effect of uncertainty propagation alone.

\section{Application to population-scale single-cell data}
\label{sec:application}

\subsection{Data and analysis setup}
\label{subsec:app_data_setup}

We examined how the association between rs1131017 dosage and BAX
expression varies across CD4$^+$ T-cell states, and how it
decomposes into a component mediated by RPS26 and a remaining direct
component. The rs1131017--RPS26 exposure--mediator pair was selected
a priori from the CD4$^+$ T-cell cross-cohort eQTL results in the
distQTL study \citep{Coulter2025distQTL}. Among the two
RPS26-associated variants in the candidate list, rs1131017 showed
stronger association evidence in both the Lifelines Deep Cohort and
OneK1K. BAX was selected using independent perturbation evidence:
RPS26 silencing in Jurkat cells increased apoptosis and upregulated
pro-apoptotic genes, including BAX \citep{Chen2021RPS26}.

The analysis used 403,087 CD4$^+$ T cells from the OneK1K study
\citep{Yazar2022}. All 961 donors in the available genotype release
were retained after matching expression and cell annotations.
We coded rs1131017 as G-allele dosage $Z=0,1,2$, corresponding to
322 C/C, 473 C/G, and 166 G/G donors. RPS26 and BAX expression
provided the mediator and outcome measurements, respectively, on
the SCTransform Pearson-residual scale
\citep{HafemeisterSatija2019,Hao2023}. Preprocessing removed cells
with zero total UMI counts and applied SCTransform before defining
the CD4$^+$ subset. Its negative-binomial regression used a
1,000-cell subsample, and no additional covariates were regressed
out. No further expression-based outlier filtering was applied.
All reported effects refer to one additional G allele on this
expression scale.

The sample comprised 181,285 naive cells and 221,802 central-memory
or effector-memory cells, with a median of 414 cells per donor
(range, 19--1,031). Following the annotation constraints in
Section~\ref{subsec:latent_state}, naive cells were assigned latent
states in $[0,0.5]$, and both memory groups were assigned states in
$[0.5,1]$. Expression measurements then informed each cell's state
within its allowed interval. Data construction, variant and
gene identifiers, and preprocessing details are given in
Supplementary Material, Section D.

We considered $K=4,\ldots,9$ using the conditional plug-in
criterion in Section~\ref{subsec:basis_selection}. The five folds
were shared across candidates and stratified within donor and
coarse state group. Each fold used 100,000 MCMC iterations,
discarded the first 50,000, and retained every 100th subsequent draw
for scoring. The criterion averages cell log predictive densities
within each donor and then across donors, giving each donor equal
weight. Table~\ref{tab:app_k_selection_main} shows the resulting
joint scores. The maximum occurred at $K=7$, exceeding the
runner-up $K=8$ by 0.091386. The mediator and conditional-outcome
score components are reported in Supplementary Material, Section D.

\begin{table}[htbp]
\centering
\small
\setlength{\tabcolsep}{4pt}
\caption{Application basis selection using donor-balanced joint
conditional log predictive scores. Each candidate uses its own
full-data posterior mean latent states. Larger values are preferred;
the selected dimension is shown in bold.}
\label{tab:app_k_selection_main}
\begin{tabular}{lrrrrrr}
\toprule
$K$ & 4 & 5 & 6 & \textbf{7} & 8 & 9\\
\midrule
Joint LPD & $-0.008$ & $-0.289$ & $-0.052$
& $\mathbf{0.160}$ & $0.069$ & $0.004$\\
\bottomrule
\end{tabular}
\end{table}

At the selected dimension, the primary analysis used the full joint BLaFM
posterior, with latent states and trajectory coefficients sampled
together. We ran one MCMC chain for 100,000 iterations and discarded
the first 50,000. All 50,000 retained population-coefficient draws
were used for the primary effect summaries.

\subsection{Primary functional analysis}
\label{subsec:app_primary}

\begin{figure}[htbp]
\centering
\includegraphics[width=\linewidth]{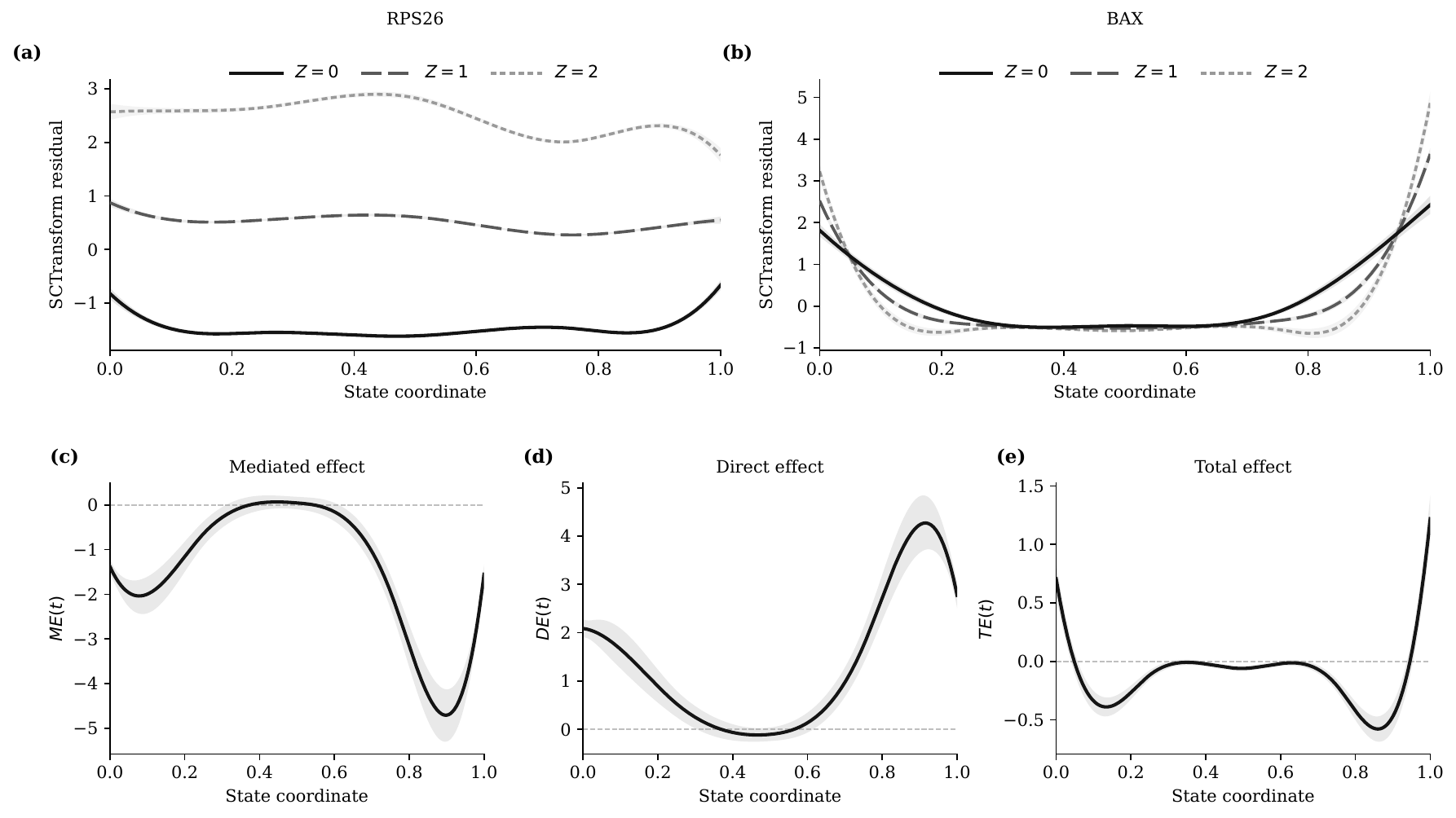}
\caption{Primary functional analysis of the rs1131017--RPS26--BAX
relationship in CD4$^+$ T cells at $K=7$, selected by conditional
plug-in cross-validation. Panels~(a)--(b) show genotype-specific
population mean RPS26 and BAX trajectories for $Z=0,1,2$.
Panels~(c)--(e) show the mediated, direct, and total effect functions
per additional G allele. Curves denote posterior means and shading
denotes pointwise, equal-tailed 95\% credible intervals from the
full joint BLaFM posterior. Horizontal
dashed lines in the effect panels denote zero.}
\label{fig:app_primary}
\end{figure}

Figure~\ref{fig:app_primary} presents genotype-specific population
mean trajectories and the corresponding effect functions from
the full BLaFM fit at $K=7$. The curves compare genotype groups at the
same latent state, and the shaded regions show pointwise 95\%
credible intervals. Effect summaries were calculated within
posterior draws as described in Section~\ref{subsec:posterior_effects}.

The posterior mean RPS26 trajectories were ordered by G-allele dosage
throughout the evaluated state grid, with G/G donors having the highest
mean expression. For BAX, higher G-allele dosage corresponded to higher
posterior mean expression near both endpoints and lower mean expression
through most of the interior. The posterior mean total effect followed
this pattern, crossing zero at approximately $t=0.048$ and $t=0.946$
and remaining negative between these points. The pointwise credible
intervals for the total effect included zero in parts of the interior.

Figures~\ref{fig:app_primary}c--e show the estimated mediated, direct,
and total effects per additional G allele. Over most of the fitted state
domain, the posterior mean mediated effect through RPS26 was negative
and the direct effect was positive. Both components had their largest
estimated magnitudes in the upper portion of the state coordinate.
Because the two components partly offset each other, the posterior mean
total effect was smaller in magnitude than either component over much
of the domain.

The fitted mediated component thus linked the higher RPS26 expression
associated with each additional G allele to lower BAX expression over
most of the state domain. This direction is consistent with the increase
in BAX expression observed after RPS26 silencing in Jurkat cells
\citep{Chen2021RPS26}.

The posterior mean IME was $-1.459$
(95\% CrI, $-1.662$ to $-1.253$), IDE was
$1.344$ (95\% CrI, $1.137$ to $1.549$), and ITE was $-0.115$
(95\% CrI, $-0.149$ to $-0.081$)
(Table~\ref{tab:app_integrated_effects}).
These quantities correspond to $\mathrm{IME}$, $\mathrm{IDE}$,
and $\mathrm{ITE}$, respectively, and assign equal weight to
each part of the standardized state domain.
Under the stated assumptions, they indicate opposing indirect
and direct contributions whose partial cancellation yields a
smaller negative integrated total effect.

\begin{table}[htbp]
\centering
\small
\caption{Integrated effect summaries from the full joint BLaFM posterior at the selected dimension $K=7$. Effects are per additional
G allele on the Pearson-residual scale, integrated uniformly over
the standardized state coordinate. IME, IDE, and ITE denote the uniformly integrated mediated, direct, and total effects, respectively. SD denotes posterior standard deviation;
CrIs are equal-tailed 95\% credible intervals conditional on the
selected dimension. ESS is the autocorrelation-adjusted effective
sample size, and MCSE is the Monte Carlo standard error of the
posterior mean, based on 50,000 retained draws from one chain.}
\label{tab:app_integrated_effects}
\smallskip
\setlength{\tabcolsep}{4pt}
\begin{tabular}{lrrcrr}
\toprule
Effect & Mean & SD & 95\% CrI & ESS & MCSE\\
\midrule
IME & $-1.459$ & $0.105$ & $(-1.662,-1.253)$ & $632.3$ & $0.00417$\\
IDE & $1.344$ & $0.106$ & $(1.137,1.549)$ & $629.3$ & $0.00421$\\
ITE & $-0.115$ & $0.017$ & $(-0.149,-0.081)$ & $1173.0$ & $0.00051$\\
\bottomrule
\end{tabular}
\end{table}

The effective sample sizes for these integrated effects ranged
from 629 to 1,173 (Table~\ref{tab:app_integrated_effects}). Their
MCSEs were approximately 2.9\%--4.0\% of the respective posterior
standard deviations, indicating that Monte Carlo error in the
estimated means was small relative to posterior uncertainty.
The supplementary trace plots show fluctuations around stable
levels without evident sustained drift. These diagnostics describe
within-chain behavior and posterior-mean precision; between-chain
convergence was not assessed because only one chain was run.

For comparison, a descriptive regression of donor-average BAX
expression on donor-average RPS26 expression and genotype dosage
gave an RPS26 coefficient of 0.046 (95\% confidence interval,
0.029--0.063; $p=2.09\times10^{-7}$). This coefficient describes
the association between the two donor averages conditional on
genotype. The functional mediated effect combines the genotype
association with the RPS26 trajectory and the relationship between
the RPS26 and BAX trajectory coefficients.
Additional functional summaries, including the fitted
mediator--outcome kernel, are reported in Supplementary Material,
Section D.

\subsection{Sensitivity to residual dependence}
\label{subsec:app_residual_dependence}

The primary BLaFM model assumes that the donor-level coefficient
residuals $\boldsymbol\eta_i$ and $\boldsymbol\xi_i$ are independent.
To assess sensitivity to unmeasured mediator--outcome confounding,
we allowed residual dependence between matching mediator and
outcome coefficient components,
\[
\operatorname{Corr}(\eta_{ik},\xi_{ik})=\rho_s,
\qquad
\rho_s\in\{-0.5,-0.3,-0.1,0,0.1,0.3,0.5\},
\]
using the same fixed correlation for all basis components
\citep{ImaiKeeleYamamoto2010,ImaiKeeleTingley2010}.
Here $\rho_s$ describes residual dependence in the donor-level
regressions. Under the joint Gaussian sensitivity model, let
\[
L_{\rho_s}=\operatorname{diag}
\left\{\rho_s\sqrt{\Lambda_{kk}/\Sigma_{kk}}\right\}.
\]
Then $\mathbb E(\boldsymbol\xi_i\mid\boldsymbol\eta_i)
=L_{\rho_s}\boldsymbol\eta_i$. Substituting
$\boldsymbol\eta_i=\boldsymbol\alpha_i-\boldsymbol\alpha_0-AZ_i$
into the outcome equation gives
\begin{equation*}
\begin{split}
\mathbb E(\boldsymbol\beta_i\mid\boldsymbol\alpha_i,Z_i)
={}&(\boldsymbol\beta_0-L_{\rho_s}\boldsymbol\alpha_0)\\
&+(C-L_{\rho_s}A)Z_i
+(B+L_{\rho_s})\boldsymbol\alpha_i.
\end{split}
\end{equation*}
Thus the conditional regression coefficients are
$\widetilde B=B+L_{\rho_s}$ and
$\widetilde C=C-L_{\rho_s}A$, which satisfy
\begin{equation}
\widetilde B A+\widetilde C=BA+C.
\label{eq:rho_total_invariance_main}
\end{equation}
Residual dependence can therefore change the allocation between
the mediated and direct paths while preserving their total
coefficient contrast. This algebraic identity explains why the
total association may be more stable across sensitivity fits;
it does not require the fitted posterior summaries to be identical
across $\rho_s$, since the parameters and variances are re-estimated.

For each value of $\rho_s$, we fixed
$\boldsymbol\alpha_i$ and $\boldsymbol\beta_i$ at their posterior
means from the full BLaFM fit at $K=7$ and refitted the remaining
coefficient-level parameters and variances. These refits assess
sensitivity conditional on the estimated trajectories.

The appropriate reference for this comparison is the conditional
$\rho_s=0$ refit. Its posterior means were
$\mathrm{IME}=-2.069$, $\mathrm{IDE}=1.955$, and
$\mathrm{ITE}=-0.115$. Even at zero residual correlation, fixing
the donor coefficients and refitting shifted the mediated and
direct means by approximately $-0.610$ and $0.610$ relative to
the primary joint BLaFM analysis, while the total mean changed by less
than 0.0001. These shifts arise from conditioning and refitting;
they cannot be attributed to allowing nonzero residual correlation.

\begin{figure}[htbp]
\centering
\includegraphics[width=\linewidth]
{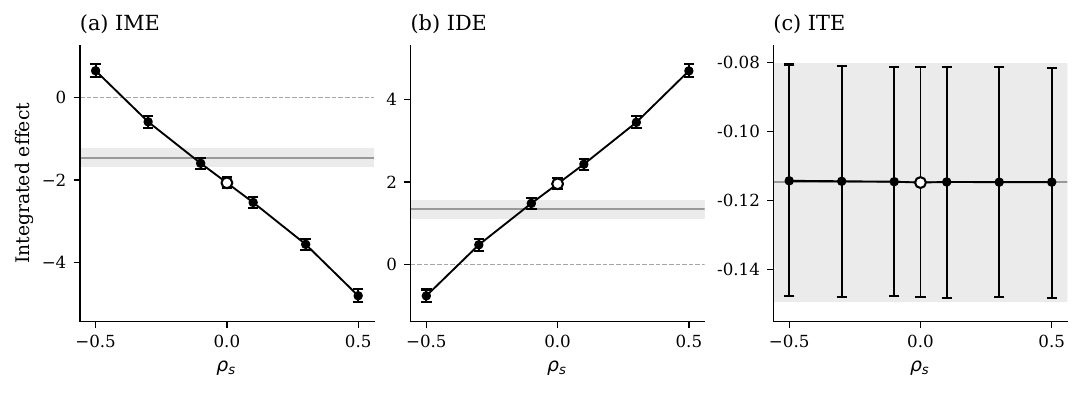}
\caption{Conditional residual-correlation sensitivity based on the
selected $K=7$ fit. Points and error bars show integrated posterior
means and equal-tailed 95\% CrIs for each fixed $\rho_s$; lines
connect the examined values. All fits condition on posterior mean
donor-specific trajectory coefficients from the full BLaFM fit. The open
point marks the conditional $\rho_s=0$ reference. Gray horizontal
lines and shaded bands give the corresponding full primary posterior
means and 95\% CrIs for comparison. The panels use separate vertical
scales; the conditional reference and the full primary posterior
target different uncertainty summaries.}
\label{fig:app_sensitivity}
\end{figure}

Relative to this conditional reference, the allocation between
mediated and direct components varied substantially across the
examined correlations (Figure~\ref{fig:app_sensitivity}). At
$\rho_s=-0.5$, the IME was 0.646 (95\% CrI, 0.494--0.799)
and the IDE was $-0.760$ (95\% CrI, $-0.916$ to $-0.606$),
reversing both component signs. At $\rho_s=0.5$, their means
were $-4.806$ and 4.691. The ITE means, however, ranged only
from $-0.1148$ to $-0.1143$, and all seven 95\% CrIs lay below
zero. This behavior is consistent with offsetting changes in
$B$ and $C$ that preserve the total coefficient contrast $BA+C$,
as in Equation~\eqref{eq:rho_total_invariance_main}.

In the five refits with $|\rho_s|\leq 0.3$, the integrated mediated component remained negative and the direct component remained positive, with conditional 95\% credible intervals excluding zero. Both signs reversed at $\rho_s=-0.5$. The integrated total association remained negative across all seven refits. Thus, conditional on the estimated donor trajectories, the sensitivity analysis supports a stable direction for the total association, while its allocation to the two components depends on the assumed residual correlation.

\section{Discussion}
\label{sec:discussion}

BLaFM connects donor-level exposures with
cell-level mediator and outcome measurements through donor-specific
expression trajectories. A central feature is joint inference of
each cell's latent state, the donor-specific trajectories, and
their mediation relationships. This construction
retains donors as the independent sampling units and propagates
uncertainty in the latent states into the effect functions. It
provides both a decomposition along the state coordinate and
integrated summaries of the mediated, direct, and total associations.

In the OneK1K application, the small negative integrated association
between rs1131017 dosage and BAX expression reflected partial cancellation
between a negative RPS26-mediated component and a positive direct
component. The state-specific curves showed where these components were
largest and where they most strongly offset one another. The aggregation
experiment illustrates why this distinction matters: averaging expression
within donors can change the mediated and direct components even when the
total association is preserved.

In the sensitivity analysis conditional on estimated donor trajectories,
all three integrated effects retained their signs and their 95\% credible
intervals excluded zero at the five examined correlations with
$|\rho_s|\leq 0.3$. The integrated total association remained negative
across all seven settings, while the mediated and direct estimates varied
substantially and reversed signs at $\rho_s=-0.5$. Fixing the trajectories
at their posterior means also shifted both component estimates appreciably
at $\rho_s=0$. A joint sensitivity analysis could assess how residual
dependence affects the estimated trajectories and the decomposition together.

Causal interpretation requires the conditions outlined in
Section~\ref{subsec:tv_effects} and detailed in Supplementary Material,
Section A. The current application did
not adjust for batch effects or population structure, so these
remain potential sources of confounding. Incorporating relevant
cell- and donor-level covariates would strengthen the analysis.
Replication in independent cohorts and perturbation studies in
relevant cellular contexts would further assess the proposed
regulatory relationships.

The functional targets also depend on the representation of cellular
state. The shared coordinate is learned from two expression
measurements and coarse annotations, and uniform integration gives
equal weight to each part of that coordinate. The effect functions
compare exposure groups at common states; an extension that also
models genotype-associated changes in cell-state composition could
capture an additional pathway. More flexible state representations
would accommodate branching or cyclic processes
\citep{Lonnberg2017,Boukouvalas2018}, while relaxing the diagonal
coefficient map and allowing multiple mediators would broaden the
regulatory relationships represented by the model.

Several aspects of posterior computation merit further development.
The basis-selection criterion evaluates conditional prediction at
states estimated from the full data, and the reported intervals
condition on the selected dimension. Accounting for uncertainty in
this choice would extend the uncertainty propagated by the joint
model. For larger gene screens, computation
and multiplicity control will also need to scale with the number
of candidate triplets.

\section*{Supplementary material}
The \hyperref[supp:overview]{supplementary material} follows the references and contains:
Causal assumptions and proof of Proposition~\ref{prop:functional_natural_effects_main};
posterior computation and the basis-selection criterion; simulation
designs and additional results; and application preprocessing,
posterior diagnostics, and sensitivity analyses.

\clearpage
\phantomsection
\pdfbookmark[1]{References}{references}
\bibliographystyle{plainnat}
\bibliography{ref}

\clearpage
\appendix
\setcounter{equation}{0}
\setcounter{figure}{0}
\setcounter{table}{0}
\setcounter{proposition}{0}
\setcounter{remark}{0}
\renewcommand{\thesection}{\Alph{section}}
\renewcommand{\thesubsection}{\thesection.\arabic{subsection}}
\renewcommand{\theequation}{S.\arabic{equation}}
\renewcommand{\thefigure}{S\arabic{figure}}
\renewcommand{\thetable}{S\arabic{table}}
\renewcommand{\theHequation}{supp.\arabic{equation}}
\renewcommand{\theHfigure}{supp.\arabic{figure}}
\renewcommand{\theHtable}{supp.\arabic{table}}
\renewcommand{\theHproposition}{supp.\arabic{proposition}}
\renewcommand{\theHremark}{supp.\arabic{remark}}
\phantomsection
\label{supp:overview}
\pdfbookmark[0]{Supplementary Material}{supplementary-material}
\begin{center}
  {\Large\bfseries Supplementary Material\par}
  \vspace{0.7em}
  {\large Bayesian Latent Functional Mediation Analysis with Application to
  Population-Scale Single-Cell Data\par}
\end{center}
\medskip

\etocdepthtag.toc{supplement}
\etocsettagdepth{main}{none}
\etocsettagdepth{supplement}{subsection}
\renewcommand{\contentsname}{Supplementary contents}

\noindent
This Supplementary Material provides the causal formulation,
posterior computation, simulation designs and evaluation definitions,
and additional application analyses. BLaFM denotes the Bayesian latent
functional mediation model proposed in the main article. Section~A distinguishes causal
interpretation from the identification of the latent measurement
model; Section~B gives posterior computation and the basis-selection
criterion; Section~C describes the
four simulation experiments and their reporting; and Section~D
provides data-construction, posterior, and sensitivity summaries.

\tableofcontents
\clearpage

\section{Causal interpretation and model-implied derivations}
\label{supp:causal}

\paragraph{Potential functional outcomes.}
Let $z_0$ and $z_1$ denote two exposure levels and write
$\Delta z=z_1-z_0$. For donor $i$, let
\begin{equation}
\bm\alpha_i(z)
=
\bm\alpha_0+Az+\bm\eta_i
\label{eq:potential_alpha}
\end{equation}
denote the mediator coefficient vector that would arise if the
exposure were set to $z$. Its corresponding potential mediator
trajectory is
\begin{equation*}
G_i^z(t)
=
\bm\phi(t)^\top\bm\alpha_i(z),
\qquad t\in[0,1].
\end{equation*}
For exposure level $z$ and mediator coefficient vector
$\bm a\in\R^K$, define
\begin{equation}
\bm\beta_i(z,\bm a)
=
\bm\beta_0+Cz+B\bm a+\bm\xi_i
\label{eq:potential_beta}
\end{equation}
and
\begin{equation*}
H_i^{z,\bm a}(t)
=
\bm\phi(t)^\top\bm\beta_i(z,\bm a).
\end{equation*}
The nested trajectory
$H_i^{z,\bm\alpha_i(z')}(t)$ is the potential outcome trajectory
under exposure $z$ with the mediator coefficient vector set to the
value that would arise under exposure $z'$.

We define the functional natural indirect, direct, and total effects
by
\begin{align*}
\operatorname{NIE}(t;z_1,z_0)
&=
\E\!\left[
H_i^{z_1,\bm\alpha_i(z_1)}(t)
-
H_i^{z_1,\bm\alpha_i(z_0)}(t)
\right],
\\
\operatorname{NDE}(t;z_1,z_0)
&=
\E\!\left[
H_i^{z_1,\bm\alpha_i(z_0)}(t)
-
H_i^{z_0,\bm\alpha_i(z_0)}(t)
\right],
\\
\operatorname{NTE}(t;z_1,z_0)
&=
\E\!\left[
H_i^{z_1,\bm\alpha_i(z_1)}(t)
-
H_i^{z_0,\bm\alpha_i(z_0)}(t)
\right].
\end{align*}

Because the coefficient-level outcome model contains no
exposure--mediator interaction, fixing the exposure in the indirect
contrast at $z_0$ instead of $z_1$ yields the same indirect-effect
expression.

\paragraph{Identification assumptions.}
The effect expressions are always defined within the regression
model. Their interpretation as natural effects requires additional
conditions linking the observed-data distribution and the potential
trajectories \citep{ImaiKeeleYamamoto2010,ImaiKeeleTingley2010}.
For the coefficient equations shown in the article, these are
stated marginally below. In an explicitly covariate-adjusted
extension, the exchangeability and support conditions would instead
be conditional on the included baseline covariates; preprocessing
alone must not be assumed to guarantee them.

\begin{enumerate}[label=(A\arabic*)]
\item \textbf{Consistency.}
Under observed exposure $Z_i=z$, the latent mediator coefficient
vector equals $\bm\alpha_i(z)$ and the latent outcome coefficient
vector equals $\bm\beta_i\{z,\bm\alpha_i(z)\}$. A specified
intervention on the mediator coefficient vector has the potential
outcome in Equation~\eqref{eq:potential_beta}.

\item \textbf{No interference.}
One donor's potential trajectories do not depend on other donors'
exposures or mediator values.

\item \textbf{Positivity and mediator support.}
The exposure levels in the contrast have positive mass, or positive
density for continuous exposure. Mediator coefficient values
required by the nested contrasts lie in the corresponding
conditional support. The Gaussian coefficient model has full
$\R^K$ support when its variances are positive.

\item \textbf{Exposure exchangeability.}
For the exposure and mediator values under consideration,
\[
\{\bm\alpha_i(z'),\bm\beta_i(z,\bm a)\}\perp Z_i.
\]
In a genetic application, the relevant population structure and
other donor-level sources of confounding must therefore be
adequately addressed for a causal interpretation.

\item \textbf{Mediator--outcome exchangeability.}
For all relevant exposure levels $z,z'$ and mediator coefficient
vectors $\bm a$, we assume
\[
\bm\beta_i(z,\bm a)\perp\bm\alpha_i(z').
\]
Under the structural equations, a sufficient condition for both
(A4) and (A5) is that $\bm\eta_i$ and $\bm\xi_i$ are mutually
independent and that the pair $(\bm\eta_i,\bm\xi_i)$ is jointly
independent of $Z_i$. Since $\bm\alpha_i$ is a function of $Z_i$
and $\bm\eta_i$ under consistency, these independence conditions also
imply $\E(\bm\xi_i\mid Z_i,\bm\alpha_i)=\bm 0$ for mean-zero
$\bm\xi_i$.

\item \textbf{No exposure-induced mediator--outcome confounder.}
There is no post-exposure variable that is a common cause of the
mediator and outcome coefficient vectors. The present natural-effect
formulation does not address a general model with such a variable.

\item \textbf{Common state coordinate.}
Potential trajectories under different exposure levels are compared
at the same standardized cellular-state coordinate. The effects
exclude an additional pathway through exposure-induced changes in
the distribution of cells along the state coordinate.

\item \textbf{Adequate functional specification.}
The measurement and coefficient-level models adequately represent
the relationships relevant to the specified contrasts.
\end{enumerate}

\paragraph{Proof of the functional natural-effect result.}
Proposition~1 in the main article connects the functional natural
effects to the coefficient-level decomposition. We establish the
result using the potential trajectories defined above.

\begin{proof}
Equation~\eqref{eq:potential_alpha} gives
\[
\bm\alpha_i(z_1)-\bm\alpha_i(z_0)=A\Delta z.
\]
Therefore
\begin{align*}
H_i^{z_1,\bm\alpha_i(z_1)}(t)
-
H_i^{z_1,\bm\alpha_i(z_0)}(t)
&=
\bm\phi(t)^\top
B\{\bm\alpha_i(z_1)-\bm\alpha_i(z_0)\}\\
&=
\bm\phi(t)^\top BA\,\Delta z.
\end{align*}
Taking expectations gives the stated natural indirect effect.
Similarly,
\[
H_i^{z_1,\bm\alpha_i(z_0)}(t)
-
H_i^{z_0,\bm\alpha_i(z_0)}(t)
=
\bm\phi(t)^\top C\,\Delta z,
\]
which gives the natural direct effect. Summing the two contrasts
and taking expectations yields the natural total effect
$\bm\phi(t)^\top(BA+C)\,\Delta z$.

It remains to link the structural coefficients to the coefficient-level
regressions in the main article. By consistency,
$\bm\alpha_i=\bm\alpha_i(Z_i)$ and
$\bm\beta_i=\bm\beta_i\{Z_i,\bm\alpha_i(Z_i)\}$.
Under the structural specification in which $\bm\eta_i$ and
$\bm\xi_i$ are mean-zero and mutually independent, with
$(\bm\eta_i,\bm\xi_i)$ jointly independent of $Z_i$, we have
\[
\E(\bm\alpha_i\mid Z_i=z)=\bm\alpha_0+Az,\qquad
\E(\bm\beta_i\mid Z_i=z,\bm\alpha_i=\bm a)
=\bm\beta_0+Cz+B\bm a.
\]
Thus the coefficients in the natural-effect formulas are the
coefficients in the corresponding regressions for the latent
trajectories.
\end{proof}

Thus, under assumptions (A1)--(A8), the model-implied
$\ME(t)$, $\DE(t)$, and $\TE(t)$ coincide with the
corresponding functional natural indirect, direct, and total effects.
Without these assumptions, the same algebra remains a regression-path
decomposition of the exposure--outcome association.

\paragraph{Integrated effects.}
The integrated summaries assign equal weight
to all locations on the standardized state domain:
\begin{align*}
\IME(z_1,z_0)
&=
\int_0^1 \ME(t;z_1,z_0)\,dt,
\\
\IDE(z_1,z_0)
&=
\int_0^1 \DE(t;z_1,z_0)\,dt,
\\
\ITE(z_1,z_0)
&=
\int_0^1 \TE(t;z_1,z_0)\,dt.
\end{align*}
Let
\begin{equation*}
\bm m_\phi
=
\int_0^1\bm\phi(t)\,dt.
\end{equation*}
Then
\begin{equation*}
\IME
=
\bm m_\phi^\top BA\,\Delta z,
\qquad
\IDE
=
\bm m_\phi^\top C\,\Delta z,
\qquad
\ITE
=
\bm m_\phi^\top(BA+C)\,\Delta z.
\end{equation*}
These quantities differ from averages weighted by the empirical cell
distribution along the state coordinate, which additionally depend on the
cell-sampling density. Uniform integration also depends on the chosen
coordinate convention: under a nonlinear transformation $u=f(t)$,
the corresponding integral in $u$ weights the original coordinate
by $f'(t)$. The reported integrals therefore refer to the standardized
scale used in the fitted model.

\paragraph{Orientation of the latent state coordinate.}
For the symmetric cubic B-spline basis used here, there exists a
permutation matrix $J$ such that
\begin{equation*}
\bm\phi(1-t)=J\bm\phi(t).
\end{equation*}
Under the simultaneous transformation
\begin{align*}
t_{ic}^{\star}&=1-t_{ic},
&
\bm\alpha_i^\star&=J\bm\alpha_i,
&
\bm\beta_i^\star&=J\bm\beta_i,
\\
\bm\alpha_0^\star&=J\bm\alpha_0,
&
A^\star&=JA,
&
\bm\beta_0^\star&=J\bm\beta_0,
\\
C^\star&=JC,
&
B^\star&=JBJ^\top,
&
\Sigma^\star&=J\Sigma J^\top,
\qquad
\Lambda^\star=J\Lambda J^\top,
\end{align*}
the fitted mediator and outcome values at the observed cells are
unchanged. The sampling model alone therefore does not determine the
orientation of the state coordinate. The cell-state support restrictions used in
the article anchor this global reflection ambiguity.

\section{Posterior computation and basis selection}
\label{supp:mcmc}

\subsection{Posterior computation}
\label{supp:posterior_computation}

For donor $i$, let
$\bm x_i=(X_{i1},\ldots,X_{iC_i})^\top$,
$\bm y_i=(Y_{i1},\ldots,Y_{iC_i})^\top$, and let $\Phi_i$ be the
$C_i\times K$ matrix with $c$th row
$\bm\phi(t_{ic})^\top$. Conditional on the current latent states,
\begin{align*}
\bm x_i\mid\bm\alpha_i,\bm t_i,\sigma_i^2
&\sim
N(\Phi_i\bm\alpha_i,\sigma_i^2I_{C_i}),
\\
\bm y_i\mid\bm\beta_i,\bm t_i,\tau_i^2
&\sim
N(\Phi_i\bm\beta_i,\tau_i^2I_{C_i}),
\\
\bm\alpha_i\mid Z_i,\bm\alpha_0,A,\Sigma
&\sim
N(\bm\alpha_0+AZ_i,\Sigma),
\\
\bm\beta_i\mid Z_i,\bm\alpha_i,\bm\beta_0,C,B,\Lambda
&\sim
N(\bm\beta_0+CZ_i+B\bm\alpha_i,\Lambda).
\end{align*}

\paragraph{Distributional parameterizations.}
We use the shape--scale inverse-gamma parameterization
\begin{equation*}
X\sim\IG(a,b)
\quad\Longleftrightarrow\quad
p(x)
=
\frac{b^a}{\Gamma(a)}
x^{-a-1}\exp(-b/x),
\qquad x>0,
\end{equation*}
and the shape--rate gamma parameterization
\begin{equation*}
X\sim\operatorname{Gamma}(a,b)
\quad\Longleftrightarrow\quad
p(x)
=
\frac{b^a}{\Gamma(a)}
x^{a-1}\exp(-bx),
\qquad x>0.
\end{equation*}
For the generalized inverse Gaussian distribution,
\begin{equation*}
X\sim\GIG(\lambda,\psi,\chi)
\quad\Longleftrightarrow\quad
p(x)\propto
x^{\lambda-1}
\exp\left\{-\frac{\psi x+\chi/x}{2}\right\},
\qquad x>0.
\end{equation*}

\paragraph{Update of latent states.}
For each cell,
\begin{equation*}
p(t_{ic}\mid-)
\propto
\mathbf 1\{t_{ic}\in\mathcal I_{ic}\}
\exp\left[
-\frac{\{X_{ic}-\bm\phi(t_{ic})^\top\bm\alpha_i\}^2}
      {2\sigma_i^2}
-\frac{\{Y_{ic}-\bm\phi(t_{ic})^\top\bm\beta_i\}^2}
      {2\tau_i^2}
\right].
\end{equation*}
Writing $\mathcal I_{ic}=[\ell_{ic},u_{ic}]$, we propose
\begin{equation*}
t_{ic}^{\ast}
\sim
\TN(t_{ic},s_t^2;\ell_{ic},u_{ic}).
\end{equation*}
The logarithm of the Metropolis--Hastings ratio is
\begin{align*}
\log r_{ic}
&=
-\frac{\{X_{ic}-g_i(t_{ic}^{\ast})\}^2
       -\{X_{ic}-g_i(t_{ic})\}^2}
      {2\sigma_i^2}
\\
&\quad
-\frac{\{Y_{ic}-h_i(t_{ic}^{\ast})\}^2
       -\{Y_{ic}-h_i(t_{ic})\}^2}
      {2\tau_i^2}
\\
&\quad
+\log
\frac{
\Phi\{(u_{ic}-t_{ic})/s_t\}
-
\Phi\{(\ell_{ic}-t_{ic})/s_t\}
}{
\Phi\{(u_{ic}-t_{ic}^{\ast})/s_t\}
-
\Phi\{(\ell_{ic}-t_{ic}^{\ast})/s_t\}
}.
\end{align*}
The proposal is accepted with probability
$\min\{1,\exp(\log r_{ic})\}$.

\paragraph{Joint update of donor-specific trajectory coefficients.}
Section~3.1 of the main article gives the joint Gaussian full
conditional of $(\bm\alpha_i^\top,\bm\beta_i^\top)^\top$, including
its block precision matrix and information vector. We sample this
$2K$-dimensional block using a Cholesky factorization. The observation
and population-level updates below complete the Gibbs steps.

\paragraph{Update of observation-level variances.}
Let
$\bm r_{Xi}=\bm x_i-\Phi_i\bm\alpha_i$ and
$\bm r_{Yi}=\bm y_i-\Phi_i\bm\beta_i$. Then
\begin{align*}
\sigma_i^2\mid-
&\sim
\IG\left(
a_1+\frac{C_i}{2},
b_1+\frac12\bm r_{Xi}^\top\bm r_{Xi}
\right),
\\
\tau_i^2\mid-
&\sim
\IG\left(
a_2+\frac{C_i}{2},
b_2+\frac12\bm r_{Yi}^\top\bm r_{Yi}
\right).
\end{align*}

\paragraph{Update of population-level mediation parameters.}
For basis component $k$, define
\[
\bm d_k^{(\alpha)}
=
(\alpha_{0k},a_k)^\top,
\qquad
G=
\begin{pmatrix}
1&Z_1\\
\vdots&\vdots\\
1&Z_n
\end{pmatrix}.
\]
With
$\bm d_k^{(\alpha)}
\mid\Sigma_{kk},\tau_a
\sim
N(\bm0,\tau_a\Sigma_{kk}I_2)$,
\begin{equation*}
\bm d_k^{(\alpha)}\mid-
\sim
N\left(
V_\alpha G^\top\bm\alpha_{\cdot k},
\Sigma_{kk}V_\alpha
\right),
\qquad
V_\alpha
=
(G^\top G+\tau_a^{-1}I_2)^{-1}.
\end{equation*}
Similarly, because $B$ is diagonal, let
\[
\bm d_k^{(\beta)}
=
(\beta_{0k},c_k,b_k)^\top,
\qquad
H_k=
\begin{pmatrix}
1&Z_1&\alpha_{1k}\\
\vdots&\vdots&\vdots\\
1&Z_n&\alpha_{nk}
\end{pmatrix}.
\]
With
$\bm d_k^{(\beta)}
\mid\Lambda_{kk},\tau_d
\sim
N(\bm0,\tau_d\Lambda_{kk}I_3)$,
\begin{equation*}
\bm d_k^{(\beta)}\mid-
\sim
N\left(
V_{\beta k}H_k^\top\bm\beta_{\cdot k},
\Lambda_{kk}V_{\beta k}
\right),
\qquad
V_{\beta k}
=
(H_k^\top H_k+\tau_d^{-1}I_3)^{-1}.
\end{equation*}

\paragraph{Update of coefficient-level variances.}
Define
\begin{align*}
S_{\alpha k}
&=
\sum_{i=1}^n
(\alpha_{ik}-\alpha_{0k}-a_kZ_i)^2,
\\
S_{\beta k}
&=
\sum_{i=1}^n
(\beta_{ik}-\beta_{0k}-c_kZ_i-b_k\alpha_{ik})^2.
\end{align*}
Then
\begin{align*}
\Sigma_{kk}\mid-
&\sim
\IG\left(
a_3+\frac n2+1,
b_3+\frac12S_{\alpha k}
+\frac{\alpha_{0k}^2+a_k^2}{2\tau_a}
\right),
\\
\Lambda_{kk}\mid-
&\sim
\IG\left(
a_4+\frac n2+\frac32,
b_4+\frac12S_{\beta k}
+\frac{\beta_{0k}^2+c_k^2+b_k^2}{2\tau_d}
\right).
\end{align*}

\paragraph{Update of global shrinkage parameters.}
Let
\[
\chi_a
=
\sum_{k=1}^K
\frac{\alpha_{0k}^2+a_k^2}{\Sigma_{kk}},
\qquad
\chi_d
=
\sum_{k=1}^K
\frac{\beta_{0k}^2+c_k^2+b_k^2}{\Lambda_{kk}}.
\]
Then
\begin{align*}
\tau_a\mid-
&\sim
\GIG(a_a-K,2b_a,\chi_a),
\\
\tau_d\mid-
&\sim
\GIG\left(a_d-\frac{3K}{2},2b_d,\chi_d\right).
\end{align*}

\paragraph{Posterior summaries.}
For retained draw $m$, the effect curves are
\begin{align*}
\ME^{(m)}(t)
&=
\bm\phi(t)^\top B^{(m)}A^{(m)},
\\
\DE^{(m)}(t)
&=
\bm\phi(t)^\top C^{(m)},
\\
\TE^{(m)}(t)
&=
\ME^{(m)}(t)+\DE^{(m)}(t).
\end{align*}
Pointwise posterior means and credible intervals are formed across
draws at each state. Integrated effects are calculated
draw by draw before posterior summarization:
\[
\IME^{(m)}
=
\int_0^1\ME^{(m)}(t)\,dt,
\qquad
\IDE^{(m)}
=
\int_0^1\DE^{(m)}(t)\,dt,
\qquad
\ITE^{(m)}
=
\IME^{(m)}+\IDE^{(m)}.
\]

\paragraph{Prior and MCMC settings.}
The application analyses and the simulation fits using the BLaFM
mediation hierarchy, including CV, oracle, and second-stage fits, use
\[
\sigma_i^2,\tau_i^2,\Sigma_{kk},\Lambda_{kk}
\sim\operatorname{IG}(2,0.1),\qquad
\tau_a,\tau_d\sim\operatorname{Gamma}(10,0.1),
\]
independently at their respective levels. The Gaussian
population-coefficient priors retain their variance scaling
by $\Sigma_{kk}$ and $\Lambda_{kk}$.

The finite-sample, full-data basis-recovery, and aggregation
analyses used 10,000 MCMC iterations, with the first 5,000
discarded as burn-in. In the aggregation experiment, this
setting applied to both BLaFM and the oracle analysis.
For basis selection in the simulations, each conditional
cross-validation fold likewise used 10,000 iterations with
5,000 discarded. Predictive densities were evaluated at every
tenth post-burn-in iteration, giving 500 scoring draws per fold.

The primary application analysis used 100,000 iterations and
discarded the first 50,000. Population regression coefficients
were retained at every subsequent iteration, giving 50,000
posterior draws from iterations 50,001--100,000. Auxiliary variance and
global-scale draws were stored every 20 post-burn-in
iterations, giving 2,500 draws at iterations
50,001, 50,021,\ldots,99,981. Joint calculations using these
auxiliary draws align them with population draws at the
recorded iteration indices. The truncated-normal proposal for each
latent state had standard deviation 1.0.

Initial values are zero for $\bm\alpha_0,\bm\beta_0$, 0.1
for each entry of $A,C,\diag(B)$, identity matrices for
$\Sigma,\Lambda$, 0.1 for the observation variances, and 10
for $\tau_a,\tau_d$. Donor coefficients start at the
hierarchical regression means, and latent states are drawn
uniformly within their annotation intervals. Cross-validation
fits use the same initialization scheme for the sampled
parameters, with states fixed at the candidate-specific
full-data posterior means. Application cross-validation
settings are given in Section~\ref{supp:app_k_selection}.

Cubic B-splines are constructed from the clamped knot vector
\[
\bigl(0,0,0,0,\,1/(K-3),\ldots,(K-4)/(K-3),\,1,1,1,1\bigr),
\]
with no interior knots when $K=4$. All $K$ basis functions
are included, with no additional intercept column. They form
a partition of unity on $[0,1]$;
the population intercept functions are represented by
$\bm\alpha_0$ and $\bm\beta_0$ within this basis. For $K=5$,
the single interior knot is 0.5. The finite-sample recovery
fits were run with $K=5$.
Conditional plug-in cross-validation
selects $K=7$ for the primary application; its interior knots
are 0.25, 0.50, and 0.75.

\subsection{Basis selection conditional on estimated states}
\label{supp:cv}

\paragraph{Scoring target.}
Our functional targets compare trajectories at specified
state coordinates. To distinguish this target from prediction
after averaging over a cell's state interval, let
$w_{ic}=(X_{ic},Y_{ic})$ and let
$p_K(w_{ic}\mid t,\theta)$ denote its joint Gaussian sampling
density. A state-marginal held-out density has the form
\begin{equation*}
p^{\mathrm{marg}}_{K,-f}(w_{ic})
=\int\left\{\frac1{|\mathcal I_{ic}|}
\int_{\mathcal I_{ic}}p_K(w_{ic}\mid t,\theta)\,dt\right\}
\pi_{K,-f}(d\theta),
\end{equation*}
where $\pi_{K,-f}$ is the posterior for the remaining parameters
from a latent-state fit to the training cells. This density
averages over both the held-out cell's unobserved location and
the training posterior. It evaluates the resulting expression
mixture, which need not distinguish different assignments of
expression levels to locations within the interval.

\paragraph{A within-interval illustration.}
Consider a scalar example conditional on one allowed state
interval $\mathcal I=[u,v]$, with $\ell=v-u>0$. Let
$T\sim\operatorname{Unif}(\mathcal I)$ and
$X=T+\epsilon$, where $\epsilon\sim N(0,\sigma^2)$ is
independent of $T$. Compare
\[
g_0(t)=t,\qquad g_P(t)=t+d,\qquad g_Q(t)=u+v-t,
\qquad d>0.
\]
Evaluating both candidates at the same realized state $T$ gives
the expected conditional log-score difference
\begin{equation}
\operatorname{E}_0\left[
\log\frac{f_N\{X;g_P(T),\sigma^2\}}
{f_N\{X;g_Q(T),\sigma^2\}}\right]
=\frac{\ell^2/3-d^2}{2\sigma^2},
\label{eq:pq_conditional_score}
\end{equation}
so $P$ is preferred when $d<\ell/\sqrt3$. Here
$f_N(x;\mu,s^2)$ is the normal density with mean $\mu$ and
variance $s^2$, and $\operatorname{E}_0$ denotes expectation
under the true joint distribution.
Let $p_0$ be the true marginal density of $X$, and define
$q_j(x)=\ell^{-1}\int_u^v f_N\{x;g_j(t),\sigma^2\}\,dt$
for $j=P,Q$. Reflection within $\mathcal I$ preserves its
uniform measure, giving $q_Q(x)=p_0(x)$ and
$q_P(x)=p_0(x-d)$. Consequently,
\begin{equation}
\operatorname{E}_0[\log q_Q(X)-\log q_P(X)]
=\operatorname{KL}(p_0\Vert q_P)>0.
\label{eq:pq_marginal_score}
\end{equation}
For $0<d<\ell/\sqrt3$, marginal scoring favors $Q$, whereas
conditional scoring at the same states favors $P$. The reflection stays within
one state interval and does not exchange coarse-state groups.
This scalar illustration motivates keeping estimated locations
fixed during scoring; it is not a guarantee of optimal
effect-function recovery with estimated coordinates.

\begin{figure}[!t]
\centering
\suppfig[width=\linewidth]{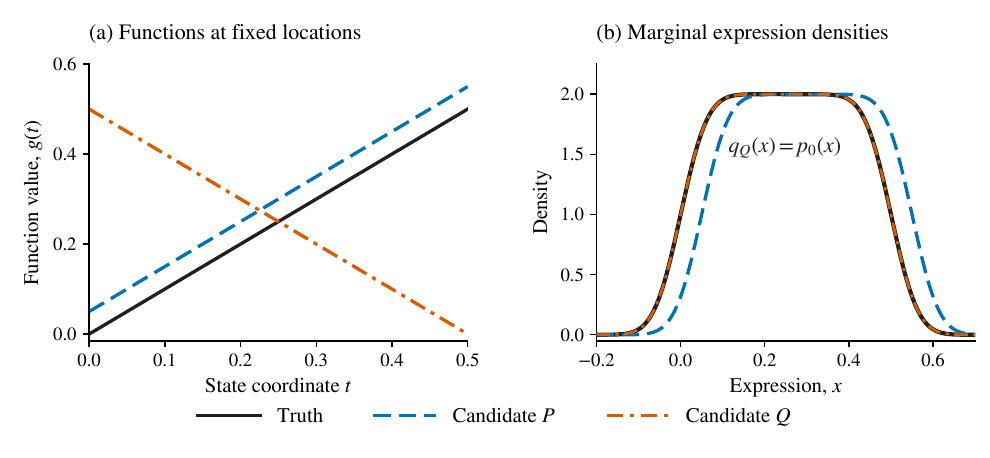}
\caption{Analytic illustration of conditional and state-marginal
prediction. We set $\mathcal I=[0,0.5]$, $d=0.05$, and
$\sigma=0.05$, with $T\sim\operatorname{Unif}(\mathcal I)$ and
$X=T+\epsilon$, $\epsilon\sim N(0,\sigma^2)$ independent of $T$.
(a) True function $g_0(t)=t$ and candidates
$g_P(t)=t+d$ and $g_Q(t)=u+v-t$. The expected conditional
log score at the same state favors $P$, as shown in
Equation~\eqref{eq:pq_conditional_score}.
(b) Expression densities after convolution with the common Gaussian
noise and integration over $T$. The density under $Q$ coincides
exactly with the truth (overlaid solid and dash-dotted curves),
whereas $P$ shifts the density by $d$. The expected marginal log
score therefore favors $Q$, as shown in Equation~\eqref{eq:pq_marginal_score}.
All curves are calculated analytically.}
\label{fig:pq_scoring_illustration}
\end{figure}

Figure~\ref{fig:pq_scoring_illustration} displays the candidate
functions and their induced marginal densities for one numerical
instantiation of this analytic example.

\paragraph{Two-stage conditional procedure.}
Let $\mathcal K$ denote the candidate basis dimensions. For each
$K\in\mathcal K$, we first fit the full latent-state model
and obtain the cell-specific posterior means
\[
\widehat t_{ic,K}
=\operatorname{E}(t_{ic}\mid\mathcal D,K).
\]
Each candidate retains its own full-data estimates.
These estimates are held fixed for training and held-out cells during
five-fold cell-level cross-validation. Fold assignments are
stratified within donor and coarse cell-state group and are
shared across the candidate dimensions.

Let $f(ic)$ denote the fold of cell $(i,c)$, and let
$\theta_{K,-f}^{(m)}$ be a posterior draw of the remaining
parameters conditional on the training expression observations
and their fixed estimated states. All remaining parameters,
including donor coefficients and variance components, continue
to be sampled; they are not replaced by posterior point estimates.
For a held-out cell,
the joint conditional density contribution is
\begin{align}
q_{ic,K}^{(m)}
&=f_N\!\left\{X_{ic};
\bm\phi_K(\widehat t_{ic,K})^\top
\bm\alpha_{i,K,-f}^{(m)},
(\sigma_{i,K,-f}^{2})^{(m)}\right\}
\notag\\
&\quad\times f_N\!\left\{Y_{ic};
\bm\phi_K(\widehat t_{ic,K})^\top
\bm\beta_{i,K,-f}^{(m)},
(\tau_{i,K,-f}^{2})^{(m)}\right\},
\label{eq:cv_fixed_time_density}
\end{align}
where $f_N(x;\mu,v)$ denotes a normal density with variance $v$.
The corresponding log score is
\begin{equation}
\widehat{\operatorname{lpd}}^{XY}_{ic,K}
=\log\left\{M_f^{-1}\sum_{m=1}^{M_f}
q_{ic,K}^{(m)}\right\}.
\label{eq:cv_logmean}
\end{equation}
The logarithm is taken after averaging densities over posterior
draws, using log-sum-exp computation. No integration or
posterior updating of latent states is performed during these
conditional fits or held-out evaluations. The score nevertheless
integrates over the remaining conditional posterior and retains
observation noise through the Gaussian sampling densities.

Every cell is scored once in its held-out fold. The primary
criterion is
\begin{equation}
L_{\mathrm{donor}}^{\mathrm{cond}}(K)
=\frac1n\sum_{i=1}^{n}\frac1{C_i}
\sum_{c=1}^{C_i}
\widehat{\operatorname{lpd}}^{XY}_{ic,K},
\qquad
\widehat K
=\min\operatorname*{arg\,max}_{K\in\mathcal K}
L_{\mathrm{donor}}^{\mathrm{cond}}(K).
\label{eq:conditional_basis_selection}
\end{equation}
Thus every donor receives equal weight, and exact score ties
are resolved by the smaller $K$. Selection uses scores at
full numerical precision. The secondary cell-weighted
summary is
\[
L_{\mathrm{cell}}^{\mathrm{cond}}(K)
=\frac{\sum_i\sum_c
\widehat{\operatorname{lpd}}^{XY}_{ic,K}}
{\sum_i C_i}.
\]
For component summaries, the marginal $X$ density is averaged
over the same posterior draws, and
\[
\widehat{\operatorname{lpd}}^{Y\mid X}_{ic,K}
=\widehat{\operatorname{lpd}}^{XY}_{ic,K}
-\widehat{\operatorname{lpd}}^{X}_{ic,K}.
\]

The reference latent states use the full expression data, including
the observations later held out in each fold. This is therefore
a conditional plug-in selection score; it does not estimate
independent predictive performance of the entire latent-state
learning procedure. Training donors remain represented by their
other cells, so this score concerns held-out cells within observed
donors rather than new-donor prediction. After selection, the reported effect
summaries use the full latent-state posterior at
$K=\widehat K$, rather than a fixed-state fold posterior. Credible intervals
condition on the selected dimension and do not include uncertainty
from choosing that dimension.

\section{Additional simulation details and results}
\label{supp:sim}

This section provides additional simulation details and results.
Section~\ref{supp:finite} gives the data-generating mechanism
and additional finite-sample recovery summaries.
Section~\ref{supp:basis} describes the basis-independent
data-generating mechanism and additional basis-complexity results.
Section~\ref{supp:aggregation} gives the construction and
population-level calculation for the aggregation-stress experiment.
Section~\ref{subsec:two_stage_supplement} compares joint and
two-stage estimation using the same $R=10$ master datasets
with $n=1000$ donors from the finite-sample experiment.

\subsection{Finite-sample recovery}
\label{supp:finite}

\paragraph{Data generation.}
For each of $R=10$ independent master replicates, we generated a
dataset containing $1000$ donors. A random permutation of the donor
indices was then used to construct the nested samples
\[
\mathcal D_{200}
\subset
\mathcal D_{500}
\subset
\mathcal D_{1000},
\]
with all cells from each selected donor retained. The exposure and
number of cells per donor were generated independently as
\[
Z_i\sim N(0,1),
\qquad
C_i\sim
\operatorname{DiscreteUniform}\{10,\ldots,200\}.
\]

Both the data-generating and fitted models used a $K=5$ cubic
B-spline basis on $[0,1]$. The data-generating coefficient model was
\begin{align*}
\bm\alpha_i
&=
\bm\alpha_0^0+A^0Z_i+\bm\eta_i,
&
\bm\eta_i
&\sim
N(\bm0,0.01I_5),
\\
\bm\beta_i
&=
\bm\beta_0^0+C^0Z_i+B^0\bm\alpha_i+\bm\xi_i,
&
\bm\xi_i
&\sim
N(\bm0,0.0001I_5),
\end{align*}
where
\begin{align*}
\bm\alpha_0^0
&=
(0.3,\;0.4,\;0.1,\;0.1,\;0.2)^\top,
&
\bm\beta_0^0
&=
(0.2,\;0.3,\;0.2,\;0.2,\;0.1)^\top,
\\
A^0
&=
(0.3,\;0.5,\;0.8,\;0.7,\;0.8)^\top,
&
B^0
&=
\operatorname{diag}(0.4,\;0.5,\;0.7,\;0.6,\;0.6),
\\
C^0
&=
(0.3,\;0.4,\;0.2,\;0.5,\;0.7)^\top.
\end{align*}
Cell-level observations were generated from
\begin{equation*}
X_{ic}
=
\bm\phi(t_{ic}^0)^\top\bm\alpha_i+\epsilon_{ic},
\qquad
Y_{ic}
=
\bm\phi(t_{ic}^0)^\top\bm\beta_i+\delta_{ic},
\end{equation*}
with independent
\[
\epsilon_{ic}\sim N(0,0.01),
\qquad
\delta_{ic}\sim N(0,0.01).
\]
For donor $i$, the true state coordinates were fixed at the
equally spaced grid
\[
t_{ic}^0
=
\frac{c-1}{C_i-1},
\qquad
c=1,\ldots,C_i.
\]
The first $\lfloor C_i/2\rfloor$ cells were assigned to the lower
state interval and the remaining cells to the upper interval. During model fitting,
the true latent states were not supplied; cells in the lower and upper intervals
were instead restricted to $[0,0.5]$ and $[0.5,1]$, respectively.

\paragraph{Evaluation.}
For coefficient component $\theta_k$, component-wise recovery was
summarized by
\[
\operatorname{Bias}(\widehat\theta_k)
=
\frac{1}{R}
\sum_{r=1}^{R}
(\widehat\theta_{rk}-\theta_k^0),
\qquad
\operatorname{RMSE}(\widehat\theta_k)
=
\left\{
\frac{1}{R}
\sum_{r=1}^{R}
(\widehat\theta_{rk}-\theta_k^0)^2
\right\}^{1/2}.
\]
Donor-specific trajectory recovery was evaluated by
\[
\operatorname{IMSE}_{g,r}
=
\frac{1}{n}
\sum_{i=1}^{n}
\int_0^1
\{\widehat g_{ir}(t)-g_{ir}^0(t)\}^2\,dt,
\]
with $\operatorname{IMSE}_{h,r}$ defined analogously. Recovery of
$\ME(t)$, $\DE(t)$, and $\TE(t)$ was assessed within replicate by
\[
\operatorname{RISE}(\widehat E,E^0)
=\left[\int_0^1\{\widehat E(t)-E^0(t)\}^2\,dt\right]^{1/2},
\]
and then averaged over the independent replicates.

\begin{figure}[htbp]
\centering
\suppfig[width=\linewidth]
{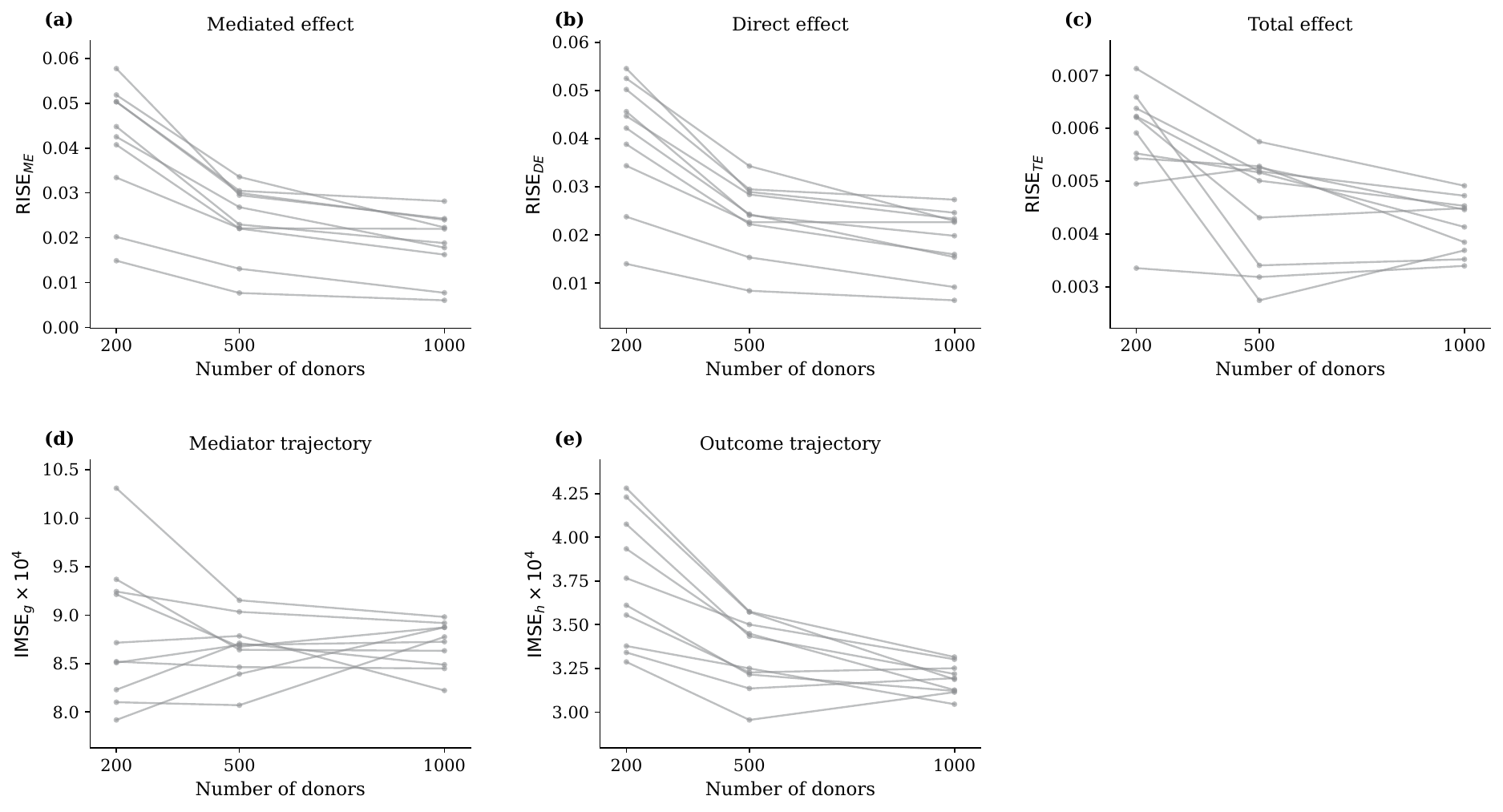}
\caption{Replicate-level finite-sample recovery across the nested
sample sizes. Panels~(a)--(c) show RISE for the mediated, direct, and
total effect functions, respectively, and panels~(d)--(e) show
donor-averaged IMSE for the mediator and outcome trajectories,
multiplied by $10^4$. RISE is shown on its original scale.
Each gray line connects results from the same master replicate;
there are $R=10$ independent master replicates. No interval
estimates are displayed.}
\label{fig:finite_additional}
\end{figure}

\paragraph{Additional results.}
Table~\ref{tab:finite_additional} reports component-wise path recovery
and donor-specific trajectory errors. Figure~\ref{fig:finite_additional}
shows the corresponding paired replicate-level functional and
trajectory recovery across the nested sample sizes.

\begin{table}[htbp]
\centering
\small
\renewcommand{\arraystretch}{1.10}
\setlength{\tabcolsep}{5pt}
\caption{Additional finite-sample recovery summaries under the fixed
basis dimension $K=5$. Panel A reports component-wise bias and RMSE
for the coefficient-level path parameters; each entry is
$\operatorname{Bias}\;(\operatorname{RMSE})$ and is multiplied by
$10^2$. Panel B reports donor-averaged trajectory IMSE, multiplied by
$10^4$, with simulation Monte Carlo standard errors in parentheses.
Results are based on $R=10$ master replicates.}
\label{tab:finite_additional}

\textit{Panel A: Component-wise path recovery}

\vspace{3pt}

\begin{tabular}{llccc}
\toprule
Parameter block
& Component
& $n=200$
& $n=500$
& $n=1000$
\\
\midrule
$A$
& $a_1$ & $-0.220\;(0.899)$ & $-0.103\;(0.574)$ & $-0.048\;(0.420)$\\
& $a_2$ & $-1.162\;(1.410)$ & $-0.856\;(1.072)$ & $-1.041\;(1.130)$\\
& $a_3$ & $-0.550\;(1.110)$ & $-0.592\;(0.884)$ & $-0.377\;(0.623)$\\
& $a_4$ & $ 0.563\;(0.907)$ & $ 0.551\;(0.840)$ & $ 0.383\;(0.511)$\\
& $a_5$ & $-0.191\;(0.896)$ & $-0.227\;(0.664)$ & $-0.067\;(0.425)$\\

\addlinespace
$\operatorname{diag}(B)$
& $b_1$ & $-0.389\;(1.897)$ & $ 0.447\;(1.491)$ & $ 0.794\;(1.854)$\\
& $b_2$ & $-1.405\;(3.011)$ & $-0.607\;(1.764)$ & $-0.205\;(1.228)$\\
& $b_3$ & $-13.062\;(13.585)$ & $-8.037\;(8.324)$ & $-6.504\;(6.872)$\\
& $b_4$ & $-2.514\;(3.457)$ & $-0.969\;(1.531)$ & $-0.380\;(1.227)$\\
& $b_5$ & $-1.411\;(4.827)$ & $-0.307\;(1.899)$ & $ 0.248\;(1.888)$\\

\addlinespace
$C$
& $c_1$ & $ 0.819\;(1.106)$ & $ 0.512\;(0.694)$ & $ 0.398\;(0.595)$\\
& $c_2$ & $-0.790\;(1.725)$ & $-0.974\;(1.338)$ & $-1.054\;(1.198)$\\
& $c_3$ & $11.779\;(12.082)$ & $ 7.710\;(7.943)$ & $ 6.315\;(6.570)$\\
& $c_4$ & $ 2.028\;(2.528)$ & $ 0.929\;(1.327)$ & $ 0.601\;(1.059)$\\
& $c_5$ & $-0.062\;(3.762)$ & $-0.789\;(1.822)$ & $-1.255\;(1.933)$\\
\bottomrule
\end{tabular}

\vspace{8pt}

\textit{Panel B: Donor-specific trajectory recovery}

\vspace{3pt}

\begin{tabular}{lccc}
\toprule
Target
& $n=200$
& $n=500$
& $n=1000$
\\
\midrule
$\operatorname{IMSE}_g\times10^4$
& $9.021\;(0.232)$
& $8.880\;(0.096)$
& $8.925\;(0.080)$
\\
$\operatorname{IMSE}_h\times10^4$
& $3.852\;(0.114)$
& $3.423\;(0.062)$
& $3.282\;(0.028)$
\\
\bottomrule
\end{tabular}
\end{table}

Block-average errors decreased with donor sample size, but some
component biases remained. At $n=1000$, the biases of $b_3$ and
$c_3$ were approximately $-0.0650$ and $0.0632$; the latter is
about 32\% of its true value 0.2. These remaining component
errors coexist with more accurate total-effect recovery.
Mean functional mediated- and direct-effect recovery improved,
whereas donor-specific trajectory errors changed more modestly.

\subsection{Sensitivity to functional basis complexity}
\label{supp:basis}

\paragraph{Data generation.}
For each of $R=10$ independent simulation replicates, we generated a
dataset containing $n=500$ donors. The exposure and donor-specific
cell counts were generated independently as
\[
Z_i\sim N(0,1),
\qquad
C_i\sim
\operatorname{DiscreteUniform}\{10,\ldots,200\}.
\]
Within each donor, $\lfloor C_i/2\rfloor$ state coordinates were
sampled independently from $\operatorname{Uniform}(0,0.5)$ and the
remaining locations from $\operatorname{Uniform}(0.5,1)$.

Unlike the finite-sample experiment, the functional truths were
defined independently of the B-spline representation used for
fitting. Specifically,
\begin{align*}
g_i^0(t)
&=
\mu_X(t)+a(t)Z_i+u_i(t),
\\
h_i^0(t)
&=
\mu_Y(t)+c(t)Z_i+\kappa g_i^0(t)+v_i(t),
\end{align*}
where $\kappa=-1$ and
\begin{align*}
\mu_X(t)
&=
-0.4+0.5t+0.25\sin(2\pi t),
\\
a(t)
&=
-0.05+0.65\sin(\pi t)+t,
\\
\mu_Y(t)
&=
0.20-0.25t+0.20\cos(2\pi t),
\\
c(t)
&=
-0.15+1.55t+0.48\sin(\pi t).
\end{align*}

Donor-specific deviations were generated from a seven-dimensional
Fourier expansion,
\[
u_i(t)
=
\sum_{\ell=1}^{7}q_{i\ell}\psi_\ell(t),
\qquad
v_i(t)
=
\sum_{\ell=1}^{7}r_{i\ell}\psi_\ell(t),
\]
with
\[
\psi_1(t)=1,
\qquad
\psi_{2m}(t)=\sqrt{2}\sin(2\pi mt),
\qquad
\psi_{2m+1}(t)=\sqrt{2}\cos(2\pi mt),
\quad m=1,2,3,
\]
and independent coefficients
\[
q_{i\ell}
\sim
N(0,\lambda_\ell^X),
\qquad
r_{i\ell}
\sim
N(0,\lambda_\ell^Y),
\]
where
\[
\lambda_\ell^X
=
0.04(0.45)^{\ell-1},
\qquad
\lambda_\ell^Y
=
0.025(0.45)^{\ell-1}.
\]
Finally,
\[
X_{ic}
=
g_i^0(t_{ic})+\epsilon_{ic},
\qquad
Y_{ic}
=
h_i^0(t_{ic})+\delta_{ic},
\]
with independent
$\epsilon_{ic},\delta_{ic}\sim N(0,0.01)$.

The resulting population-level effect functions are
\begin{align*}
\ME^0(t)
&=
0.05-t-0.65\sin(\pi t),
\\
\DE^0(t)
&=
-0.15+1.55t+0.48\sin(\pi t),
\\
\TE^0(t)
&=
-0.10+0.55t-0.17\sin(\pi t).
\end{align*}
Their uniform integrals are
\[
\IME^0=-0.8638,
\qquad
\IDE^0=0.9306,
\qquad
\ITE^0=0.0668.
\]
Thus, none of the candidate B-spline dimensions corresponds to a
data-generating value of $K$.

Each generated dataset was fitted separately using cubic B-spline
bases with
\[
K\in\{4,5,6,7,8,9\}.
\]
The same dataset was used for all candidate dimensions within each
replicate. Recovery was evaluated on the functional scale using the
RISE and donor-averaged IMSE criteria defined in Section~\ref{supp:finite}.
Five-fold held-out-cell cross-validation was also conducted using the
same fold assignments across $K$ within each replicate.

\paragraph{Conditional plug-in basis selection.}
For each of the $R=10$ simulation replicates, we applied
the procedure in Section~\ref{supp:cv} to
$K\in\mathcal K=\{4,5,6,7,8,9\}$. Each candidate used its
own full-data posterior mean states. Cells were assigned
to five approximately balanced folds within donor and coarse-state
strata, with the same assignments across candidates within
each replicate. All remaining parameters were refitted by MCMC
using the training expression observations at the fixed locations.

Held-out joint densities and their log scores were computed
using Equations~\eqref{eq:cv_fixed_time_density}
and~\eqref{eq:cv_logmean}. Each cell was scored once.
The donor-balanced score in
Equation~\eqref{eq:conditional_basis_selection} selected
$\widehat K_r$ separately for each replicate; cell-weighted
scores were secondary, and exact ties were resolved by the
smaller dimension. Effect and trajectory recovery were then
evaluated from the selected full latent-state fit.
The conditional interpretation and distinction from new-donor
prediction are as described in Section~\ref{supp:cv}.

\paragraph{Additional results.}
All six candidate dimensions had complete five-fold scores
in each replicate. Table~\ref{tab:basis_additional}
reports donor-level trajectory recovery.

\begin{table}[htbp]
\centering\small
\caption{Donor-trajectory integrated mean squared errors (IMSEs)
for $R=10$ paired simulation replicates. IMSE is multiplied by
$10^2$; entries are means with simulation Monte Carlo standard
errors in parentheses.}
\label{tab:basis_additional}
\begin{tabular}{ccc}
\toprule
$K$ & $\operatorname{IMSE}_g$ & $\operatorname{IMSE}_h$\\
\midrule
4 & $5.69\;(0.24)$ & $6.48\;(0.21)$\\
5 & $3.70\;(0.21)$ & $4.20\;(0.17)$\\
6 & $5.37\;(0.37)$ & $5.25\;(0.23)$\\
7 & $7.08\;(0.38)$ & $5.98\;(0.16)$\\
8 & $4.55\;(0.41)$ & $4.65\;(0.21)$\\
9 & $8.76\;(0.76)$ & $6.43\;(0.33)$\\
\bottomrule
\end{tabular}
\end{table}

For a replicate-level quantity $Q_r$, the reported
simulation Monte Carlo standard error is
$s_Q/\sqrt{10}$, with $s_Q$ the sample standard
deviation over the ten independently generated datasets.
This measures simulation variability, not posterior
uncertainty or variation from treating cells as independent
simulation replicates.

Across the ten replicates, full latent-state fits at the
conditionally selected dimensions $\widehat K_r$ yielded
donor-trajectory IMSE means of 0.04714 and 0.04720 for the
mediator and outcome, respectively. Each IMSE was evaluated
at the dimension selected in the same replicate.

Table~\ref{tab:basis_integrated} reports recovery of the
integrated effects. The effects are estimated from the
full latent-state models rather than the fixed-state
cross-validation fits.

\begin{table}[htbp]
\centering
\small
\caption{Recovery of the integrated mediation effects in the
basis-complexity experiment. Entries report the mean estimate across
the $R=10$ paired simulation replicates, with RMSE in parentheses.
The true values are $\IME^0=-0.8638$, $\IDE^0=0.9306$, and
$\ITE^0=0.0668$.}
\label{tab:basis_integrated}
\begin{tabular}{cccc}
\toprule
$K$ & IME & IDE & ITE \\
\midrule
$4$ & $-0.799\;(0.071)$ & $0.873\;(0.065)$ & $0.074\;(0.012)$ \\
$\mathbf{5}$ &
$\mathbf{-0.832\;(0.040)}$ &
$\mathbf{0.900\;(0.040)}$ &
$\mathbf{0.068\;(0.008)}$ \\
$6$ & $-0.790\;(0.078)$ & $0.861\;(0.075)$ & $0.071\;(0.010)$ \\
$7$ & $-0.792\;(0.074)$ & $0.861\;(0.073)$ & $0.070\;(0.010)$ \\
$8$ & $-0.769\;(0.102)$ & $0.840\;(0.099)$ & $0.072\;(0.011)$ \\
$9$ & $-0.763\;(0.103)$ & $0.834\;(0.100)$ & $0.071\;(0.010)$ \\
\bottomrule
\end{tabular}
\end{table}

Across the paired replicates, $K=5$ attained the smallest
RISE for $\ME(t)$, $\DE(t)$, and $\TE(t)$ in 8, 8, and 7
of the 10 replicates, respectively, and the smallest
trajectory IMSE for both $g_i(t)$ and $h_i(t)$ in 7 of
the 10 replicates. It also gave the smallest RMSE for
IME, IDE, and ITE in Table~\ref{tab:basis_integrated}.
These recovery comparisons describe the accuracy of the
fitted functional targets and are distinct from the
conditional predictive criterion used for selection.

\begin{figure}[htbp]
\centering
\suppfig[width=\linewidth]
{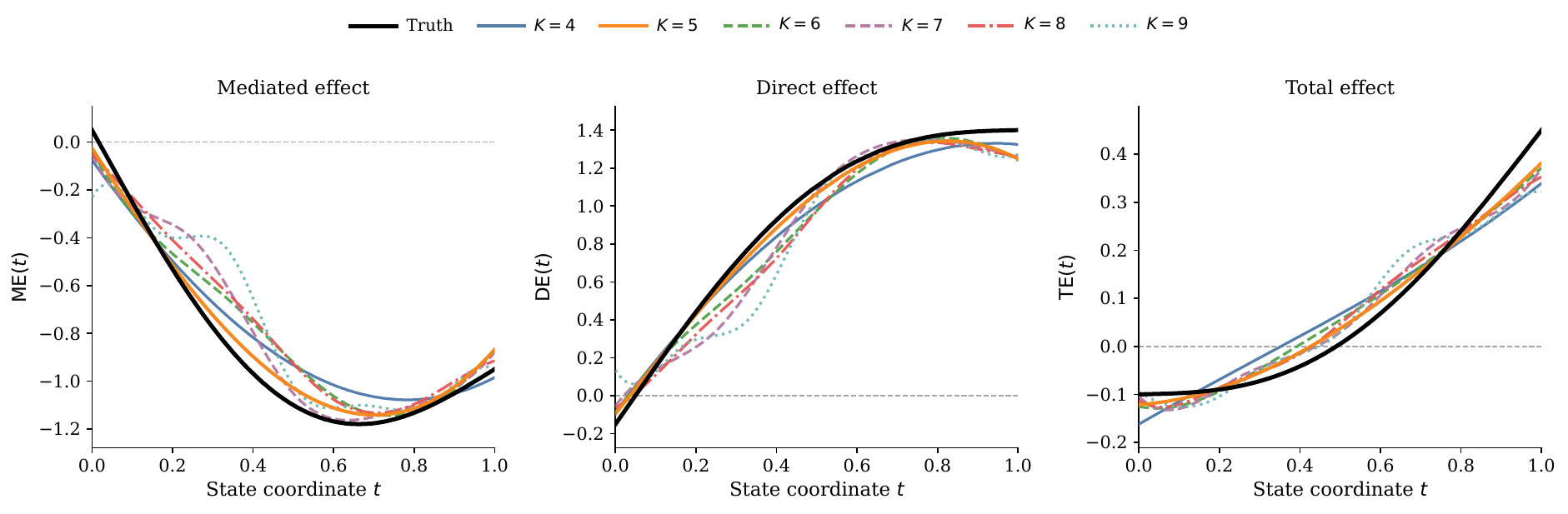}
\caption{Recovery of the state-specific mediated, direct, and total
effect functions across all candidate working basis dimensions.
The black curve denotes the true function, and the remaining curves
denote posterior mean effect functions averaged across the $R=10$
paired simulation replicates for $K=4,\ldots,9$.}
\label{fig:basis_effect_curves_all}
\end{figure}

Figure~\ref{fig:basis_effect_curves_all} compares the
posterior mean effect functions across the candidate
working dimensions. The $K=5$ fit most closely recovered
the component-specific effects on average, whereas larger
basis dimensions exhibited more local variation in the
mediated and direct effects. The total-effect function was
comparatively less sensitive across intermediate dimensions.

\subsection{Consequences of donor-level aggregation}
\label{supp:aggregation}

\paragraph{Data generation.}
The aggregation-stress experiment used $R=10$ independently generated
datasets, each containing $n=1000$ donors. We generated
\[
Z_i\sim N(0,1),
\qquad
C_i\sim
\operatorname{DiscreteUniform}\{10,\ldots,200\},
\]
and placed the true state coordinates on an equally spaced grid
over $[0,1]$. The first $\lfloor C_i/2\rfloor$ cells were assigned
to the lower state interval and the remaining cells to the upper interval.

Data were generated using a cubic B-spline basis with $K=4$ and
\begin{align*}
\bm\alpha_i
&=
\bm\alpha_0+A Z_i+\bm\eta_i,
&
\bm\eta_i
&\sim N(\bm0,0.01I_4),
\\
\bm\beta_i
&=
\bm\beta_0+C Z_i+B\bm\alpha_i+\bm\xi_i,
&
\bm\xi_i
&\sim N(\bm0,0.0001I_4),
\end{align*}
where
\begin{align*}
\bm\alpha_0
&=
(0.3,\;0.4,\;0.1,\;0.1)^\top,
&
\bm\beta_0
&=
(0.2,\;0.3,\;0.2,\;0.2)^\top,
\\
A
&=
(0.4,\;0.4,\;-0.4,\;-0.4)^\top,
&
B
&=
\operatorname{diag}(1,\;1,\;0.25,\;0.25),
\\
C
&=
(0.2,\;0.2,\;0.2,\;0.2)^\top.
\end{align*}
Cell-level observations were generated as
\[
X_{ic}
=
\bm\phi(t_{ic})^\top\bm\alpha_i+\epsilon_{ic},
\qquad
Y_{ic}
=
\bm\phi(t_{ic})^\top\bm\beta_i+\delta_{ic},
\]
with independent
\[
\epsilon_{ic},\delta_{ic}\sim N(0,0.01).
\]

The corresponding integrated functional targets were
\[
\IME^0=0.15,\qquad
\IDE^0=0.20,\qquad
\ITE^0=0.35.
\]
The donor-average scalar analysis instead targets
\[
(0,\;0.35,\;0.35)
\]
for the indirect, direct, and total effects, respectively.

\paragraph{Comparison procedures.}
Each dataset was analyzed using BLaFM, an oracle analysis in which the true
latent states were treated as known, and a conventional
product-of-coefficients mediation analysis based on donor-averaged
mediator and outcome measurements.

\paragraph{Fitted basis and scalar procedure.}
BLaFM and the oracle analysis both used cubic B-splines with
$K=4$, matching the data-generating basis dimension. The oracle
analysis holds latent states at their true values throughout MCMC,
whereas BLaFM updates them within the
annotation intervals. Both analyses use the same coefficient
and variance prior hyperparameters as the other functional fits.

The scalar procedure first calculates ordinary within-donor
cell averages $\bar X_i$ and $\bar Y_i$. It then fits, by
unweighted ordinary least squares with an intercept,
\[
\bar X_i=a_0^{(s)}+a^{(s)}Z_i+u_i^{(s)},\qquad
\bar Y_i=b_0^{(s)}+c^{(s)}Z_i+b^{(s)}\bar X_i+v_i^{(s)}.
\]
Each donor contributes one equally weighted row, regardless
of cell count. The estimated indirect, direct, and total
components are $\widehat a^{(s)}\widehat b^{(s)}$,
$\widehat c^{(s)}$, and their sum. A separate regression of
$\bar Y_i$ on $Z_i$ checks the marginal total slope.
The scalar procedure includes conventional and HC3 coefficient
covariance estimates and a donor-level percentile bootstrap with
2,000 resamples for within-dataset uncertainty assessment. These
quantities are distinct from the reported simulation MCSEs,
which summarize variability across the ten independently
generated datasets.

\subsection{Comparison with two-stage estimation}
\label{subsec:two_stage_supplement}

\paragraph{Comparison design.}
We reused the $R=10$ master datasets with $n=1000$ donors from
Section~\ref{supp:finite}. Each dataset was analyzed using BLaFM, two-stage estimation with estimated states,
two-stage estimation with estimated coefficients, and an oracle
procedure with known states. All four procedures used the same
$K=5$ cubic B-spline basis. The two-stage procedures shared a
single first-stage fit within each master replicate, and comparisons
were paired within replicate.

\paragraph{First-stage model and computation.}
The first stage used the cell-level observation likelihood in
Section~\ref{supp:posterior_computation}, with conditionally
independent observation errors and shared states $t_{ic}$ for the
mediator and outcome measurements. States had uniform priors on
the annotation-specific intervals $[0,1/2]$ or $[1/2,1]$, and
the observation variances had independent
$\operatorname{IG}(2,0.1)$ priors. The donor-level coefficient
model was
\begin{align*}
\bm\alpha_i\mid\bm\mu_\alpha,D_\alpha
&\sim N_5(\bm\mu_\alpha,D_\alpha),&
D_\alpha&=\operatorname{diag}(v_{\alpha1},\ldots,v_{\alpha5}),\\
\bm\beta_i\mid\bm\mu_\beta,D_\beta
&\sim N_5(\bm\mu_\beta,D_\beta),&
D_\beta&=\operatorname{diag}(v_{\beta1},\ldots,v_{\beta5}).
\end{align*}
Independent hyperpriors were
\[
\mu_{\alpha k},\mu_{\beta k}\sim N(0,10^2),\qquad
v_{\alpha k},v_{\beta k}\sim\operatorname{IG}(2,0.1),
\quad k=1,\ldots,5,
\]
with the inverse-gamma parameterization given in
Section~\ref{supp:posterior_computation}. The population-mean
priors were independent of the population variances. Conditional
on these population parameters, coefficient vectors were
independent across donors and between the mediator and outcome
blocks. Exposure and the mediation regressions did not enter
this model, although the shared unknown states could induce
posterior dependence between the two coefficient blocks.
$D_\alpha$ and $D_\beta$ describe first-stage population variation,
not the conditional residual variation in the mediation regressions.

The sampler used Gibbs updates for the coefficient vectors,
population means, population variances, and observation variances,
together with the annotation-constrained Metropolis--Hastings
state updates used in BLaFM. Let $\widehat t_{ic}$,
$\widehat{\bm\alpha}_i$, and $\widehat{\bm\beta}_i$ denote the
first-stage posterior means after burn-in.

\paragraph{Second stage with estimated states.}
This procedure fixed $t_{ic}=\widehat t_{ic}$ and re-estimated all remaining parameters of the
BLaFM model specified in Section~\ref{supp:posterior_computation}, with
diagonal $B$, $\Sigma$, and $\Lambda$. Both sets of donor-specific
trajectory coefficients and the observation variances were
re-estimated jointly with the mediation parameters, using the BLaFM priors and Gibbs updates. The mediation regressions supplied
the coefficient prior structure; the first-stage population means
and variances did not enter this stage.

\paragraph{Second stage with estimated coefficients.}
This procedure instead treated the first-stage posterior mean
coefficient vectors as observed inputs to the donor-level regressions,
\begin{align*}
\widehat{\bm\alpha}_i
&=\bm\alpha_0+AZ_i+\bm\eta_i,
&\bm\eta_i&\sim N_5(\bm0,\Sigma),\\
\widehat{\bm\beta}_i
&=\bm\beta_0+CZ_i+B\widehat{\bm\alpha}_i+\bm\xi_i,
&\bm\xi_i&\sim N_5(\bm0,\Lambda).
\end{align*}
The matrices $B$, $\Sigma$, and $\Lambda$ were diagonal. The
regression parameters, residual variances, and shrinkage
hyperparameters were updated using the corresponding mediation-regression
priors and Gibbs updates from BLaFM. The coefficient vectors remained fixed:
the cell-level likelihood, state updates, and observation-variance
updates were not used in this stage. The first-stage population
parameters did not enter either second-stage model. This procedure
conditions on estimated trajectories without propagating their
first-stage posterior uncertainty.

\paragraph{Oracle analysis.}
The oracle used the same conditional fitting procedure as the
analysis with estimated states, replacing $\widehat t_{ic}$ by
the true states and re-estimating all remaining parameters of the BLaFM model.
It therefore provides an oracle reference for conditioning on
states, rather than on trajectory coefficients.

\paragraph{Evaluation.}
Evaluation focused on recovery of the mediated, direct, and total
effect functions and their integrated summaries. For a unit
exposure contrast, posterior mean effects were calculated from
the retained draws as in Section~\ref{supp:posterior_computation},
with $B^{(m)}A^{(m)}$ formed within each draw before averaging.
Effect-function RISE and donor-averaged trajectory IMSE were
defined as in Section~\ref{supp:finite}. Integrals were approximated
by the trapezoidal rule on the saved grid over $[0,1]$.
RISE was computed within replicate before averaging. Paired RISE
differences were calculated for each two-stage procedure minus
BLaFM within the same replicate; positive differences favor
BLaFM.
Integrated-effect recovery was summarized by the mean estimate,
bias, and RMSE across replicates.

Trajectory and path-coefficient recovery provided supporting
diagnostics under the specified spline representation.
Trajectory IMSE weighted donors equally and was then averaged
across replicates. The procedure with estimated states used
coefficients re-estimated in the second stage, whereas the procedure
with estimated coefficients used the fixed first-stage posterior
means. For a path-coefficient block
$\vartheta\in\{A,\operatorname{diag}(B),C\}$, we used posterior
mean estimates $\widehat\vartheta_{rk}$ to calculate the pooled
component RMSE,
\[
\operatorname{RMSE}_{\vartheta}
=\left\{\frac{1}{RK}\sum_{r=1}^{R}\sum_{k=1}^{K}
  (\widehat\vartheta_{rk}-\vartheta_k^0)^2\right\}^{1/2}.
\]

Simulation Monte Carlo standard errors (MCSEs) for means, biases,
and paired differences were calculated as the between-replicate
sample standard deviation divided by $\sqrt{R}$, with $R=10$.
MCSEs for integrated-effect RMSEs and pooled component RMSEs
used a delete-one-replicate jackknife. These simulation MCSEs
are distinct from MCMC standard errors and posterior uncertainty.

\paragraph{Additional results.}
For the procedure with estimated states, mean paired RISE
differences relative to BLaFM were
$0.09139$ ($0.01275$), $0.13607$ ($0.00428$), and
$0.17739$ ($0.00588$) for $\ME$, $\DE$, and $\TE$, respectively.
For the procedure with estimated coefficients, the corresponding
differences were $0.16036$ ($0.00850$), $0.18371$ ($0.00868$), and
$0.04257$ ($0.00206$). Parentheses contain simulation MCSEs.
All six mean differences favor BLaFM.

Table~\ref{tab:two_stage_integrated} reports integrated-effect
recovery. Both BLaFM and the oracle analysis had smaller RMSEs
than either two-stage procedure for all three integrated effects.
Both two-stage procedures overestimated $\IME$ and underestimated
$\IDE$. For the procedure with estimated coefficients, the
opposing biases of $0.0917$ and $-0.1043$ partially canceled,
giving an $\ITE$ bias of $-0.0126$. Its smaller total-effect bias
therefore did not indicate accurate recovery of the decomposition.
Opposing biases in $\IME$ and $\IDE$ also remained under the
oracle analysis.

\begin{table}[htbp]
\centering
\small
\renewcommand{\arraystretch}{1.10}
\setlength{\tabcolsep}{4pt}
\caption{Integrated-effect recovery across $R=10$ paired
replicates at $n=1000$ and $K=5$, on the original scale. The true
values are approximately $(\IME^0,\IDE^0,\ITE^0)
=(0.3825,0.4000,0.7825)$. Parentheses following bias and RMSE
contain simulation MCSEs.}
\label{tab:two_stage_integrated}
\begin{tabular}{lrrr}
\toprule
Target & Mean & Bias & RMSE \\
\midrule
\multicolumn{4}{l}{\textit{BLaFM}} \\
$\IME$ & $0.3677$ & $-0.0148\;(0.0021)$ & $0.0160\;(0.0018)$ \\
$\IDE$ & $0.4136$ & $0.0136\;(0.0021)$ & $0.0149\;(0.0018)$ \\
$\ITE$ & $0.7813$ & $-0.0012\;(0.0002)$ & $0.0014\;(0.0003)$ \\
\addlinespace[4pt]
\multicolumn{4}{l}{\textit{Two-stage with estimated states}} \\
$\IME$ & $0.4786$ & $0.0961\;(0.0098)$ & $0.1006\;(0.0089)$ \\
$\IDE$ & $0.3436$ & $-0.0564\;(0.0096)$ & $0.0633\;(0.0088)$ \\
$\ITE$ & $0.8223$ & $0.0398\;(0.0019)$ & $0.0402\;(0.0019)$ \\
\addlinespace[4pt]
\multicolumn{4}{l}{\textit{Two-stage with estimated coefficients}} \\
$\IME$ & $0.4742$ & $0.0917\;(0.0086)$ & $0.0952\;(0.0078)$ \\
$\IDE$ & $0.2957$ & $-0.1043\;(0.0086)$ & $0.1074\;(0.0080)$ \\
$\ITE$ & $0.7699$ & $-0.0126\;(0.0005)$ & $0.0127\;(0.0005)$ \\
\addlinespace[4pt]
\multicolumn{4}{l}{\textit{Oracle with known states}} \\
$\IME$ & $0.3672$ & $-0.0153\;(0.0014)$ & $0.0159\;(0.0013)$ \\
$\IDE$ & $0.4149$ & $0.0149\;(0.0015)$ & $0.0156\;(0.0013)$ \\
$\ITE$ & $0.7822$ & $-0.0003\;(0.0003)$ & $0.0009\;(0.0003)$ \\
\bottomrule
\end{tabular}
\end{table}

Table~\ref{tab:two_stage_additional} reports supporting trajectory
and path-coefficient diagnostics. Both two-stage procedures had
larger trajectory IMSEs and path-coefficient RMSEs than BLaFM. Compared with the procedure with estimated states, the
procedure with estimated coefficients had smaller trajectory
IMSEs and a smaller RMSE for $A$, but larger RMSEs for
$\operatorname{diag}(B)$ and $C$. These comparisons reflect the
first-stage model and the quantities held fixed in the second
stage; they do not isolate uncertainty propagation alone.

\begin{table}[htbp]
\centering
\small
\renewcommand{\arraystretch}{1.10}
\setlength{\tabcolsep}{4pt}
\caption{Additional recovery summaries for $R=10$ paired
replicates at $n=1000$ and $K=5$. Panel A reports mean
trajectory IMSEs with donors weighted equally and simulation
MCSEs in parentheses; both are multiplied by $10^2$.
The procedure with estimated states uses second-stage coefficient
estimates; the procedure with estimated coefficients uses fixed
first-stage posterior means. Panel B reports pooled
component RMSE on the original scale, with delete-one-replicate
jackknife MCSEs in parentheses.}
\label{tab:two_stage_additional}

\textit{Panel A: Trajectory recovery ($\times 10^2$)}

\vspace{3pt}
\begin{tabular}{lrr}
\toprule
Method & $\operatorname{IMSE}_g$ & $\operatorname{IMSE}_h$ \\
\midrule
BLaFM & $0.0893\;(0.0008)$ & $0.0328\;(0.0003)$ \\
Two-stage with estimated states & $9.7342\;(0.3325)$ & $14.9891\;(0.4763)$ \\
Two-stage with estimated coefficients & $0.3012\;(0.0129)$ & $0.6473\;(0.0386)$ \\
Oracle with known states & $0.0446\;(0.0003)$ & $0.0201\;(0.0002)$ \\
\bottomrule
\end{tabular}

\vspace{9pt}
\textit{Panel B: Path-coefficient recovery (original scale)}

\vspace{3pt}
\begin{tabular}{lrrr}
\toprule
Method & $A$ & $\operatorname{diag}(B)$ & $C$ \\
\midrule
BLaFM & $0.0068\;(0.0004)$ & $0.0338\;(0.0026)$ & $0.0316\;(0.0020)$ \\
Two-stage with estimated states & $0.1253\;(0.0051)$ & $0.2353\;(0.0085)$ & $0.2775\;(0.0112)$ \\
Two-stage with estimated coefficients & $0.0387\;(0.0013)$ & $0.3919\;(0.0145)$ & $0.3257\;(0.0124)$ \\
Oracle with known states & $0.0041\;(0.0003)$ & $0.0285\;(0.0023)$ & $0.0206\;(0.0021)$ \\
\bottomrule
\end{tabular}
\end{table}

\section{Additional real-data results}
\label{supp:application}

\subsection{Data construction and candidate selection}
\label{supp:data}

The OneK1K data release used in this analysis contained genotype
data for 961 donors and expression measurements and metadata
for 929,817 cells from those donors. All 961 donors were retained
in the CD4$^+$ T-cell subset, which contained 403,087 cells.
The published OneK1K cohort contains 982 donors
\citep{Yazar2022}; the difference in donor counts reflects
the data release used here rather than additional donor exclusions.
OneK1K study data are accessible through the Human Cell Atlas
Data Portal.\footnote{OneK1K data collection:
\url{https://explore.data.humancellatlas.org/projects/f2078d5f-2e7d-4844-8552-f7c41a231e52}.}

Expression measurements were matched to cell metadata by barcode
and to rs1131017 genotypes by donor identifier. Genotype calls
$0/0$, $0/1$, and $1/1$ occurred in 322, 473, and 166 donors,
respectively, and were analyzed as dosage $Z=0,1,2$.
The analyzed rs1131017 C/G record is reported on GRCh37/hg19
at chromosome 12, position 56,435,929, with REF=C and ALT=G.
NCBI lists this allele as \texttt{NC\_000012.11:g.56435929C>G}
on GRCh37 and \texttt{NC\_000012.12:g.56042145C>G}
on GRCh38.\footnote{NCBI dbSNP rs1131017 genomic placements:
\url{https://www.ncbi.nlm.nih.gov/snp/rs1131017}.}
The assembly labels distinguish two reference placements
of the same variant. Thus $Z=0,1,2$ counts G alleles,
corresponding to C/C, C/G, and G/G, and a unit contrast
represents one additional G allele.

Cells were included according to the annotations in
Table~\ref{tab:cell_annotation_map}. Naive cells were assigned
to the lower state interval ($S=0$), and central-memory and
effector-memory cells to the upper interval ($S=1$). Other cell
types, including separately annotated regulatory and cytotoxic
T cells, were excluded.

\begin{table}[htbp]
\centering\small
\caption{CD4$^+$ T-cell annotations and coarse state groups.}
\label{tab:cell_annotation_map}
\begin{tabular}{p{0.64\linewidth}cc}\toprule
Exact metadata annotation&$S_{ic}$&State support\\\midrule
naive thymus-derived CD4-positive, alpha-beta T cell&0&$[0,0.5]$\\
central memory CD4-positive, alpha-beta T cell&1&$[0.5,1]$\\
effector memory CD4-positive, alpha-beta T cell&1&$[0.5,1]$\\
\bottomrule\end{tabular}
\end{table}

\begin{table}[htbp]
\centering
\caption{Final analytic sample for the OneK1K application.
State 0 contains naive cells; state 1 combines central-memory
and effector-memory cells.}
\label{tab:application_sample}
\begin{tabular}{lr}\toprule
Quantity&Value\\\midrule
Available genotype donors&961\\
Retained donors&961\\
CD4$^+$ T cells&403,087\\
Donors with $Z=0$&322\\
Donors with $Z=1$&473\\
Donors with $Z=2$&166\\
Cells with $S=0$&181,285\\
Cells with $S=1$&221,802\\
Cells per donor, median (range)&414 (19--1,031)\\
Selected primary basis dimension&$K=7$\\
\bottomrule\end{tabular}
\end{table}

\paragraph{Expression preprocessing.}
Expression preprocessing started from the RNA-assay UMI count matrix
in the available Seurat object. SCTransform was then applied
to the retained count matrix before construction of the CD4$^+$ T-cell
analytic subset \citep{HafemeisterSatija2019,Hao2023}. The
negative-binomial regression step used a subsample of 1,000 cells
(\texttt{ncells=1000}), and Pearson residuals were requested for
RPS26 (ENSG00000197728) and BAX (ENSG00000087088) through
\texttt{residual.features}; these residuals, obtained from the SCT
assay, were used as the mediator and outcome measurements in the
subsequent analysis. No additional covariates were regressed out
during SCTransform.

Cells were subsequently matched to metadata by barcode and to
genotypes by donor identifier, and the CD4$^+$ subset was defined
using the cell annotations in Table~\ref{tab:cell_annotation_map}.
No additional expression-based outlier filtering was applied after
normalization. All 961 donors were retained after data matching.

\paragraph{Candidate selection.}
The rs1131017--RPS26 exposure--mediator pair was selected
a priori from CD4$^+$ T-cell cross-cohort eQTL results reported
in the distQTL study \citep{Coulter2025distQTL}. Among the two
RPS26-associated variants represented in the candidate list,
rs1131017 showed stronger association evidence in both the
Lifelines Deep Cohort and OneK1K. The corresponding
$\log_{10}p$ values are reproduced in
Table~\ref{tab:app_candidate_selection}. Candidate selection
was completed independently of the fitted BLaFM effect estimates.

\begin{table}[htbp]
\centering
\small
\caption{Cross-cohort association evidence for RPS26 variants
in CD4$^+$ T cells. Entries are $\log_{10}p$ values for the
Lifelines Deep Cohort (LDC) and OneK1K; the OneK1K distQTL
values are null-corrected. Source: author-provided distQTL
cross-cohort results, Table S9 \citep{Coulter2025distQTL}.}
\label{tab:app_candidate_selection}
\begin{tabular}{llrr}
\toprule
Gene&SNP&LDC&OneK1K distQTL\\
\midrule
RPS26&rs1131017&$-16.836$&$-1194.449$\\
RPS26&rs10876864&$-13.672$&$-915.012$\\
\bottomrule
\end{tabular}
\end{table}

RPS26 has Ensembl identifier ENSG00000197728. The rs1131017
association was stronger in both cohorts. BAX was selected
using independent perturbation evidence concerning RPS26
silencing and pro-apoptotic gene expression
\citep{Chen2021RPS26}, rather than from fitted BLaFM effect curves.

\paragraph{Donor-average descriptive benchmark.}
The scalar benchmark was
\begin{equation*}
\bar Y_i
=
\theta_0+\theta_ZZ_i+\theta_X\bar X_i+e_i.
\end{equation*}
The coefficient for donor-average RPS26 expression was
\[
\widehat\theta_X
=
0.045849,
\qquad
95\%\ {\rm CI}
=
(0.028644,0.063054),
\qquad
p=2.086\times10^{-7}.
\]
This regression is descriptive and does not retain cell-state
variation.

\subsection{Selection of the application basis dimension}
\label{supp:app_k_selection}

We applied the conditional plug-in rule in Section~\ref{supp:cv}
to $K=4,\ldots,9$. The same five-fold assignment was used for
all candidates, with cells stratified within donor and coarse
cell-state group. Each fold used 100,000 MCMC iterations,
discarded the first 50,000, and retained every 100th subsequent
draw for predictive scoring, giving 500 scoring draws per fold.
Training and held-out latent states remained fixed at the
candidate-specific full-data posterior means.

The primary criterion averaged out-of-fold scores within each
donor and then across the 961 donors. All 403,087 cells were
included once in this aggregation. The maximal donor-balanced
joint score was attained at $K=7$; $K=8$ ranked second, with
a score difference of 0.091386. The full joint BLaFM
posterior at $K=7$ was used for the primary functional analysis.

\begin{table}[htbp]
\centering\small
\caption{Application basis selection using donor-balanced
conditional plug-in log predictive scores. Each candidate uses
its own full-data posterior mean states. Larger joint
scores are preferred; selection uses unrounded values.}
\label{tab:app_k_selection}
\begin{tabular}{rrrrr}
\toprule
$K$ & LPD$_X$ & LPD$_{Y\mid X}$ & Joint LPD & Rank\\
\midrule
4 & -0.995825 & 0.987652 & -0.008173 & 4\\
5 & -1.014505 & 0.725442 & -0.289064 & 6\\
6 & -1.058319 & 1.006655 & -0.051664 & 5\\
7 & -0.997978 & 1.158285 & \textbf{0.160307} & 1\\
8 & -1.009689 & 1.078610 & 0.068921 & 2\\
9 & -0.963648 & 0.967289 & 0.003640 & 3\\
\bottomrule
\end{tabular}
\end{table}

\subsection{Primary posterior diagnostics and additional functional summaries}
\label{supp:primary_extra}

\paragraph{Genotype-specific population trajectories and intervals.}
For each retained posterior draw $m$ and dosage
$z\in{0,1,2}$, we evaluated the population mean trajectories
\begin{align*}
\bar g_z^{(m)}(t)
&=\bm\phi(t)^\top\bigl(\bm\alpha_0^{(m)}+A^{(m)}z\bigr),
\\
\bar h_z^{(m)}(t)
&=\bm\phi(t)^\top\Bigl\{\bm\beta_0^{(m)}+C^{(m)}z
+B^{(m)}\bigl(\bm\alpha_0^{(m)}+A^{(m)}z\bigr)\Bigr\}.
\end{align*}
The mediated, direct, and total effect curves were calculated
analogously from the draw-specific quantities
$B^{(m)}A^{(m)}$, $C^{(m)}$, and
$B^{(m)}A^{(m)}+C^{(m)}$, respectively. All nonlinear
combinations were formed within posterior draw before
summarization. Posterior mean curves and equal-tailed pointwise
95\% credible intervals were obtained from the empirical
posterior summaries at each state. These intervals
refer to population mean trajectories and effect functions and
do not include additional donor-specific or cell-level predictive
variation; they are pointwise intervals rather than simultaneous
credible bands. They condition on the selected dimension $K=7$.

\paragraph{Primary MCMC diagnostics.}

The selected full $K=7$ application fit used a single MCMC chain
of 100,000 iterations, with the first 50,000 discarded as burn-in.
Table~\ref{tab:app_ess_K7} reports posterior summaries and Monte
Carlo precision diagnostics for the three integrated effects and
the integrated same-state mediator--outcome coupling, based on
the 50,000 retained draws.

For each scalar functional $\theta$, the Monte Carlo standard
error (MCSE) quantifies simulation uncertainty in the estimated
posterior mean. It is related to the autocorrelation-adjusted
effective sample size (ESS) through
\[
\widehat{\operatorname{MCSE}}(\bar\theta)
\simeq
\frac{\widehat{\operatorname{SD}}(\theta\mid\mathcal D)}
     {\sqrt{\widehat{\operatorname{ESS}}(\theta)}}.
\]
For IME, IDE, and ITE, the ESS values were 632.3, 629.3, and
1,173.0, respectively. Their MCSEs were approximately
2.9\%--4.0\% of their posterior standard deviations, indicating
that Monte Carlo error in the estimated means was small relative
to posterior uncertainty.

\begin{table}[htbp]
\centering
\small
\setlength{\tabcolsep}{3pt}
\caption{Posterior summaries and MCMC precision diagnostics for
four integrated functionals from the selected full $K=7$
application fit, based on 50,000 retained draws from one chain.
The first three rows report effects per additional G allele,
integrated uniformly over $[0,1]$; the final row reports the
integral of the same-state mediator--outcome coupling.
SD denotes posterior standard deviation, and CrIs are equal-tailed
95\% credible intervals conditional on the selected dimension.
ESS is the autocorrelation-adjusted effective sample size, and
MCSE is the Monte Carlo standard error of the posterior mean.
These MCSEs are distinct from the simulation-replicate MCSEs
in Section~C.}
\label{tab:app_ess_K7}
\begin{tabular}{lrrcrr}
\toprule
Quantity & Mean & SD & 95\% CrI & ESS & MCSE \\
\midrule
$\IME$
& $-1.458930$ & $0.104724$
& $(-1.662,-1.253)$
& $632.3204$ & $0.004165$ \\
$\IDE$
& $1.344225$ & $0.105520$
& $(1.137,1.549)$
& $629.2633$ & $0.004206$ \\
$\ITE$
& $-0.114700$ & $0.017487$
& $(-0.149,-0.081)$
& $1173.0010$ & $0.000511$ \\
$\int_0^1\gamma(t)\,dt$
& $-6.654941$ & $0.370610$
& $(-7.375,-5.928)$
& $651.1708$ & $0.014523$ \\
\bottomrule
\end{tabular}
\end{table}

\begin{figure}[htbp]
\centering
\suppfig[width=0.90\linewidth,height=0.75\textheight,keepaspectratio]
{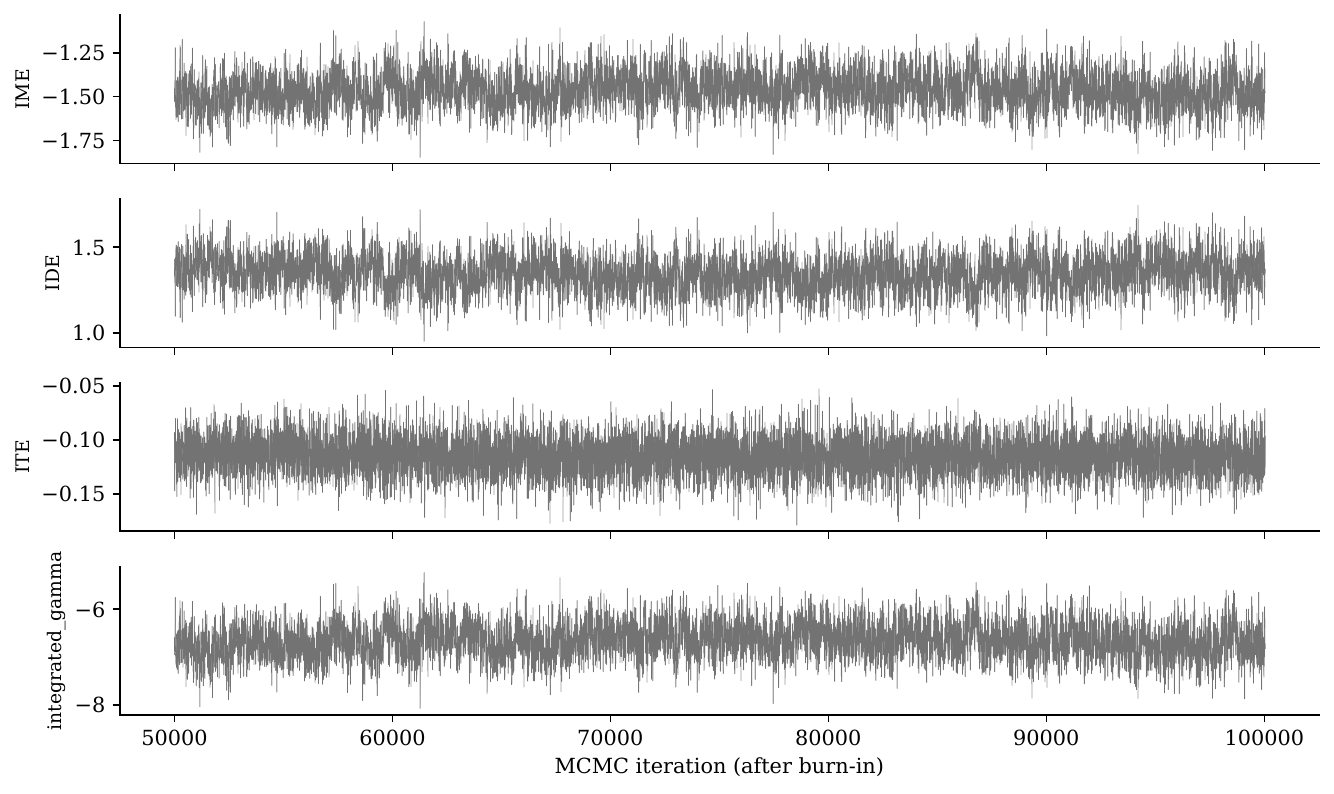}
\caption{Retained MCMC traces of the integrated mediated (IME),
direct (IDE), and total (ITE) effects and the uniform integral
of the same-state mediator--outcome coupling from the selected
full $K=7$ fit. The first 50,000 of 100,000 iterations were discarded;
these panels display the remaining iterations.}
\label{fig:trace_integrated_K7}
\end{figure}

\begin{figure}[htbp]
\centering
\suppfig[width=0.90\linewidth,height=0.75\textheight,keepaspectratio]
{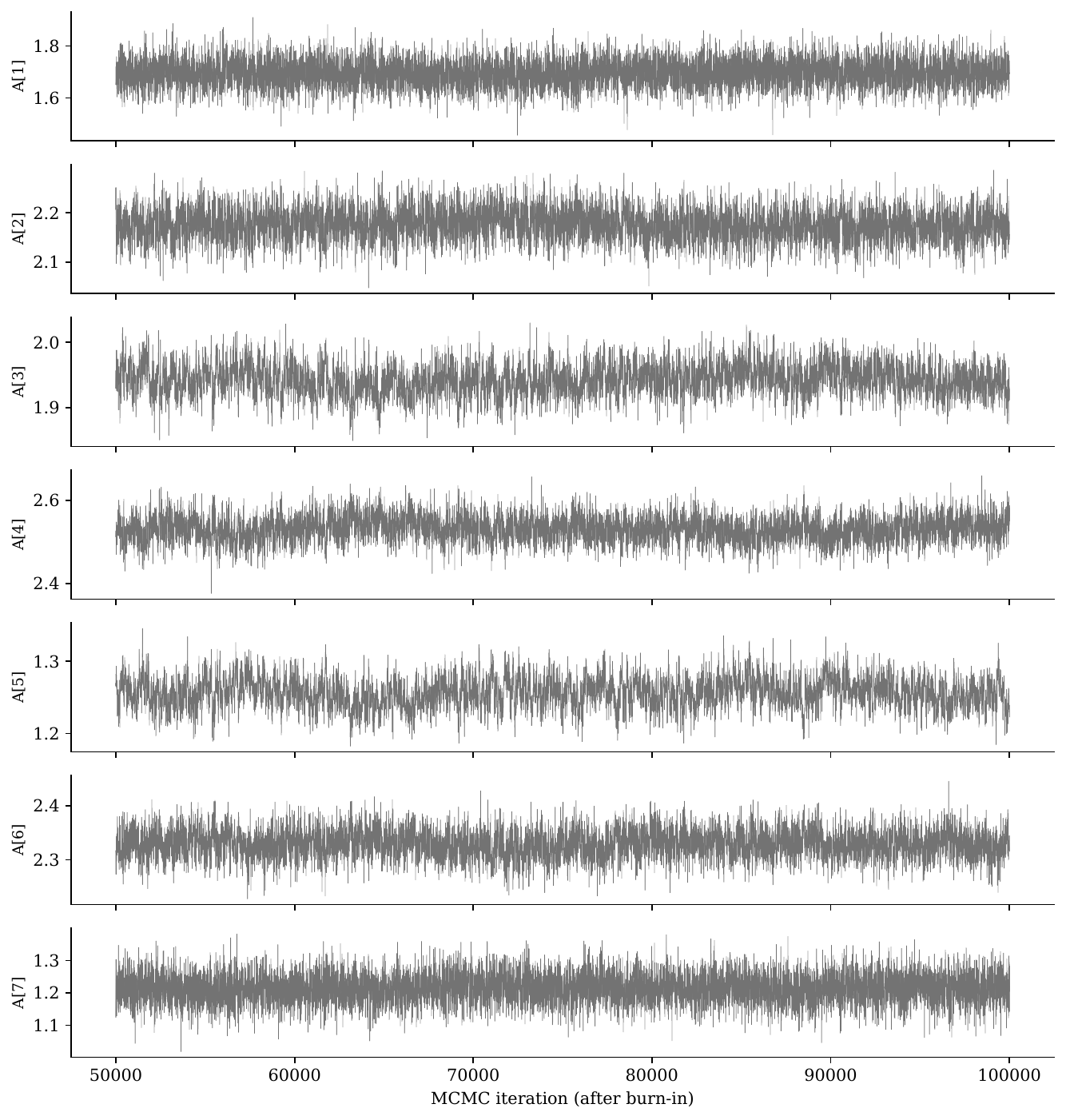}
\caption{Selected full $K=7$ traces for $A$ after
removal of the first 50,000 iterations. Panels correspond to the
seven coefficients. Panels display the 50,000 retained iterations.}
\label{fig:trace_A_K7}
\end{figure}

\begin{figure}[htbp]
\centering
\suppfig[width=0.90\linewidth,height=0.75\textheight,keepaspectratio]
{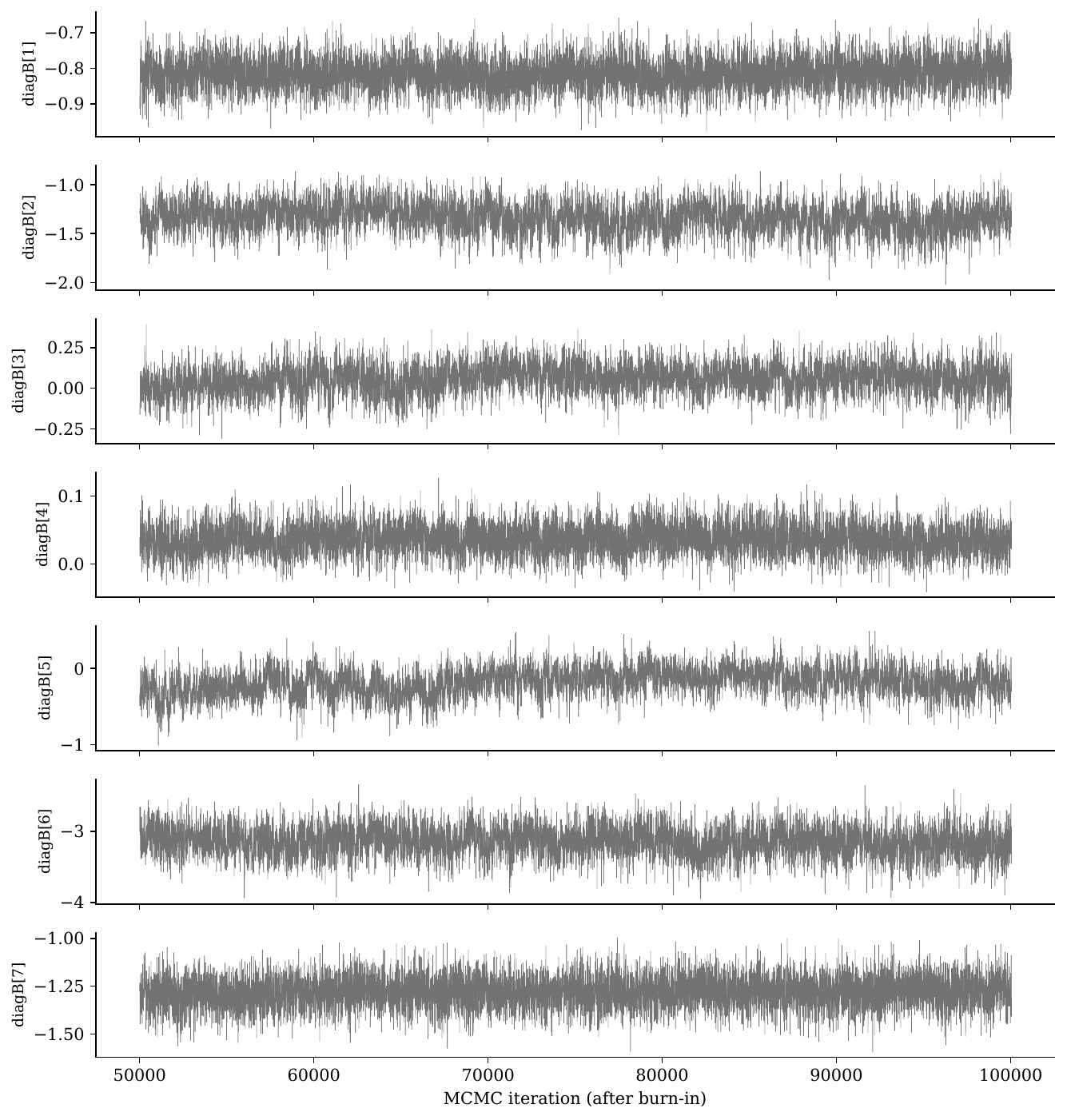}
\caption{Selected full $K=7$ traces for $\diag(B)$ after
removal of the first 50,000 iterations. Panels correspond to the
seven coefficients. Panels display the 50,000 retained iterations.}
\label{fig:trace_B_K7}
\end{figure}

\begin{figure}[htbp]
\centering
\suppfig[width=0.90\linewidth,height=0.75\textheight,keepaspectratio]
{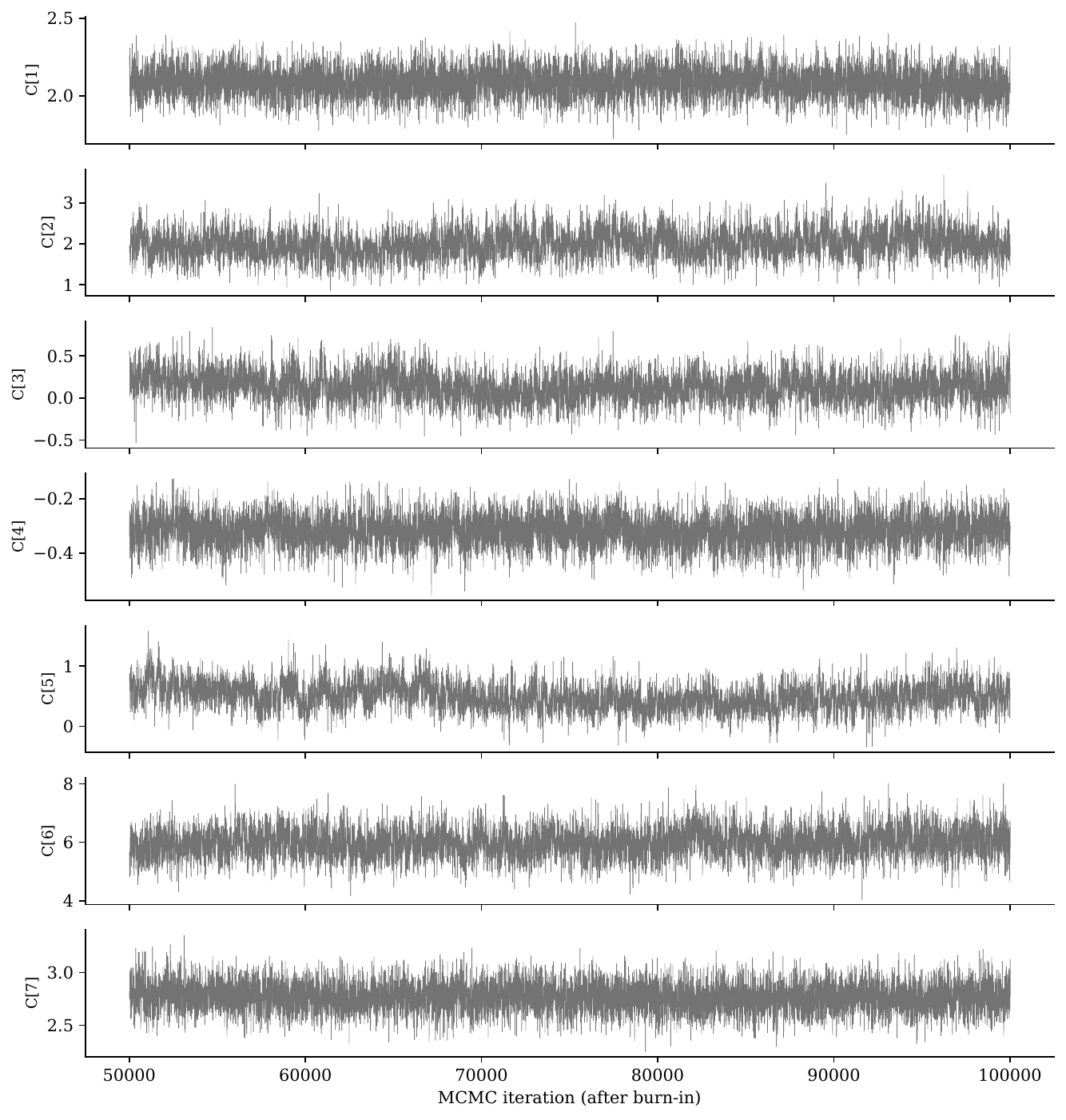}
\caption{Selected full $K=7$ traces for $C$ after
removal of the first 50,000 iterations. Panels correspond to the
seven coefficients. Panels display the 50,000 retained iterations.}
\label{fig:trace_C_K7}
\end{figure}

The trace plots for the integrated functionals and for $A$,
$\diag(B)$, and $C$ are presented in
Figures~\ref{fig:trace_integrated_K7}--\ref{fig:trace_C_K7}.
Across the displayed retained iterations, the traces fluctuate
around comparatively stable levels, without evident sustained
linear drift. This visual assessment and the numerical diagnostics
in Table~\ref{tab:app_ess_K7} do not constitute a between-chain
convergence check, since only one chain is available. In particular,
the numerical diagnostics concern the four listed integrated
functionals, not every spline coefficient or latent state
coordinate.

\clearpage
\paragraph{Implied mediator--outcome regression kernel.}

The function-on-function representation of the coefficient model
is given in Section~2.3 of the main article. For computing kernel
summaries, define the Gram matrix and kernel coefficients by
\begin{equation}
\bm G_\phi=\int_0^1\bm\phi(s)\bm\phi(s)^\top\,ds,
\qquad
\bm D=\bm G_\phi^{-1}B^\top.
\label{eq:D_from_B}
\end{equation}
Then $\zeta(s,t)=\bm\phi(s)^\top\bm D\bm\phi(t)$, and its
same-state slice is
\begin{equation*}
\gamma(t)
=
\zeta(t,t)
=
\bm\phi(t)^\top\bm D\bm\phi(t).
\end{equation*}

For each retained posterior draw,
\[
q^{(m)}
=\int_0^1\gamma^{(m)}(t)\,dt
=\operatorname{tr}\{B^{(m)}\}.
\]
Posterior summaries and MCMC diagnostics for the integrated
coupling were computed from these scalar draws.

This slice summarizes the kernel along matching mediator and outcome
states. It is distinct from the mediated effect $\mathrm{ME}(t)$,
which also depends on the exposure-to-mediator pathway through $A$.

\subsection{Sensitivity to coefficient-level residual dependence}
\label{supp:rho}

\paragraph{Sensitivity model.}
The primary BLaFM model assumes independent donor-level coefficient
residuals $\bm\eta_i$ and $\bm\xi_i$. To assess sensitivity to this
assumption, we introduce a fixed residual-correlation parameter
$\rho_s$ for matching mediator and outcome coefficient components:
\begin{equation*}
\begin{pmatrix}
\eta_{ik}\\
\xi_{ik}
\end{pmatrix}
\sim
N\left[
\begin{pmatrix}0\\0\end{pmatrix},
\begin{pmatrix}
\Sigma_{kk}
&
\rho_s\sqrt{\Sigma_{kk}\Lambda_{kk}}
\\
\rho_s\sqrt{\Sigma_{kk}\Lambda_{kk}}
&
\Lambda_{kk}
\end{pmatrix}
\right],
\qquad k=1,\ldots,K.
\end{equation*}
Here $\rho_s$ represents residual dependence between the mediator
and outcome coefficient processes, rather than between the
cell-level measurement errors. We considered
\[
\rho_s
\in
\{-0.5,-0.3,-0.1,0,0.1,0.3,0.5\}.
\]

For fixed $\rho_s$, define
\[
r_{\alpha,ik}
=
\alpha_{ik}-\alpha_{0k}-a_kZ_i,
\qquad
r_{\beta,ik}
=
\beta_{ik}-\beta_{0k}-c_kZ_i-b_k\alpha_{ik}.
\]
Conditional on the donor-specific trajectory coefficients, the
log-likelihood contribution for component $k$, up to an additive
constant, is
\begin{align}
\ell_k
&=
-\frac n2
\log\{\Sigma_{kk}\Lambda_{kk}(1-\rho_s^2)\}
\notag\\
&\quad
-\frac{1}{2(1-\rho_s^2)}
\sum_{i=1}^n
\left[
\frac{r_{\alpha,ik}^2}{\Sigma_{kk}}
+
\frac{r_{\beta,ik}^2}{\Lambda_{kk}}
-
\frac{
2\rho_s r_{\alpha,ik}r_{\beta,ik}
}{
\sqrt{\Sigma_{kk}\Lambda_{kk}}
}
\right].
\label{eq:rho_loglik}
\end{align}

\paragraph{Conditional residual distribution.}
Define
\[
L_{\rho_s}
=
\diag\left\{
\rho_s\sqrt{\Lambda_{kk}/\Sigma_{kk}}
\right\}.
\]
The joint normal specification implies
\[
\E(\bm\xi_i\mid\bm\eta_i)
=
L_{\rho_s}\bm\eta_i,
\qquad
\Var(\bm\xi_i\mid\bm\eta_i)
=
(1-\rho_s^2)\Lambda.
\]
These identities yield the reparameterization in the main article
and provide the conditional moments used in the updates below.

\paragraph{Conditional refitting strategy.}
For each fixed value of $\rho_s$, the donor-specific trajectory
coefficients are fixed at their posterior means from the full BLaFM fit at the
selected dimension $K=7$,
\[
\bm\alpha_i=\widehat{\bm\alpha}_i,
\qquad
\bm\beta_i=\widehat{\bm\beta}_i.
\]
Conditional on these estimated trajectories, we refit
\[
\bm\alpha_0,\ A,\ \bm\beta_0,\ B,\ C,\ \Sigma,\ \Lambda,\
\tau_a,\ \tau_d
\]
under Equation~\eqref{eq:rho_loglik}. Because
$\Sigma_{kk}$ and $\Lambda_{kk}$ are coupled through
$\rho_s$, each variance pair is updated jointly using a
two-dimensional random-walk Metropolis step on the log scale;
the remaining coefficient-level parameters are updated from
their corresponding conditional distributions. This analysis
therefore quantifies sensitivity of the coefficient-level
decomposition conditional on the donor trajectories estimated
under the primary joint BLaFM fit.

\paragraph{Coefficient-block and variance-block updates.}
For component $k$, write
\[
s_k=\rho_s\sqrt{\Lambda_{kk}/\Sigma_{kk}},
\qquad
v_k=(1-\rho_s^2)\Lambda_{kk}.
\]
Let $W$ have rows $(1,Z_i)$, let $H_k$ have rows
$(1,Z_i,\alpha_{ik})$, and define
\[
\bm a_k=(\alpha_{1k},\ldots,\alpha_{nk})^\top,
\qquad
\bm b_k=(\beta_{1k},\ldots,\beta_{nk})^\top.
\]
With
\[
\bm d_{\alpha k}=(\alpha_{0k},a_k)^\top,
\qquad
\bm d_{\beta k}=(\beta_{0k},c_k,b_k)^\top,
\]
the population-level coefficient blocks have Gaussian full
conditionals with precision matrices and precision-weighted means
\begin{align*}
Q_{\alpha k}
&=
\left(\Sigma_{kk}^{-1}+s_k^2/v_k\right)W^\top W
+
(\tau_a\Sigma_{kk})^{-1}I_2,
\\
q_{\alpha k}
&=
\Sigma_{kk}^{-1}W^\top\bm a_k
-
(s_k/v_k)W^\top
\left(
\bm b_k-H_k\bm d_{\beta k}-s_k\bm a_k
\right),
\\
Q_{\beta k}
&=
v_k^{-1}H_k^\top H_k
+
(\tau_d\Lambda_{kk})^{-1}I_3,
\\
q_{\beta k}
&=
v_k^{-1}H_k^\top
\left\{
\bm b_k-s_k(\bm a_k-W\bm d_{\alpha k})
\right\}.
\end{align*}
Each block is sampled from $N(Q^{-1}q,Q^{-1})$.

For the variance update, let
\[
u_k=\log\Sigma_{kk},
\qquad
w_k=\log\Lambda_{kk}.
\]
A symmetric bivariate normal random-walk proposal was used on
$(u_k,w_k)$, with marginal proposal standard deviations 0.12 and
zero proposal correlation. If
$\ell_k^{\rm post}(\Sigma_{kk},\Lambda_{kk})$ denotes the log
posterior density with respect to the two variances, including their
inverse-gamma priors and the variance-scaled Gaussian coefficient
priors, the transformed target is
\[
\widetilde\ell_k(u_k,w_k)
=
\ell_k^{\rm post}(e^{u_k},e^{w_k})+u_k+w_k.
\]
The Metropolis acceptance probability is
\[
\min\left\{
1,
\exp\left(
\widetilde\ell_k^\ast-\widetilde\ell_k
\right)
\right\}.
\]
The global variance multipliers retain their GIG full conditionals.

\begin{figure}[htbp]
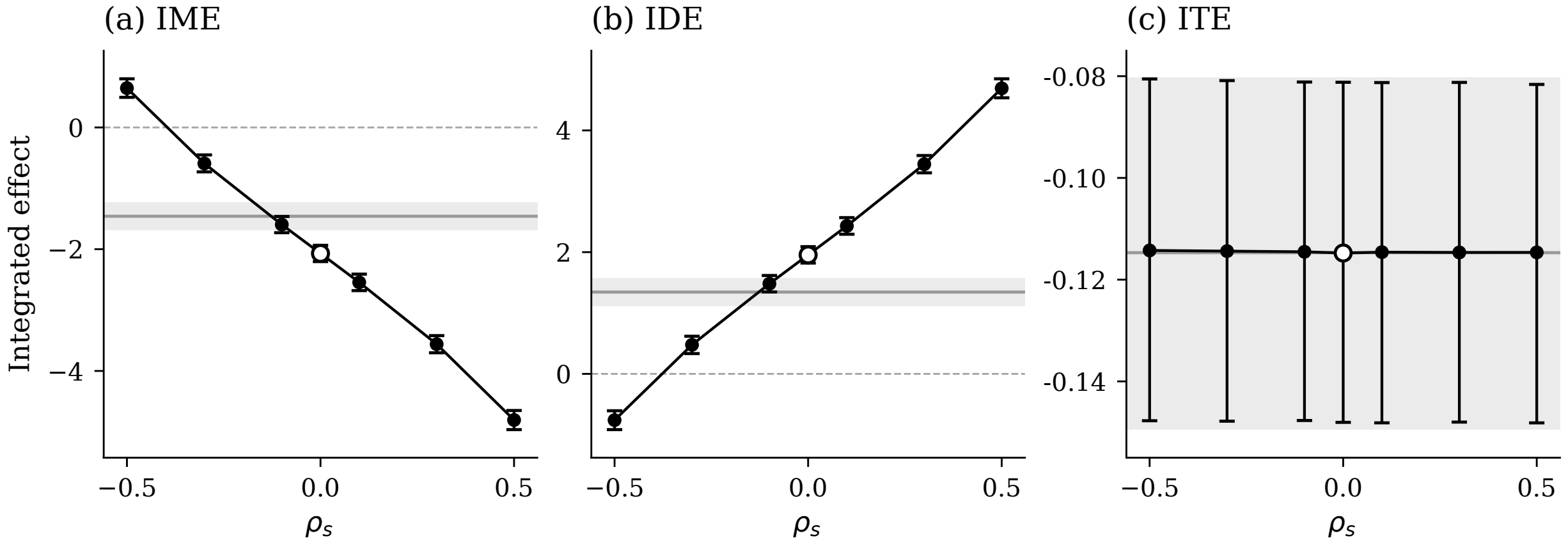

\centering
\suppfig[width=\linewidth]
{figures/rho_sensitivity_integrated_effects_K7.pdf}
\caption{Conditional coefficient-level sensitivity of the
integrated mediated (IME), direct (IDE), and total (ITE) effects at
$K=7$, across fixed $\rho_s\in\{-0.5,-0.3,-0.1,0,0.1,0.3,0.5\}$.
Dots are conditional posterior means and vertical bars are
equal-tailed 95\% credible intervals. Each refit
conditions on the donor trajectories estimated from the full BLaFM fit at the
selected dimension $K=7$; these are not full joint BLaFM posteriors.}
\label{fig:rho_integrated}
\end{figure}

\paragraph{Integrated sensitivity results.}
Figure~\ref{fig:rho_integrated} presents the integrated effect
summaries for the seven fixed residual correlations. The conditional
posterior mean IME decreases as $\rho_s$ increases, whereas IDE
increases. These changes are substantial relative to the total
association and exhibit the compensating behavior described by
the reparameterization in the main article. At the most negative setting,
$\rho_s=-0.5$, the plotted integrated ME and DE even have the
opposite signs from their conditional $\rho_s=0$ counterparts;
the mediated/direct signs are therefore not robust across the
entire sensitivity range. In contrast, the conditional posterior
mean ITE remains close to the same negative value across the
sensitivity grid. The displayed intervals also illustrate greater sensitivity of
the decomposition than of the integrated total effect.

\begin{figure}[htbp]
\centering
\suppfig[width=\linewidth]
{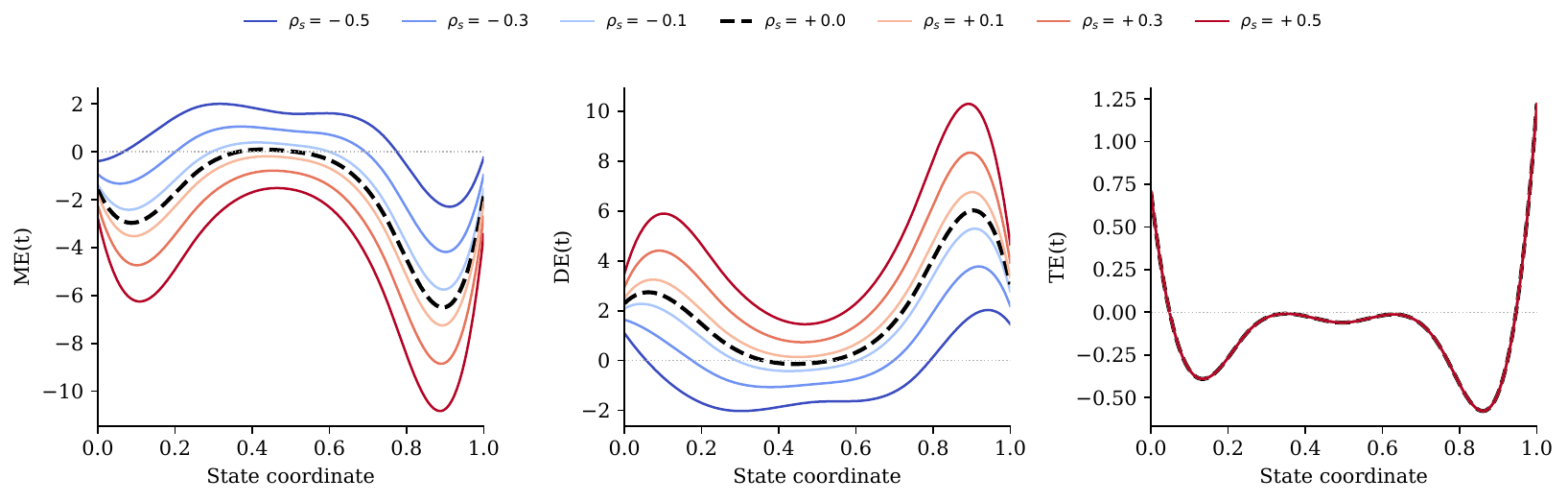}
\caption{Conditional posterior mean mediated, direct, and total
effect curves for
$\rho_s\in\{-0.5,-0.3,-0.1,0,0.1,0.3,0.5\}$.
The thick black dashed curve denotes the conditional $\rho_s=0$ reference.
Sensitivity fits condition on the donor-specific trajectory
coefficients estimated from the full BLaFM fit at the
selected dimension $K=7$.}
\label{fig:rho_curves}
\end{figure}

\paragraph{Pointwise sensitivity results.}
Figure~\ref{fig:rho_curves} shows that changes in $\rho_s$
primarily shift the conditional ME and DE curves in opposite
directions. For small positive $\rho_s$, the ME curve is more
negative and the DE curve more positive than at $\rho_s=0$;
negative $\rho_s$ produces the reverse pattern. The curve shapes
remain nonconstant: pronounced deviations occur in the lower and
upper regions of the state coordinate. The TE curves are nearly superimposed
throughout the interval, even where the component curves are well
separated. The pointwise comparison therefore corroborates the
integrated total-effect stability visible in
Figure~\ref{fig:rho_integrated}. Because the component paths can
cross zero, this does not imply pointwise sign invariance of
ME or DE.

\paragraph{Conditional reference and primary analysis.}
The $\rho_s=0$ conditional refit is the appropriate internal
reference for comparisons across the fixed sensitivity grid.
It is not identical to the full joint BLaFM posterior at $K=7$, since the latter
jointly updates the latent states and trajectory coefficients,
whereas all sensitivity refits hold the donor coefficients at their
estimated values. Thus differences between the full BLaFM estimates
in Table~\ref{tab:app_ess_K7} and the conditional
$\rho_s=0$ figure mix the effects of conditioning and posterior
refitting; they should not be attributed to nonzero residual
correlation. The conspicuous shifts in IME and IDE under varying
$\rho_s$ indicate that allocation between pathways is sensitive
to the residual-dependence assumption, even while ITE is stable
in these conditional refits.

\begin{figure}[htbp]
\centering
\suppfig[width=\linewidth]
{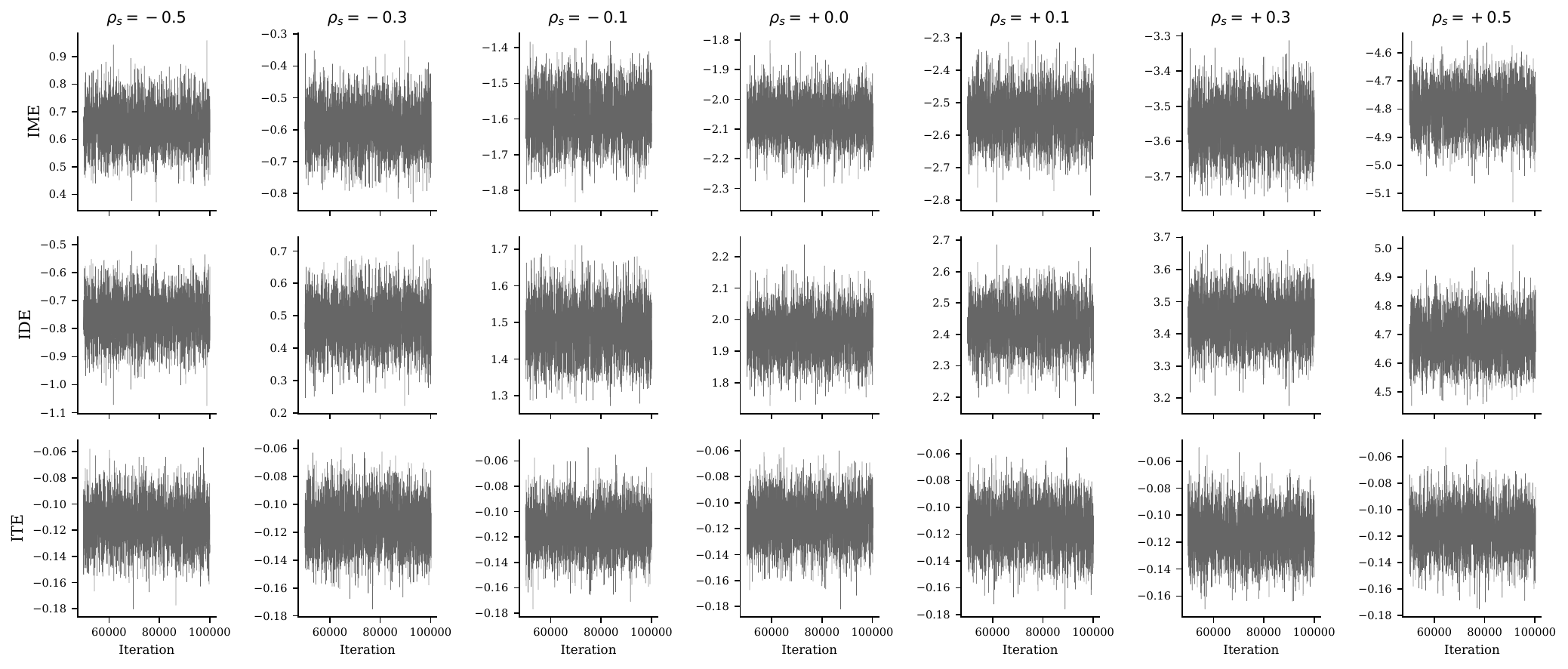}
\caption{Post-burn-in conditional MCMC traces for integrated IME,
IDE, and ITE across the seven fixed residual correlations $\rho_s$.
Each panel contains retained iterations 50,001--100,000 from the
corresponding coefficient-level sensitivity refit. The rows show
the three integrated effects and the columns the correlation values.}
\label{fig:rho_traces}
\end{figure}

\paragraph{MCMC diagnostics for the sensitivity fits.}
The conditional $K=7$ refits used 100,000 iterations and discarded
the first 50,000. Figure~\ref{fig:rho_traces} shows the retained
IME, IDE, and ITE traces separately for each fixed $\rho_s$.
The conditional traces fluctuate around levels consistent with the
integrated sensitivity plot, without an obvious systematic drift
in the displayed intervals.

\end{document}